\newcommand{\mr}{\mathrm}

\documentclass[12pt,a4paper,twoside]{report}
\usepackage{graphicx}
\usepackage{lscape}
\usepackage[bf]{caption}
\usepackage{subfigure}
\usepackage{amssymb}
\usepackage{amsmath}
\usepackage{float}
\usepackage{fancyhdr}
\usepackage{longtable}
\usepackage{epsfig}
\usepackage{color}
\usepackage{graphicx}
\usepackage{url}
\usepackage{epstopdf}
\usepackage{amsmath}
\usepackage{natbib}
\usepackage{wrapfig}
\usepackage{sidecap}
\usepackage{rotating}
\usepackage[nottoc]{tocbibind}
\usepackage{breakcites}

\renewcommand{\vec}[1]{\mathbf{#1}}
\fancypagestyle{plain}{%
   \fancyhead{} %
}

\newcommand{\be}{\begin{equation}}
\newcommand{\ee}{\end{equation}}
\newcommand{\ba}{\begin{eqnarray}}
\newcommand{\ea}{\end{eqnarray}}

\newcommand{\apjl}{Astrophysical Journal, Letters}
\newcommand{\apjs}{Astrophysical Journal, Supplement}
\newcommand{\ao}{Applied Optics}

\newcommand{\aap}{Astronomy and Astrophysics}

\renewcommand{\vec}[1]{\ensuremath{\underline{#1}}}	

\newcommand{\ppt}[2]					
	{\frac{\partial^{2}\! #1}{\partial#2^{2}}}
\newcommand{\ppm}[3]						
	{\frac{\partial^{2}\! #1}{\partial#2\partial#3}}	

\newcommand{\grumpy}				
{\mbox{\begin{picture}(20,20)(0,7)
\put(10,10){\circle{15}}
\put(12.5,12.5){\circle*{2}}
\put(7.5,12.5){\circle*{2}}
\put(6,1){\shortstack[l]{$\mbox{}^{\frown}$}}
\end{picture}}}

\newcommand{\happy}				
{\mbox{\begin{picture}(20,20)(0,7)
\put(10,10){\circle{15}}
\put(12.5,12.5){\circle*{2}}
\put(7.5,12.5){\circle*{2}}
\put(6,1){\shortstack[l]{$\mbox{}^{\smile}$}}
\end{picture}}}

\begin{document}
%

%
%

\begin{titlepage}

\begin{center}

\vspace*{2cm}

\huge{\textbf{Electron Beam Evolution and Radio Emission in the Inhomogeneous Solar Corona}}

\vspace{1.2cm}
\large{\textbf{Heather Ratcliffe, BSc, MMathPhys}}
\vspace{1.2cm}

\small Astronomy and Astrophysics Group\\
Department of Physics and Astronomy\\
Kelvin Building\\
University of Glasgow\\
Glasgow, G12 8QQ\\
Scotland, U.K.
\setlength{\belowcaptionskip}{0.cm}
\vspace{1.cm}
   \begin{center}
   \centering
     \includegraphics[width=0.5\textwidth]{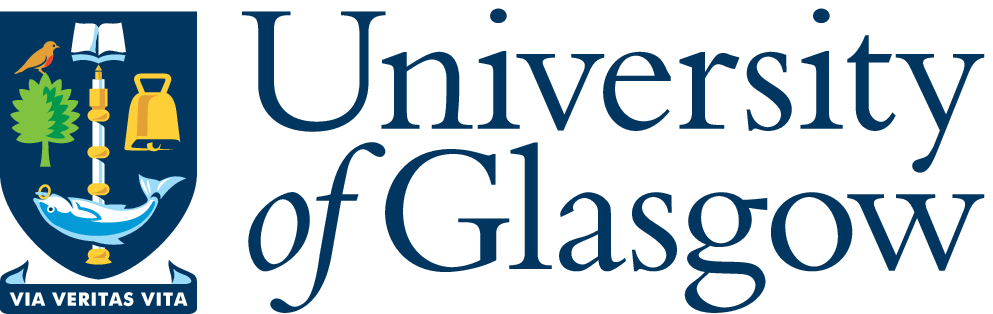}
   \end{center}

\large        Presented for the degree of\\
        Doctor of Philosophy\\
        The University of Glasgow\\
        March 2013

\end{center}
\setlength{\belowcaptionskip}{0.3cm}
\end{titlepage}


\cleardoublepage

\vspace*{8cm}
\emph{For my parents, who surely started it all by naming me after an astronomer. }


%

\thispagestyle{empty}

\newpage\markboth{}{}

\vspace*{4cm}
\begin{flushright}
\parbox{130mm}{
\hrulefill

This thesis is my own composition except where indicated in
the text. No part of this thesis has been submitted elsewhere for any other degree
or qualification.
\vspace*{1cm}

{\bf Copyright \copyright ~2013 by Heather Ratcliffe}
\vspace*{0.4cm}

27th March 2013

\hrulefill  }
\end{flushright}

\cleardoublepage
\phantomsection

\pagenumbering{roman}

\tableofcontents
\listoffigures

\thispagestyle{empty}

\chapter*{Abstract}

\addcontentsline{toc}{chapter}{Abstract}

This thesis considers the propagation of accelerated electron beams in plasma. We consider the wave particle interactions these undergo which cause their evolution, the effects of plasma density inhomogeneities on these interactions, and the effects this may have on the production of hard X-ray and radio emission by the beam. 

\emph{CGS units are used throughout.}

Chapter \ref{ref:Chapter1} introduces the important background material on the Sun and solar flares, and some basic plasma physics. We discuss the acceleration and propagation of electrons beams and their production of hard X-ray emission, and the various observed types of radio emission from the Sun. We end by discussing details of the mechanism by which radio emission can be produced by beam generated Langmuir waves at GHz frequencies.

Chapter \ref{ref:Chapter2} contains the mathematical derivation of the effects of plasma density fluctuations on Langmuir waves. This is found to be described by a diffusion of the waves in wavenumber space. We consider the situation in both one and three dimensions, for elastic and inelastic scattering of the Langmuir waves, discussing how our model expands on that previously considered in the literature, and develop a model for the fluctuations applicable to the electron beams we consider in this thesis. We derive the relevant diffusion coefficients for a few commonly observed density fluctuation spectra, then end with a brief discussion of the expected effects of the Langmuir wavenumber diffusion on the waves and electrons for a few representative cases. 

Chapter \ref{ref:Chapter3} uses the model derived in Chapter \ref{ref:Chapter2} in quasi-linear simulations of electron beam evolution. We consider two initial electron beam distributions, either a Maxwellian or a power law, and simulate the Langmuir wave generation and evolution, and the back-reaction of this on the electron beam. We find an electron acceleration effect to occur, and explore the parameters for which this is strongest. In addition we consider the production of hard X-ray emission from an initially power law beam, and the effects of the electron acceleration on this.

Chapter \ref{ref:Chapter4} considers the radio emission from an electron beam via the generation of Langmuir waves. We first derive an angle-averaged model for emission at the second harmonic of the plasma frequency, and combine this with the simulations from the previous chapters. We include the effects of density fluctuations on the Langmuir waves, and discuss how this affects the radio emission produced. 

Chapter \ref{ref:Chapter5} concludes the thesis with a summary of the effects of density inhomogeneity on Langmuir waves, and consequently on fast electron beams and their hard X-ray and radio emissions in the solar corona. 

Appendix \ref{ref:App1} contains the derivation of a mathematical model for radio emission from an electron beam at the fundamental of the local plasma frequency, which is unimportant in the parameter ranges considered in Chapter \ref{ref:Chapter4}, but essential for radio bursts in the higher corona and solar wind. 

\thispagestyle{empty}


\chapter*{Acknowledgements}

\addcontentsline{toc}{chapter}{Acknowledgements}

I gratefully acknowledge an STFC Studentship, and the University of Glasgow for the opportunity to undertake this PhD, and the Astro department for suggesting a project which has taken me to some great places and introduced me to great people. I also thank our group secretary, Ms Rachael McLauchlan, for invaluable help with travel, finances and admin.

To my supervisor Eduard, and to Nic Bian I shall be eternally grateful, not only for the in-depth discussions of physics both in the office and the pub, but for supererogatory support and guidance. 

Significant parts of the Introduction were composed during the ISSI international team meeting on \emph{X-ray and Radio Diagnostics of Energetic Electrons in Solar Flares} led by E.P. Kontar, and I thank all of the team members, and a few in particular, for the enlightening discussions.  

On a personal note, I am indebted to all the various friends and colleagues, particularly the denizens of 604, who listened to a lot of bad jokes, and answered many stupid questions. I owe much to the Stevenson building for the powerlifting, and Colin, Gail, Michele and Marina for the climbing and running that have kept me from going completely mad. 

Finally, I thank again my supervisor as this has certainly been an E.P.{\scriptsize i.}K few years. 

%
\vspace{-8cm}
\chapter*{Table of Symbols}
\addcontentsline{toc}{chapter}{Table of Symbols}
\renewcommand{\vec}[1]{\mathbf{#1}}
\vspace{-2cm}
\begin{table}[thp]
 \begin{tabular}{|l|c|}\hline 
$\omega_{pe}$&(Electron) plasma frequency (Equation \ref{Eq:omegaPe})
\\$\Omega_{ce}$&(Electron) cyclotron frequency (Equation \ref{Eq:omegaCE})
\\$\lambda_{De}$&Debye length 
\\$k_{De}=1/\lambda_{De}$&Debye wavenumber\\
$f(\mathrm{v}, t), f(\vec{v}, t)$& Electron distribution function (1D and 3D) (e.g. Equation \ref{Eq:fNorm})\\
$W(k, t), W(\vec{k}, t)$&Wave spectral energy density (1D and 3D) (e.g. Equation \ref{Eq:wNorm})
\\
$\omega$, $\Omega$ &Frequency   \\
$k, \vec{k}, (k, \theta, \phi)$ & Langmuir wavenumber, wavevector and \\ & wavevector in spherical polars (Chapter \ref{ref:Chapter2})\\
$q, \vec{q}, (q, \bar{\theta}, \bar{\phi})$ & Density fluctuation wavenumber, wavevector and \\&wavevector in spherical polars (Chapter \ref{ref:Chapter2})\\
$\mathcal{FT}[X]$ & Fourier transform of $X$ \\
$F(x,t), F(q, \Omega)$ & ``Force'' on waves (momentum rate of change) (Equation \ref{eqn:WKB1}) \\
$\tilde{n}(x, t), \tilde{n}(q, \Omega)$& Fractional density fluctuation (Equation \ref{Eq:nTilde})\\
$\langle X X \rangle$ & Autocorrelation of $X$ in real or Fourier space \\
$S_X(x ,t), S_X(q, \Omega)$ &Spectrum of $X$ in real or Fourier space (Equation \ref{Eqn:Wiener})\\
$D(k), D_{ij}(\vec{k})$& Diffusion coefficient (1D and 3D) (Equations \ref{eqn:diff_1d} and \ref{eqn:genDiff})
\\$\sqrt{\langle\tilde{n}^2\rangle}$& RMS mean of density fluctuations (Equation \ref{Eq:RMS})\\
$q_0, \Omega_0, \mathrm{v}_0:=\Omega_0/q_0$&Characteristic wavenumber, frequency and \\
&velocity of density fluctuations

\\\hline
\end{tabular}
\end{table}
\pagebreak

\pagenumbering{arabic} 
\chapter{Introduction}
\label{ref:Chapter1}
\newcommand{\rv}{{\mathrm v}}
\newcommand{\rd}{{\mathrm d}}
\newcommand{\hilight}[1]{\colorbox{green}{#1}}
In this chapter, we first outline the structure of the Sun, focusing on the solar corona and solar wind, and give a brief description of solar flares. We then introduce some basic concepts and definitions from plasma physics which are necessary to model these phenomena.  We summarise previous theoretical and simulation work on the propagation and evolution of fast electron beams due to wave generation which is the topic of Chapters \ref{ref:Chapter2} and \ref{ref:Chapter3} of this thesis, then discuss the radiation which fast electrons produce in the corona and solar wind, in particular hard X-rays and radio emission. The general nature of flare associated radio emission is summarised, followed by a more detailed description of the emission from beams propagating in the corona and solar wind, and a description of the observations and emission mechanism behind beam-generated radio emission at high frequencies, which forms the focus of Chapter \ref{ref:Chapter4}.

\section{The Sun}
\subsection{Structure}
The visible surface of the Sun is termed the photosphere, and is relatively cool, with an average temperature of 5600~K. Above the photosphere is the chromosphere, a plasma composed mostly of hydrogen and some helium, with an average temperature of around $10^4$~K, rapidly varying density (from approximately $10^{17}$~cm$^{-3}$ to $10^{11}$~cm$^{-3}$) and varying ionisation. It is approximately 2000~km thick, and named for its appearance as a red flash at the beginning and end of solar eclipses, due to the strong H$\alpha$ emission line in its spectrum. Between the chromosphere and the overlying corona is the very narrow transition region, less than 300~km thick, over which the density drops by several orders of magnitude, while the temperature rises to over $10^6$~K. This temperature rise is still not fully understood, and the coronal and chromospheric heating problem is the topic of much speculation. The temperature and density of the corona mean it is almost completely ionised.

The corona is in a constant state of outflow, forming the solar wind. Open magnetic field lines, dragged by the solar rotation, spiral outwards in a pattern known as the Parker Spiral. The solar wind has two components: fast, with velocity around $800$km~s$^{-1}$, arising from regions of open magnetic field, where the field lines extend into space; and slow, with velocity around $400$km~s$^{-1}$, arising from regions of closed magnetic field, which only connect to the heliosphere at much higher altitudes. The influence of the outflowing solar wind extends to the Earth and beyond, eventually terminating at the heliopause. 

\subsection{Coronal Density Profiles}
A simple model for the coronal density profile may be derived by considering the case of hydrostatic equilibrium, in which the gas pressure is balanced by the gravitational force, and the whole system is unchanging in time. The resulting density profile as a function of height can be easily derived \citep[e.g.][]{aschwanden2006physics} and leads to an exponential profile $n(x) = n_0 \exp(-x/H)$ where $x$ is radial distance measured from the reference point at density $n_0$, and $H$ is the scale height, typically around $10^9$~cm in the corona. 

\begin{figure}
\begin{center}
 \includegraphics[width=10cm]{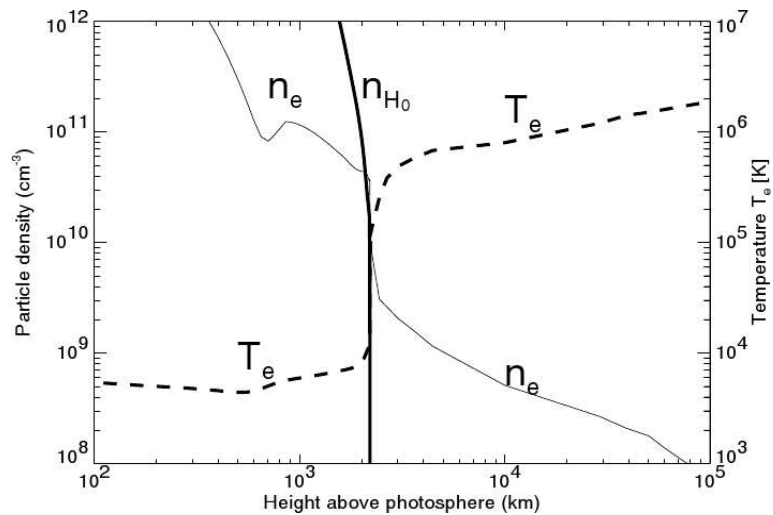}
\caption[Temperature and density profiles of the chromosphere, transition region and low corona.]{Simulated temperature and density profiles of the chromosphere, transition region and low corona. From \citet{aschwanden2006physics}.}\label{fig:TempDens}
\end{center}
\end{figure}

Measurements of the density have been performed using several methods. The radio bursts know as Type III and described below (Section \ref{sec:TIIIs}) are one, as their frequency traces the plasma frequency, and hence the electron density. They are observed over a wide frequency, and therefore density range \citep[e.g.][]{1985srph.book.....M}, and so can be used to find the plasma density from the corona to the Earth and beyond \citep[e.g][]{1998SoPh..183..165L}. White light radiating from the corona, resulting from Thomson scattering of photospheric radiation, and EUV or soft X-ray observations can give line-of-sight integrated densities, while certain emission lines can return an absolute density in some regions \citep{aschwanden2006physics}. 

Figure \ref{fig:TempDens} shows a model of the temperature and density of the chromosphere, transition region and low corona. The electron density is roughly exponential between $2\times 10^3$ and $10^5$~km height. Various models exist, from the original exponential profile of \citet{1961ApJ...133..983N} to several empirical power-law fits, such as the Baumbach-Allen \citep{1947MNRAS.107..426A} or Saito \citep{1977SoPh...55..121S} models. In addition to this smooth gradation, there may be large scale fibrous structuring, as the low plasma-beta implies transport across field lines is inhibited; and also discrete structures. 

\subsection{The Solar Wind}
While the hydrostatic approximation gives a reasonable density estimate for the quiet corona, it was realised by \citet{1958ApJ...128..664P} that the corona could not be in equilibrium, and must instead be constantly streaming outwards. For an isothermal ideal gas, assuming radial expansion (expansion factor proportional to distance squared) the hydrodynamic equations can be solved \citep[e.g.][]{1958ApJ...128..664P,1999A&A...348..614M}, and imply outwards flows of several hundred km/s, in good agreement with the observed solar wind speeds. The solar wind model of \citet{1958ApJ...128..664P} also gives rise to the Parker spiral, as the Sun's rotation drags the open field lines into a spiral. The distance from the Sun to the Earth along a magnetic field line is therefore approximately 1.2~AU. 

\subsection{Density Fluctuations}\label{sec:flucs}
Both the corona and the solar wind show local inhomogeneities on a variety of scales. Interplanetary scintillation (IPS) measurements, where a compact radio source is seen to ``twinkle'' due to density fluctuations along the line of sight, have been used to measure these in the solar wind \citep[e.g.][]{1972ApJ...171L.101C, 1979ApJ...233..998S} and closer to the Sun \citep[e.g.][]{1989ApJ...337.1023C,1995GeoRL..22..329W}. The density dependence of radio-wave propagation velocity in plasma was used by \citet{1983A&A...126..293C} to find the density between two closely spaced satellites in the solar wind and its variations over time, again giving the spectrum of density fluctuations. In general, fractional density fluctuations of up to around $10^{-3}$ have been observed at scales down to $100$~km, which will be considered in this thesis.

\subsection{Solar Activity}
Solar activity follows an 11 year cycle, with the next maximum predicted for late 2013. Sunspots, cooler darker areas of the photosphere with strong magnetic fields, begin to occur after the minimum of the cycle, drifting to smaller latitudes as the cycle proceeds. Every cycle, the polarity of the global magnetic field of the Sun reverses.

Solar flares occur in sunspot groups because of their association with complex magnetic field structure. Flares are transient brightenings, lasting from tens of minutes to a few hours, and releasing enormous amounts of energy, from $10^{28}$ to $10^{32}$ ergs or more depending on their size. Historically, coronal mass ejections (CMEs), large clouds of plasma released from the corona, were considered to be a separate phenomenon from flares, but more recently their interconnectedness has become clear. The term Solar Eruptive Event has been coined to tie together both of these, and also the energetic particles that can result from flare energy releases. Electromagnetic radiation, CMEs and accelerated particles each carry significant fractions of the released energy. 

The standard flare classification uses the emission as measured by GOES (the Geostationary Orbiting Environmental Satellite). The total flux in soft X-Rays (wavelengths between 1 and 8 \AA) is used, and the classes run from A through C then M, X, X10, each with a flux 10 times that of the previous class. In general the occurrence rate decreases as the class increases, with X-class flares being relatively uncommon, while smaller flares are observed far more often. 

The electromagnetic emission produced in flares covers the entire spectrum, from radio waves through the visible to X-Rays and gamma rays. The bulk of this emission is in visible and UV wavelengths \citep{2006JGRA..11110S14W}.  Radio emission and hard X-ray emissions are produced by fast electrons, while soft X-rays may be produced during the rise of a flare due to plasma heating, although this ``pre-heating'' is not seen in all events.

\section{Essential Plasma Physics}
Plasmas are the fourth and commonest state of ordinary matter in the universe, and are characterised by collective effects over large spatial scales, due to the ionisation of the constituent atoms. We consider here fully ionised, single species plasma, consisting of protons and electrons. This may be described using two temperatures, the \emph{electron} and the \emph{ion temperatures} $T_e, T_i$ respectively. From these temperatures we obtain two thermal speeds, $\rv_{Te}$ and $\rv_{Ti}$, and the \emph{sound speed} $\rv_s$ given by \begin{equation}\rv_{Te}=\sqrt{\frac{k_B T_e}{m_e}}, \rv_{Ti}=\sqrt{\frac{k_B T_i}{m_p}},\rv_{s}=\sqrt{\frac{k_B T_e(1+3 T_i/T_e)}{m_p}} \end{equation} where $k_B$ is Boltzmann's constant and $m_e, m_p$ are the masses of an electron and a proton respectively. 

The characteristic frequency of electron oscillations is given by the \emph{(electron) plasma frequency} $\omega_{pe}$, and that for the ions by the \emph{ion plasma frequency} $\omega_{pi}$ which are \begin{equation}\label{Eq:omegaPe}\omega_{pe}=\sqrt{\frac{4\pi e^2 n_e}{m_e}}\;,\;\omega_{pi}=\sqrt{\frac{4\pi e^2 n_i}{m_i}}.\end{equation} Here $e$ is the electronic charge and $n_e, n_i$ are the (number) density of electrons and ions respectively. In magnetised plasma, electrons gyrate around the magnetic field lines at a frequency given by the \emph{electron cyclotron frequency}
\begin{equation}\label{Eq:omegaCE}\Omega_{ce}=\frac{eB}{m_e c} \end{equation} where $B$ is the magnetic field strength, and $c$ the speed of light. Finally, the \emph{(electron) Debye wavelength} $\lambda_{De}=\rv_{Te}/\omega_{pe}$ or its reciprocal, the \emph{Debye wavenumber} $k_{De}=\omega_{pe}/\rv_{Te}$, are important as they give the maximum spatial scale over which charge separation can occur. 

\subsection{Langmuir Waves}    
Plasmas display \emph{quasineutrality}, meaning there can be no charge separation on scales larger than the Debye length, due to the significant restoring force of the electric field. Because the ions are much more massive than the electrons, the electrons can oscillate while the ions remain stationary, giving the stationary version of a plasma oscillation, or Langmuir wave, which has a frequency of $\omega_{pe}$. The travelling version, with non-zero wavenumber $0 < k \le k_{De}$, is ubiquitous in plasma, giving its alternative name of ``plasma wave''. These are weakly dispersive, with $\omega\simeq \omega_{pe}+3 \rv_{Te}^2 k^2/(2 \omega_{pe})$, and so have frequencies very close to the plasma frequency. Plasma waves at very different wavenumbers can therefore have very similar energies. The group velocity of Langmuir waves is ${\mr v}_g=3{\mr v}_{Te}^2 k/\omega_{pe}$, which is generally far less than the electron thermal speed, so for many purposes their propagation can be neglected. 

\subsection{Magnetoionic Modes}\label{sec:MagIModes}
In plasma with an ambient magnetic field, electromagnetic waves have two possible modes of circular polarisation, depending on whether the electric field vector rotates in the same direction as an electron spirals around the field lines (O-mode), or the opposite (X-mode) \citep[e.g.][]{ratcliffe1959magneto}. Unpolarised radiation is the sum of these two circular polarisation states. The O-mode propagates in plasma only above $\omega_{pe}$, and has dispersion relation $\omega^2=\omega_{pe}^2 + c^2 k^2$, identical to that of an electromagnetic wave in unmagnetised plasma, while the X-mode cutoff is at $\omega_X=\omega_{pe}+\Omega_{ce}/2$ and the dispersion relation is \begin{equation}\frac{c^2k^2}{\omega^2}= 1- \frac{\omega_{pe}^2}{\omega^2}\frac{\omega^2-\omega_{pe}^2}{\omega^2 -\omega_{pe}^2-\Omega_{ce}^2}.\end{equation} Because of the different group velocities, initially unpolarised radiation can acquire a non-zero polarisation entirely due to propagation, as waves in one mode arrive before the other.

In magnetised plasma, the Langmuir wave dispersion relation becomes \begin{equation}\omega^2\simeq \omega_{pe}^2 + 3k^2 \rv_{Te}^2 +\Omega_{ce}^2 \sin^2\theta \end{equation} where $\theta$ is the angle between $k$ and the magnetic field. These can be considered as modified Z-mode waves \citep[e.g.][]{1976SoPh...46..511M}, which are the lower-frequency ($\omega < \omega_X$) branch of the X-mode waves just discussed.

\subsection{Plasma Modelling}\label{sec:PlasTh}
The self-consistent modelling of a plasma is a rather difficult problem. Moving charged particles generate currents and thus electromagnetic fields, which in turn act upon the particles, so the motions of the particles and evolution of the fields must be considered in parallel. Further complications arise due to collective effects over large distances, which characterise the plasma state. Ignoring self-consistency and using a single-particle description, i.e. following the motion of a ``test particle'' in prescribed fields is helpful in some cases, for example when deriving the trajectory of a particle in an external magnetic field. In general however, we cannot follow the exact motion of each particle, and must instead use a statistical description. This is the basis of plasma kinetic theory. 

The problem addressed in this work of an electron beam in unmagnetised plasma is well described by quasilinear theory \citep[e.g.][]{drummond1964nucl, vedenov1962quasi}, which considers the particle and wave distributions in the limit where the background plasma contains far more energy than the beam. Non-linear effects may be easily included, and in 1-D these equations are straightforward to simulate. Such simulations may be extended to large particle numbers and large spatial and time scales without requiring excessive amounts of computing power, and so we use this approach here. 

On the other hand, particle-in-cell (PIC) simulations address the problem of self-consistency directly, following the motion of the particles in electromagnetic fields, and updating the fields according to the particle motions. Macroscopic quantities, such as the electric and magnetic field and the plasma density, are calculated on a discrete grid, while the particles are tracked individually as they move through this grid. Such methods are very demanding computationally, especially for large space or time scales or large numbers of particles. However as will be mentioned in Section \ref{Sec:angdiff} they are useful as confirmation of the effects seen using our simulation method. 

\section{Reconnection and Particle Acceleration}\label{sec:accIntro}
\subsection{Acceleration Mechanisms}
The reconfiguration of magnetic field to a lower energy state, which may be facilitated by magnetic reconnection, is the source of the vast amounts of energy released during solar flares, with perhaps $20\%$ of the free magnetic energy being released \citep{2004JGRA..10910104E}. This energy leads to particle acceleration, either in the energy release region itself, or its vicinity. 

DC electric field acceleration can occur in, for example, current sheets. Wave turbulence, on the other hand, can also lead to acceleration due to the oscillating electric fields of the waves. Wave particle interactions, such as those discussed below, can transfer energy to and from particles, and at particular velocities the particles can experience a net energy gain \citep{1992ApJ...398..350H}. The interactions are resonant, and so high frequency waves such as Langmuir waves or whistlers are required to accelerate electrons.

Shock acceleration may be either first order Fermi acceleration, usually requiring multiple shock crossings to reach the high energies observed \citep{1980SSRv...26..157T}, or second order Fermi, if the particles are reflected by stochastically moving magnetic mirrors, and statistically more likely to gain energy than lose it. This is important in some contexts, for example in association with the shocks preceding CMEs and associated with radio bursts \citep[e.g.][]{Mann_Classen}. 

\subsection{Acceleration Region Diagnostics}
The acceleration region lies at a height of ten to a few tens of Mm above the photosphere, in the cusp of the flaring loop in the corona. Radio and HXR emission from accelerated electrons provides several methods by which to derive this height and other parameters of the acceleration region \citep[reviewed by e.g.][]{2002SSRv..101....1A}. 

\subsubsection{Electron Time of Flight Inferences}
For example, electrons at a range of velocities, simultaneously injected into the loop from the acceleration region, will reach the footpoints of the loop at different times. Assuming no time dependence of the injection, the spectrum of the HXR emission produced at the footpoints (discussed in the next section) can be used to infer the emitting electron distribution, and therefore the arrival times of electrons at different velocities, which can be used to find the distance travelled by the electrons. \citet{1996ApJ...464..985A} found heights of 40~Mm for one flare event, known as the ``Masuda flare'', after  \citet{1994Natur.371..495M}, and the first event in which HXR emission was directly observed above the loop-top, now ascribed to emission from a trapped electron population. The height of this emission was measured to be close to 20~Mm, suggesting this trapping is below the acceleration region and the point of reconnection. This geometry also supports a stochastic acceleration mechanism, rather than large scale electric fields. 

\subsubsection{Bidirectional Radio Bursts}
Another extremely informative diagnostic of acceleration region height is offered by bidirectional electron beams and their radio emission (Section \ref{sec:UJRSBi}). Here the accelerated electrons propagate both upwards along open field lines and downwards along the closed field lines \citep[e.g.][]{2000SoPh..194..345R}, and produce radio emission at the local plasma frequency. The point at which the two bursts join is then the location of the accelerating region. 

\subsubsection{Radio Hard X-ray Correlations}
Finally, one-to-one associations between HXR emission and single Type III bursts were seen by \citet{1982ApJ...263..423K}, with further studies performed by e.g. \citet{1985ApJ...299.1027R}. It was concluded that the two kinds of emission were produced by similar electron populations again propagating upwards or downwards along magnetic field lines. Correlated HXR and radio observations can therefore be used to deduce the height of the acceleration region \citep{2011A&A...529A..66R}. However, it should be noted that while individual HXR and radio bursts may be correlated the overall numbers do not match, with only around 1/3rd of observed HXR events showing associated Type III emission \citep{2005SoPh..226..121B}.

\subsection{X-Ray Emission}\label{sec:HXR}
Hard X-Rays themselves account for only a small fraction of the energy released during flares, but because of the relative simplicity of the bremsstrahlung cross-section and the low optical depth of the solar atmosphere at HXR wavelengths, they are a very powerful diagnostic for accelerated electrons, which carry large amounts of energy. The recovery of the electron spectrum from the observed emission is by no means a simple procedure, but reasonable models now exist which allow this to be done. 

During flares, the HXR emission tends to follow a ``soft-hard-soft'' evolution \citep{1970ApJ...162.1003K}, where its spectral index evolves in time: it is large (steep) in the initial phase of the flare, decreases during the peak and then softens again in the decay phase. Occasionally the pattern is rather ``soft-hard-harder''. This evolution may be due to particle trapping or to changes in the acceleration efficiency or behaviour. 

A large fraction of the released HXR energy is deposited in the chromospheric plasma, leading to heating and therefore thermal bremsstrahlung emission in soft X-rays \citep{1989SoPh..121..105E}. This is the likely origin of the \emph{Neupert effect} \citep{1968ApJ...153L..59N, 1993SoPh..146..177D} where the time-integral of the HXR emission follows the time profile of the thermal SXR emission, because the coronal plasma ``integrates'' the non-thermal energy deposition.

\subsection{Bremsstrahlung and the Thick Target Model}\label{sec:brems}\label{sec:NumProb}

Bremsstrahlung emission by electrons in plasma is thermal when the emitting electrons are at the ambient plasma temperature, and can also be thick or thin target, a thick target being one in which the emitting electron loses its entire energy collisionally while emitting. Thermal bremsstrahlung can produce soft X-rays from very hot plasma, such as that generated during the fast electron bombardment of the chromosphere occurring during flares. 

A comprehensive review of bremsstrahlung cross-sections was given by \citet{1959RvMP...31..920K}, including those for thick-target emission. The application of the thick-target model to solar flare HXR emission was considered by various authors, for example \citet{1968ApJ...151..711A} and was extended to include the effects of partial ionisation \citep[e.g.][]{1973SoPh...28..151B}, and additional evolution of the emitting electrons. 

However, assuming the simple thick target model can lead to overestimates of the required electron spectrum in order to produce the observed HXR emission. In some large flares, this overestimate can be particularly problematic as more electrons seem to be required than are present in the entire acceleration region, with the discrepancy historically referred to as the \emph{electron number problem}. 

It is clear that any processes which lead to additional energy losses during beam propagation will increase the number of electrons that must initially be accelerated. Langmuir wave generation by the beam (discussed in the next section) was thought to fall into this category, but simulations showed that the net effect of wave-generation on the time-averaged electron flux was negligible \citep{1987ApJ...321..721H,1987A&A...175..255M, 2009ApJ...707L..45H,2011A&A...529A.109H}. In Chapter \ref{ref:Chapter3} we will discuss one possible means for Langmuir wave generation to instead reduce the required initial number of accelerated electrons, due to interactions between the waves and ambient plasma density fluctuations. 

\section{Electron Beam Propagation and Evolution}
Once a fast electron population has been produced, the exact acceleration mechanism remaining unimportant for our purposes, they must propagate thorough the coronal plasma. For a relatively weak beam, the magnetic field will guide this propagation, and the beam can remain highly collimated \citep[e.g.][]{1990SoPh..130..201M}. However, its evolution due to collisional energy loss and pitch angle scattering are important in very dense plasma, along with such effects as wave-particle interactions. 

\subsection{Langmuir Wave Generation and Sturrock's Dilemma}\label{sec:introLGen}
An electron population whose distribution has a positive velocity gradient at some velocity \footnote{The ``Penrose criterion''  implies that this must be $3\rv_{Te}$ or larger for instability.} (a ``bump''), will be unstable to Langmuir wave generation (Section \ref{sec:qleqs}). We note that the probability for this emission is independent of the magnetic field in the weak field limit $\Omega_{ce} \ll \omega_{pe}$. 

Historically, it was believed that the loss of energy to Langmuir waves would cause an accelerated electron stream to decelerate within only a few thousand metres \citep{1964NASSP..50..357S}, yet these streams were known, from observations of Type III radio bursts (Section \ref{sec:TIIIs}), to propagate over distances up to at least 1~AU. This apparent conflict was known as Sturrock's dilemma and appeared to be a significant problem. 

Most solutions involved suppression of the beam-plasma instability causing the wave generation \citep[e.g.][]{1985srph.book..253G, 1990SoPh..130..201M}, and effects such as wave-wave interactions or density gradients were invoked in order to achieve this, including the example of elastic angular scattering of the waves discussed in Chapter \ref{ref:Chapter2}. However for a spatially limited electron beam, it was found that the waves emitted by the electrons at the front of the beam could be reabsorbed by those at the back, and energy loss would thus be limited \citep{1977SoPh...55..211M, 1990SoPh..130..201M}. 

The numerical simulations by \citet{1998PlPhR..24..772K, 1999SoPh..184..353M} showed that this indeed occurs, and such a beam propagates as a ``Beam-Plasma Structure'' (BPS), where the electrons are accompanied by a wave distribution, emitted at the front and absorbed at the back. Such a structure has a fixed maximum propagation velocity \citep{2003SoPh..212..111M}, dependent on density, and consistent with the observed velocities of Type III radio burst producing streams (Section \ref{sec:TIIIs}). Suppression of the beam-plasma instability remains of critical importance in considerations of the beam-plasma wave interactions, for example in the theory of radio bursts, but is not required in order for the beam to persist. 

\subsection{Beam Propagation in Inhomogeneous Plasma}\label{sec:Inhom}
For a BPS propagating in inhomogeneous plasma, some of the waves will be lost from resonance with the beam, and therefore not all of the energy can be reabsorbed \citep{2001SoPh..202..131K}. Some energy will be left behind in the form of Langmuir wave turbulence, which will damp collisionally, or may interact with other wave modes, and can lead to electromagnetic emission (specifically Type V bursts, see Section \ref{sec:TVs}). 

In addition, the simulations of \citet{2010ApJ...721..864R, 2012SoPh..tmp..109R} following electron beams propagating from the Sun outwards into the solar wind, found that wave particle interactions in inhomogeneous plasma could substantially change the electron distribution found at 1~AU, with an initially injected power law flattening at low velocities due to the generation and subsequent evolution of the Langmuir waves. This was considered also for a beam propagating downwards into the deep corona and chromosphere by \citet{2013A&A...550A..51H} who found that in this case additional high energy electrons were produced, which could lead to enhanced HXR emission at some energies. A similar effect will be shown in Chapter \ref{ref:Chapter3} of this thesis for a beam evolving due to collisional effects. 

\section{Radio Emission}
\begin{figure}
 \center
\includegraphics[width=12cm]{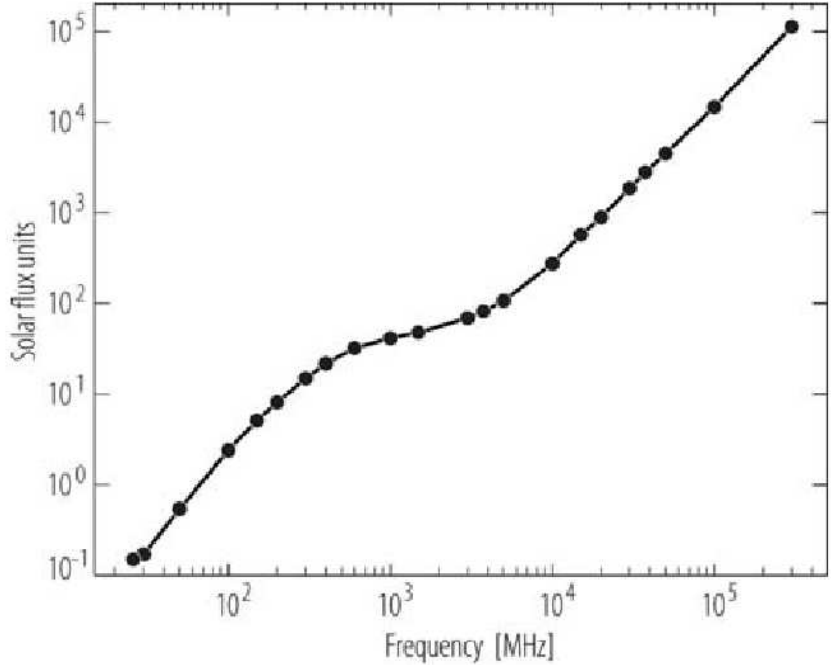}
\caption[The radio emission of the quiet Sun]{The radio emission of the quiet Sun, from \citet{2009LanB...4B..103B}.}\label{fig:quietSFU}
\end{figure}

Radio emission from the quiet Sun follows the Rayleigh Jeans law, \begin{equation} B_\omega(T)=\frac{2\omega^2 k_BT}{(2\pi)^2c^2},\end{equation} where we have taken the classical limiting case $\hbar \omega \ll k_B T$. The observed whole Sun emission in Solar Flux Units (sfu, 1~sfu=$10^{-19}$erg s$^{-1}$cm$^{-2}$Hz$^{-1}$) is shown in Figure \ref{fig:quietSFU}.

\begin{figure}
 \centering
\includegraphics[width=0.98\textwidth]{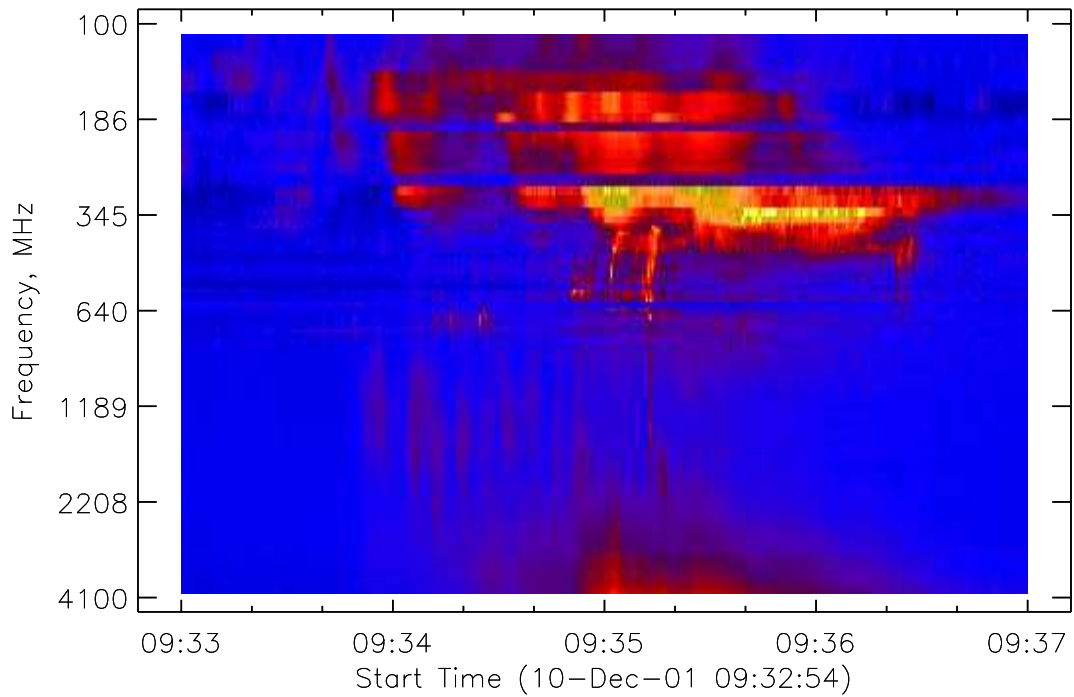}
\caption[An example radio dynamic spectrum.]{An example radio dynamic spectrum from the Phoenix-2 spectrometer. After \citet{2004ASSL..314..203B}. The horizontal bands around 186, 300 and 600 MHz are from terrestrial interference.}\label{fig:spectro} 
\end{figure}

Many kinds of bursty radio emission are also seen in association with solar flares and active events, many of which are signatures of fast electron populations. An example of a radio dynamic spectrum is shown in Figure \ref{fig:spectro} which displays many of these types simultaneously \citep{2004ASSL..314..203B}, discussed briefly in the following sections. The high frequency (2-4~GHz) continuum is gyrosynchrotron emission, while the very faint vertical stripes around 1-2~GHz at 9:34:00 are some form of quasi-periodic pulsations. Groups of bright Type IIIs\footnote{Type classifications are explained in Section \ref{sec:types}.} occur at 9:35:00 and 9:36:20 between 300 and 600~MHz, and a reverse drift burst may be seen at 9:35:20 around 2~GHz. At low frequencies the emission is a combination of Type II (emission associated with shock accelerated electron beams) and Type V (smooth, long duration emission which follows Type III bursts), although these are indistinct in this figure. The bright, short duration emission around 640~MHz at 9:34:20 may be radio spikes. 

In this work we focus on plasma emission at high frequencies from around 500~MHz up to several GHz. However, the Type III bursts are observed to span the entire frequency range from several GHz down to low kHz, and observations tend to focus on coronal (10~MHz to a few hundred MHz) and Interplanetary (IP; kHz and low MHz) bursts, as bursts at higher frequencies are much more infrequent. Important implications for the higher frequency bursts are inferred from the lower frequency behaviour, but the parameter regimes are very disparate and some features differ significantly. 

At metric and decimetric wavelengths, there are also multiple varieties of short duration emission. Some of these may be ascribed to a particular emission mechanism, but in many cases the only classification is morphological. In general, the factors considered are the bandwidth, duration, drift rate, substructures and general shape \citep{2004ASSL..314..203B}.

\subsection{Coherent and Incoherent Emission}
\begin{figure}\begin{center}
 \includegraphics[width=10cm]{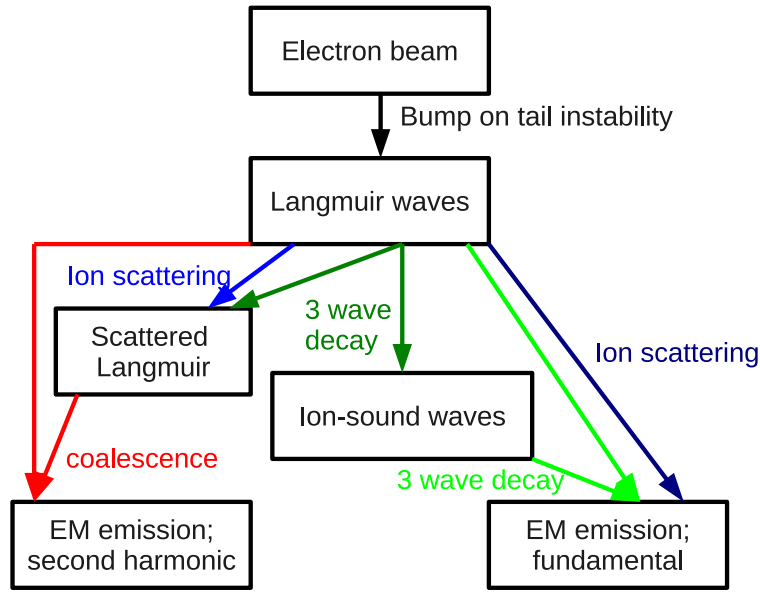}
  \end{center}
  \caption[A flowchart of the plasma emission mechanism]{Flowchart of the plasma emission mechanism (adapted from \citet{2009IAUS..257..305M}).}\label{fig:flow}
\end{figure} 

Incoherent emission refers to emission by individual particles independently. The thermal free-free emission which produces the quiet Sun background is of this type, as is another important mechanism, gyrosynchrotron emission from mildly relativistic electrons \citep[e.g.][]{1985ARA&A..23..169D}. This is particularly important in regions of strong magnetic fields, and at frequencies of a few to a few tens of GHz, as seen in Figure \ref{fig:spectro}. 

Coherent emission, in contrast, arises from plasma collective effects, and therefore emission occurs at either the electron plasma frequency $\omega_{pe}$, the electron cyclotron frequency $\Omega_{ce}$, the ionic equivalents $\omega_{pi}, \Omega_{ci}$, or their harmonics.  The exciting waves are generally produced due to a plasma instability, and so coherent emission indicates the presence of fast electrons, shocks, etc. The high frequency cutoff for observable coherent emission associated with flares is probably around 10~GHz \citep{2004ASSL..314..203B}.
\subsection{The Plasma Emission Mechanism}\label{sec:plasEmm}

Plasma emission is a coherent emission mechanism in which electromagnetic radiation is produced from Langmuir waves, and is the topic of Chapter \ref{ref:Chapter4} of this thesis. The key feature is the involvement of Langmuir waves, generated by fast electrons, and so the emission occurs at the plasma frequency or its second harmonic. The proposed plasma emission mechanism is summarised in the books by \citet{1980MelroseBothVols, 1995lnlp.book.....T}, although refinements and modifications are being continually suggested \citep[e.g.][]{1994ApJ...422..870R,2009IAUS..257..305M,2011PhPl...18e2903T}. Reviews of the currently accepted mechanism etc are given by e.g. \citet{1985ARA&A..23..169D, 1998ARA&A..36..131B, 2008SoPh..253....3N,2009IAUS..257..305M}. 

The outline of the mechanism is summarised in the flowchart in Figure \ref{fig:flow}. An electron beam propagating though the coronal or solar wind plasma generates Langmuir waves, and these in turn evolve, either decaying to ion-sound waves or scattering off plasma ions, both of which generate backwards propagating (negative wavenumber) Langmuir waves. These may then coalesce with a wave from the forwards spectrum to produce an electromagnetic wave at twice the plasma frequency. Alternatively, an ion sound wave and a Langmuir wave may coalesce, or a Langmuir wave may be scattered by an ion directly into an electromagnetic wave, both of which produce emission at the local plasma frequency.

\subsection{Radio Burst Classifications}\label{sec:types}
\begin{figure}
 \centering
\includegraphics[width=16cm]{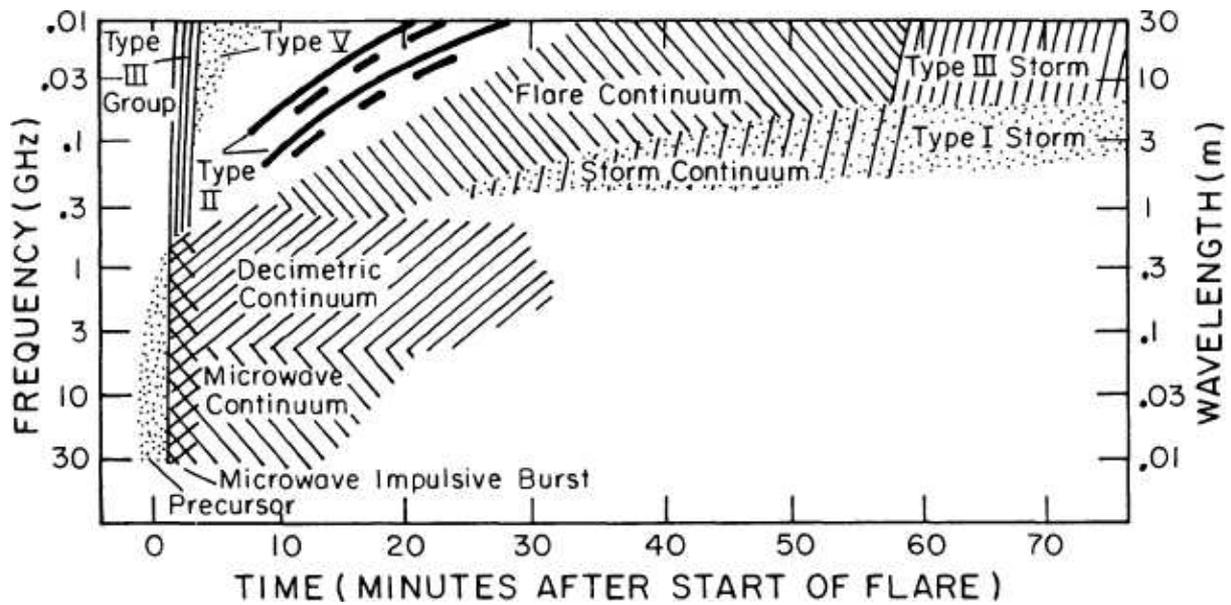}
\caption[An schematic radio dynamic spectrum.]{A schematic radio dynamic spectrum, showing the common radio bursts and broadband, long duration emission. From \citet{1985ARA&A..23..169D}.}\label{fig:schm} 
\end{figure}

Historically, three types of bright, transient radio bursts were known, and these were labelled Types I through III in order of increasing rate of frequency drift $df/dt$. A schematic dynamic spectrum is shown in Figure \ref{fig:schm}, which also includes the later added Type IV and V bursts, and some continuum emissions. 

The origins of Type I emission are not yet certain, while Types II, III and V are due to propagating accelerated electrons. The Type II exciting beams are accelerated in shocks, so the emission drifts slowly as the shock front moves. Type III exciting beams are accelerated in association with flares and move very rapidly, giving fast drifting bursts. Type V emission may be directly associated with a slow electron beam, or may be due to Langmuir waves ``left behind'' after the beam has passed \citep{2001SoPh..202..131K}. Type IV emission has several subtypes, all due to emission from trapped electrons, either via the plasma mechanism or incoherent gyrosynchrotron emission. 

\subsection{Type III Bursts}\label{sec:TIIIs}

\begin{figure}
 \centering
\includegraphics[width=13cm]{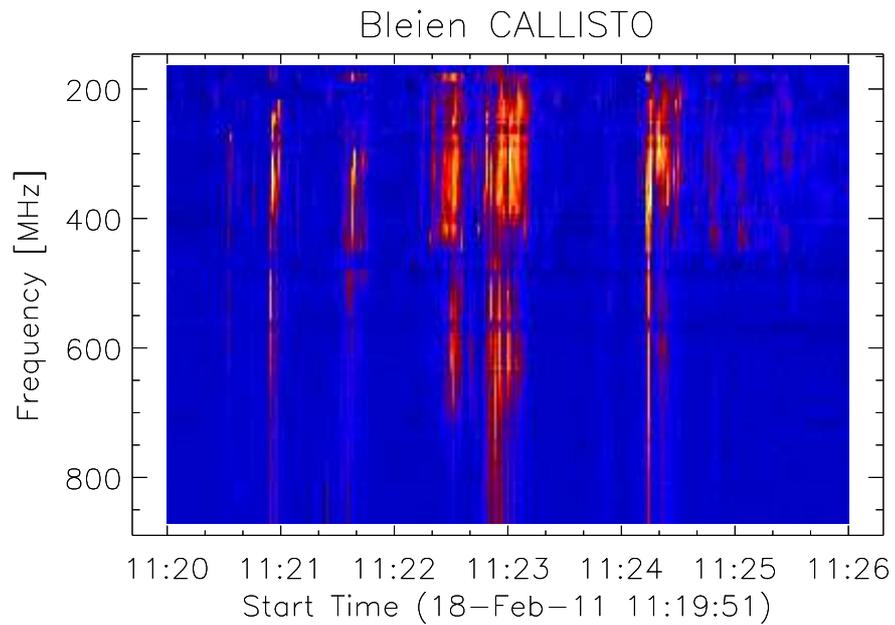}
\caption[A Type III burst group from Bleien Callisto.]{An example dynamic spectrum of a group of Type IIIs, from the Callisto instrument in Bleien.}\label{fig:III} 
\end{figure}

\begin{figure}
 \centering
\includegraphics[width=13cm]{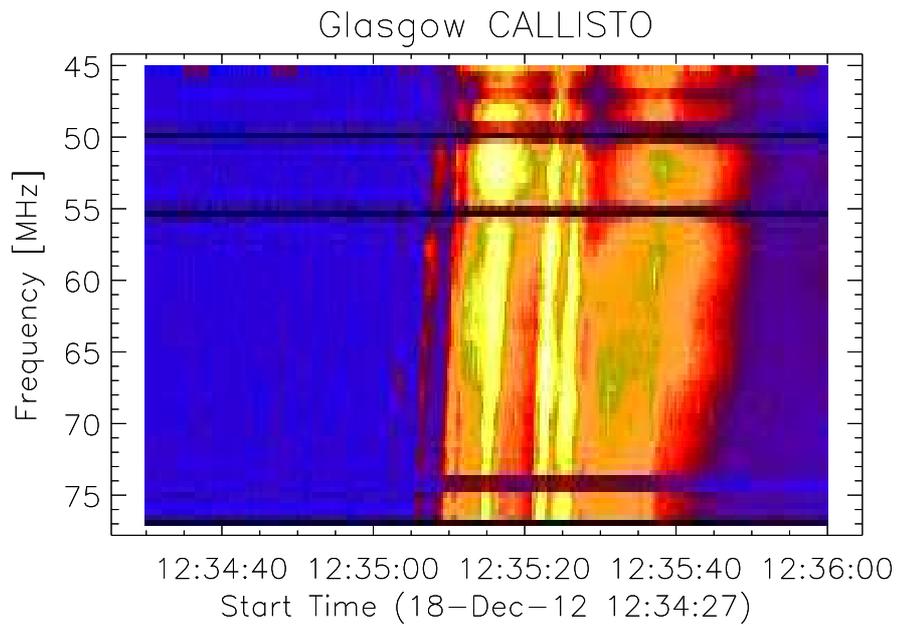}
\caption[A Type III burst from Glasgow Callisto.]{An example dynamic spectrum of a low MHz frequency Type III burst with possible harmonic structure, and diffuse following emission, possibly Type V, from the Callisto instrument in Glasgow.}\label{fig:III2} 
\end{figure}

Type III radio bursts are the classic example of plasma emission from a propagating electron stream, and display many characteristic features, which are discussed in the following sections. Figure \ref{fig:III} shows a radio dynamic spectrum of a group of Type IIIs which displays several of these features, and Figure \ref{fig:III2} an expanded view of a Type III around 80~MHz showing the frequency drift.

\subsubsection{Frequency Drift}
The electron beams which generate Type III bursts propagate at a significant fraction of the speed of light, generally between 0.1 and 0.6 $c$. The beam therefore encounters plasma with a density, and hence plasma frequency, which is rapidly changing. An empirical relation for the frequency drift rates of Type IIIs is $df/dt \simeq -0.01 f^{1.84}$ for $f$ the frequency in MHz and $df/dt$ the drift in MHz~s$^{-1}$ \citep{1973SoPh...30..175A, 1985srph.book.....M} for bursts between 550~MHz and 74~kHz. This corresponds to a value of about 900~MHz~s$^{-1}$ at 500~MHz and 170~MHz~s$^{-1}$ at 200~MHz. 

\subsubsection{Brightness}

The classic Type III bursts are characterised by very high brightness temperatures, with the maximum observed $T_b$ being of the order of $10^{15}$~K and occurring at 1~MHz. For higher frequencies, typical measured brightness temperatures at 40~MHz are $10^{12}$~K, dropping to only $10^8$~K by 169~MHz, while the brightest are $10^{10}$~K and $10^9$~K respectively. For lower frequency IP bursts typical brightnesses at 100~kHz are $10^{13}$~K and the brightest are around $10^{15}$~K \citep[e.g.][]{1985srph.book.....M}. At very high frequencies bursts are rarer and less bright, reaching perhaps $10^8$~K at a few GHz. 
\subsubsection{Source Sizes}
Source sizes steadily increase with decreasing frequency from a few hundred MHz down to kHz frequencies, measuring around $5^\prime$ at 169~MHz, $20^\prime$ at 43~MHz \citep[e.g.][]{1985srph.book.....M} and around $5^\circ$ at 1~MHz and $50^\circ$ at 100~kHz \citep{1985A&A...150..205S}. At 432 MHz, \citet{2013ApJ...762...60S} find sizes of $1.9\pm 0.8^\prime$. 

\subsubsection{Harmonic Structure}

At frequencies in the MHz and kHz ranges, the bursts may be observed in Fundamental-Harmonic (F-H) pairs, occurring very close to $\omega_{pe}$ and $2\omega_{pe}$ respectively. The exact observed F-H ratio is generally slightly below $1:2$, explained by the time delay of the fundamental due to its lower group velocity. For example, \citet{1954AuJPh...7..439W} found a ratio between 1.85 and 2.0, while \citet{1974SoPh...39..451S} found a range from 1.6 to 2.0 with an average of $1.80 \pm 0.14$. The F component is almost never observed above $500$~MHz, and often has a lower starting frequency than the harmonic, so that the ratio of their onset frequencies is closer to $1:3$ or $1:5$ \citep{1977R&QE...20..989S}.  

\subsubsection{Time Profile}
This difference in starting frequency suggests that there are significant differences in the emission mechanisms that produce the F and H components, which is confirmed by measurements of their time profiles and polarisation. At a single frequency, the emission shows a characteristic rapid rise and slow fall off, with different time constants for the fundamental and harmonic components. Originally, the decay phase was thought to be due purely to the collisional damping of the Langmuir waves. However, \citet{1972A&A....19..343A} studied the rise-decay profiles of multiple bursts and found the two times were correlated, implying that the excitation continues alongside the decay. 

For the harmonic component, collisional decay of Langmuir waves may be less important than their spectral evolution \citep{1975A&A....39..107Z}, as the emission can only occur for waves with sufficiently large wavenumber, and for fundamental emission involving ion-sound waves the damping rate of these will also be important. Empirical estimates of the decay rate were given by \citet{1973SoPh...30..175A} for kilometric wavelengths and confirmed for higher frequencies by e.g. \citet{1992SoPh..141..335B}. 

\subsubsection{Polarisation}
Fundamental emission is observed to be O-mode polarised (see Section \ref{sec:MagIModes}) with a degree of polarisation anywhere from 0 up to $70\%$. Polarisations below $10\%$ are rarer, and completely polarised emission is never seen \citep[e.g][]{1977R&QE...20..989S, 1980A&A....88..203D, 1984SoPh...90..139W}. The harmonic is generally weakly O-mode polarised, between 0 and 30$\%$ and most commonly around 10 \% \citep[e.g.][]{1977R&QE...20..989S}. In F-H pairs the degrees of polarisation of the two components are weakly correlated, with the harmonic always lower \citep[e.g][]{1977R&QE...20..989S,1980A&A....88..203D}. The degree of polarisation is not seen to correlate to the emission frequency, but does vary with the source location on the solar disk \citep[e.g.][]{1980IAUS...86..315S}. 

\subsubsection{Reverse Slope and U or J Type Bursts}\label{sec:UJRSBi}

\begin{figure}
 \centering
\includegraphics[width=13cm]{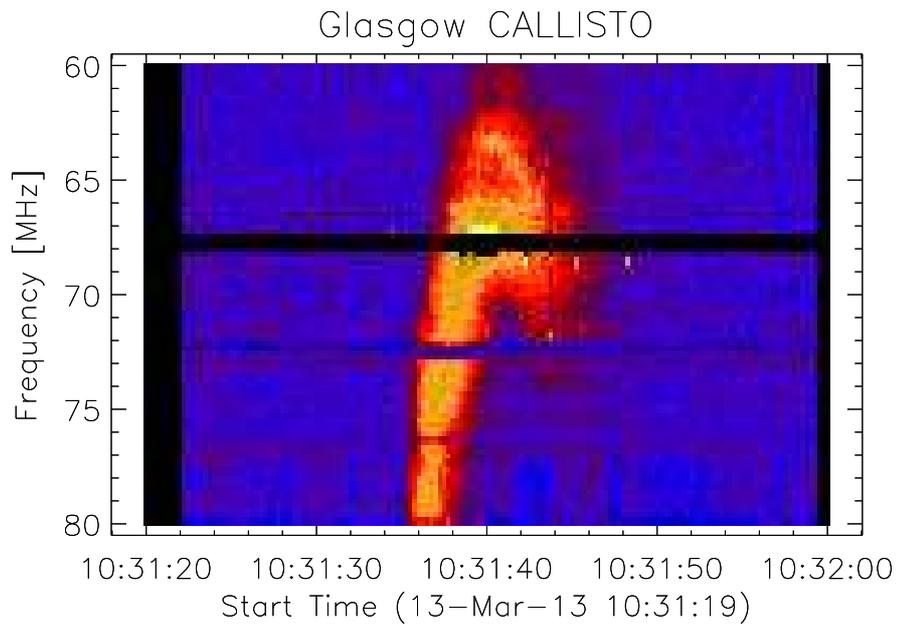}
\caption[A U-type burst from Glasgow Callisto.]{An example dynamic spectrum of a U-type radio burst, from the Callisto instrument in Glasgow. }\label{fig:U} 
\end{figure}

In a normal Type III burst, the beam is accelerated and propagates along an open magnetic field line outwards from the Sun, and therefore into plasma of decreasing density. Reverse slope (RS) bursts arise when the electron beam propagates downwards along a field line, and thus travels into plasma of increasing density. More complicated magnetic structures can produce J or U type bursts, as seen in Figure \ref{fig:U}, where the beam travels along a closed loop and the plasma density gradient, and thus the frequency drift, reverses \citep[e.g.][]{1970PASAu...1..316L}. N bursts, named by \citet{1987ApJ...319..503C} are also occasionally seen, when a second reversal of the gradient occurs probably due to magnetic mirroring in the loop footpoints, as seen in the beam simulations by \citet{1996A&A...314..303K}. 

As noted in Section \ref{sec:accIntro} the location of the acceleration region of the generating electron beams can also be inferred from simultaneous observations of normal and RS Type III bursts. Occasionally, false pairs can arise due to observational effects, for example terrestrial interference leading to a perfectly symmetric pair \citep{2001A&A...366..326B}. True pairs occur where two beams are produced, one moving upwards, the other downwards, and the emission is observed with opposite drifts and very similar starting frequency \citep[e.g.][]{2000SoPh..194..345R}. The exact background density of the acceleration region is then known, from which the location can be found using a model density profile. 

\subsection{Type V Bursts}\label{sec:TVs}
At lower frequencies, 100~MHz and below, Type III bursts are often followed by Type V emission, diffuse emission in the tail of the burst. \citet{1980A&A....88..218D} summarise the observed properties of these bursts. They are low frequency, rarely starting above 120~MHz, and long duration, lasting around 200 seconds at 20~MHz. They generally have a low degree of polarisation, below 10$\%$, but in contrast to the accompanying Type IIIs this is usually in the X-mode. The source sizes are usually comparable to Type III sources, and increase rapidly as frequency decreases, while the heights are also similar to those of Type IIIs. However their locations are often, although not always, different. Finally, their brightness temperatures are also comparable to the accompanying Type III bursts. Plasma emission at the second harmonic appears to be the origin of Type V bursts, with the difference in polarisation between the Type Vs and the preceding Type IIIs explained by differences in the emission geometry, as discussed in Section \ref{sec:Polaris} and \citet{1997SoPh..171..393W}.

\section{Observations of High Frequency Plasma Radio Emission}\label{sec:CoronalIII}

At frequencies of a few hundred MHz or more, Type III bursts occur that are similar in many respects to the bursts at lower frequencies. The most significant difference is in respect to the harmonic structure, with F-H pairs rarely observed above a few hundred MHz, and never above 500~MHz. However the differences in density and temperature, and specifically the ratio of $T_i/T_e$ mean that the mechanism proposed for IP bursts, and confirmed through observations of Langmuir, ion-sound and electromagnetic waves, cannot simply be extrapolated to higher frequencies \citep{2004ASSL..314..203B}. In addition, there are several kinds of short duration emission also ascribed to the plasma mechanism that show important differences from the lower frequency observations. 

In Chapter \ref{ref:Chapter4} of this thesis we consider the simulation of plasma emission at GHz frequencies, and so in this section we give more specific details of the features of GHz radio emission. We highlight where this differs from that at lower frequencies, and what can be inferred about the emission mechanism. 

\subsection{Classic Type III Radio Bursts}\label{sec:ClassicIIIs}

The statistical study of Type III bursts extending above 1000~MHz by \citet{1999SoPh..187...77M} is the most comprehensive of its type, and provides key information about burst behaviour at high frequencies. 160 bursts were analysed, two-thirds of which had an RS component, and 5$\%$ of which were bidirectional (both normal and reverse drift components). The peak fluxes were generally around 10~sfu (Solar Flux Units, 1~sfu=$10^{-19}$erg s$^{-1}$cm$^{-2}$Hz$^{-1}$), and rarely over 100~sfu. By 7~GHz even the maximum fluxes are barely 100~sfu \citep[e.g.][]{1992SoPh..141..335B} and the emission very rarely exceeds the background. The durations, averaging a few 100~ms, were shorter than expected from the extrapolation of the data for metric bursts, with correspondingly low bandwidths of around 200~MHz. 

Source sizes are difficult to obtain at the very high frequencies. At 432~MHz, \citet{2013ApJ...762...60S} find values of $1.9\pm 0.8^\prime$, and extrapolating their observed trends in the size from 151 to 432 MHz suggests sizes at 1~GHz of around $1^\prime$. For comparison, the VLA observations of \citet{1990ApJ...350..856W} of unclassified bursts at 1.4~GHz give source sizes around $2^\prime$. 

For comparison, bursts at 169~MHz have typical fluxes between 10 and a few hundred sfu, which we note is significantly over the thermal level due to the frequency dependence of this, shown in Figure \ref{fig:quietSFU}. They have durations of a few seconds and source sizes around $4^\prime$ \citep{1970A&A.....6..406B}. 

The empirical formula for the burst durations given by \citet{1973SoPh...30..175A} for kilometric wavelengths, and for higher frequencies by \citet{1992SoPh..141..335B} was shown by \citet{1999SoPh..187...77M} not to apply at very high frequencies. Instead the relation is given empirically by $\tau=17000 f^{-0.6}$ for $\tau$ the duration in ms and $f$ the emitting frequency in MHz. The relation found by \citet{1972A&A....16....1E} for bursts at a few 100~MHz also overestimates the duration by a factor of 2 or more. 

The lack of F-H pairs above 500~MHz and the difference in starting frequency of the two components at lower frequencies suggest that that there are significant differences in the emission mechanisms producing the two components. Further evidence is offered by the polarisation of the emission. Generally plasma emission occurs in regions where the magnetic field is weak, and thus we can ignore its effects on the emission rates. However, because in magnetised plasma the right and left hand circular polarisations have different group velocities and low-frequency cutoffs (see Section \ref{sec:MagIModes}), polarised emission can be generated due to propagation effects. 

As noted in the previous section, fundamental emission is observed to be O-mode polarised with a degree of polarisation anywhere from 0 up to $70\%$ in some cases, while the harmonic is generally weakly O-mode polarised, between 0 and 30$\%$. Taking this together with the polarisation statistics from \citet{2012ChA&A..36..175M} in the range 600-1500~MHz which suggest average values of around 20$\%$, suggests that the only observed component at high frequencies is the harmonic. This supports the discussion in Section \ref{sec:collisDamp} where it is found that only harmonic emission is able to escape dense parts of the corona. A final interesting feature of polarisation is a tendency at high frequencies for polarisation degree to increase with frequency, which can be explained by varying magnetic field strength \citep{1990SoPh..130..119M}.

\subsection{Other Plasma Emission}
Various spike-type emissions, characterised generally by their short duration and narrow bandwidths, are observed in the corona at high MHz and GHz frequencies \citep[e.g.][]{1988A&AS...75..243G, 1990SoPh..130..183A, 1994A&AS..104..145I}. The emission mechanism is inferred from the observed characteristics and their correlations with better known types of emission. For example, one class of spikes at decimetric wavelengths (DCIM), are ascribed to electron-cyclotron maser emission primarily on the basis of their frequency and polarisation, and occasionally observed harmonic structure \citep[e.g.][]{1990A&A...231..202G, 1994SoPh..154..361F,1998ARA&A..36..131B}. 

Some of these spike events may be generated by plasma emission from Langmuir waves, which provides another interesting direction for simulations of plasma emission. Theoretically spikes are even more challenging than Type III bursts due to their more complex structures. On the other hand, if we have fast electrons and an instability to Langmuir wave generation we may expect that we should observe plasma emission, and can give analytical estimates of its minimum brightness. The absence of emission is therefore also informative.

Two examples of high frequency, short duration emission which may be easily ascribed to the plasma emission mechanism are given by the ``blips'' of \citet{1983ApJ...271..355B} at frequencies from 500~MHz to the low GHz range, which are similar in several respects to Type IIIs at MHz frequencies and are now known as ``Narrowband Type IIIs'' \citep{1988A&AS...75..243G}, and the ``microwave Type IIIs'' observed at 3-5~GHz by \citet{1987A&A...175..271S} again similar to Type III emission. The latter have durations of only 20-200~ms, which is however consistent with the empirical result of \citet{1973SoPh...30..175A} when adjusted for frequency. Both of these have similar bandwidths, polarisations and frequency drifts. 

Narrow-band Type IIIs at frequencies of 1-2~GHz were also studied by \citet{2001A&A...375..243J}. These have bandwidths of around 150~MHz and total duration approximately 1 second, and are observed equally often as single bursts or in groups. They are identical to Type IIIs in drift rates, brightnesses, duration at a single frequency etc, differing only in bandwidth and therefore total duration. The statistical study of \citet{1999SoPh..187...77M} discussed above also includes these narrow bandwidth emissions.

\chapter{Langmuir Wave Diffusion in Inhomogeneous Plasma}
\label{ref:Chapter2}

\section{Introduction}

Fast electron beams occur in many plasma physics contexts, from lab experiments to solar flares. These beams can become unstable to the generation of Langmuir waves, in which case energy is transferred from the beam into the waves and the instantaneous beam distribution will be changed. If no other evolution occurs, these Langmuir waves are later reabsorbed by electrons at the same velocity as they were emitted, so there is no net effect on the particle distribution \citep{1987ApJ...321..721H,1987A&A...175..255M}. However, as we will show in this chapter and the next, redistribution of the Langmuir waves in wavenumber space can act to redistribute energy between electrons at different velocities, which can lead to electron acceleration.

In the Sun, the fast electron beams accelerated during reconnection in solar flares are known to produce Langmuir waves if the distribution attains a reverse-slope in velocity (see Section \ref{sec:introLGen}).
Spectral evolution of these Langmuir waves can occur due to density variations in the plasma, due to either wavemodes with wavelengths comparable to the Debye length \citep[e.g][]{1967PlPh....9..719V,1982PhFl...25.1062G,2005PhRvL..95u5003Y}, or longer wavelength inhomogeneities \citep[e.g.][]{1969JETP...30..131R,1976JPSJ...41.1757NF,1979ApJ...233..998S, 2001SoPh..202..131K, KRB}. Here we consider the treatment of the effects of long wavelength plasma density fluctuations using a new method based on a diffusion equation.

When an electron beam propagates into plasma of increasing or decreasing density, the generated Langmuir waves will be shifted towards smaller or larger wavenumbers respectively. A randomly fluctuating density will lead to alternating shifts of the Langmuir waves to smaller and larger wavenumbers, and as we will show, when this is averaged over the ensemble of fluctuations it results in diffusion of the Langmuir waves in wavenumber space. 

In this chapter we develop a mathematical treatment of this diffusion process in both one and three dimensions. We first show the derivation of the diffusion equation and diffusion coefficients, then give examples of their functional form for different density fluctuation spectra. We end with a brief theoretical discussion of the effects of wavenumber diffusion on the Langmuir waves and electrons. This work has been published in \citet{RBK}.

\section{Beam-Wave Interactions}\label{sec:qleqs}

In this work we address the problem of beam-plasma wave interactions in the solar corona and wind. As discussed in Section \ref{sec:introLGen}, the effects of the magnetic field on the electron-wave interactions are negligible as we are generally in the weak-field limit, $\Omega_{ce}\ll \omega_{pe}$, but the electron beam must propagate along the magnetic field lines. Moreover, the beam densities observed are often relatively small and so the situation is handled well by the quasilinear approximation. We use equations based on those given in e.g. \citet{drummond1964nucl, vedenov1962quasi,1963JETP...16..682V, 1967PlPh....9..719V, 1995lnlp.book.....T}, and add the effects of collisions, and wave-wave interactions, as necessary. 

In this chapter, we will consider the Langmuir wave evolution in both one and three dimensions. However, in our simulations of the beam-plasma interaction we use a 1-dimensional model, and neglect the effects of particle transport. The physical situations where this model is applicable are discussed in Section \ref{sec:collRelax}.

In this 1-dimensional model we write the electron distribution function as $f({\mr v},t)$ [electrons cm$^{-3}$ (cm/s)$^{-1}$] and the spectral energy density of Langmuir waves at wavenumber $k$ as $W(k,t)$ [ergs cm$^{-2}$] respectively. These are normalised so that \begin{equation}\label{Eq:fNorm}\int f(\rv,t) \rd\rv =n_e\end{equation} with $n_e$ the plasma density in cm$^{-3}$ and \begin{equation}\label{Eq:wNorm}\int W(k,t)\, \rd k=E_L,\end{equation} the total energy density of the waves in erg cm$^{-3}$. The equations describing the electron-Langmuir wave interaction are
\begin{equation}
\frac{\partial f(\rv,t)}{\partial t}= \frac{4\pi^2 e^2}{m_e^2}\frac{\partial}{\partial {\mr v}}\left(
\frac{W(k, t)}{\rv}\frac{\partial f(\rv,t)}{\partial {\mr v}}\right) \Bigg|_{\omega_{pe}=k\rv}\label{eqn:ql1}
\end{equation}
and 
\begin{equation}
\frac{\partial W(k,t)}{\partial t}-\frac{\partial \omega_{pe}}{\partial x}
\frac{\partial W(k,t)}{\partial k} = \frac{\omega_{pe}^3 m_e}{4\pi n_e}{\mr v} \ln\left(\frac{{\mr v}}{{\mr v}_{Te}}\right) f(\rv,t)
+\frac{\pi\omega_{pe}^3}{n_ek^2}W(k,t)\frac{\partial f(\rv,t)}{\partial {\mr v}} \Bigg|_{\omega_{pe}=k\rv} \label{eqn:ql2},
\end{equation} where $\omega_{pe}$ is the local plasma frequency and $n_e$ the local plasma density. 

The two terms on the right-hand side of the second equation correspond to spontaneous and stimulated emission of Langmuir waves respectively. This emission (and the corresponding absorption) is resonant, meaning an electron at velocity $\rv$ interacts only with a wave at wavenumber $k =\omega_{pe}/\rv$, or alternately that the particle velocity $\rv$ and the phase speed of the wave, given by $\rv_{ph}=\omega/k$, are equal. This is the Cerenkov or Vavilov-Cherenkov condition, and the equations are derived in the limit of a particle in constant rectilinear motion. We note that this requires that the particles are unmagnetised, i.e the effects of magnetic field on their motion are negligible.

In addition, we are restricted to the weak turbulence limit, where the energy in Langmuir waves is far less than the energy of the background plasma, i.e. \begin{equation}\label{eqn:weakTurb}\frac{W(k,t)}{n_e T_e} \ll (k\lambda_{De})^2.\end{equation} For beam generated Langmuir waves this leads to a restriction on the beam density, but this is easily satisfied for the beam densities we consider here. 

It is easily seen that the electron distribution becomes unstable to the generation of Langmuir waves when $\partial f/\partial {\mr v} >0$. This may occur due to time of flight effects, where fast electrons overtake the slower ones, or it can be due to collisional relaxation (discussed later in Section \ref{sec:collRelax}), as the collision rate is roughly proportional to ${\mr v}^{-3}$, and a ``gap'' distribution \citep{1975SoPh...43..211M, 1985ApJ...296..278W} can be produced from an initially power-law beam \citep[e.g.][]{1984ApJ...279..882E}. 

The timescale for this beam-wave interaction can be found from Equation \ref{eqn:ql2}, and is known as the quasilinear time. We use the condition $\omega_{pe}=k \rv$ and approximate \begin{equation}\frac{\partial W}{\partial t}\simeq \frac{\omega_{pe}}{n_e}\rv^2 W\frac{\partial f}{\partial \rv}. \end{equation} Assuming a Maxwellian beam, with velocity $\rv_b$, width $\Delta \rv_b$ and density $n_b$, given by \begin{equation} f(\rv)=\frac{n_{b} }{\sqrt{\pi} \Delta {\mr v}_b} \exp\left(-\frac{({\mr v}-{\mr v}_b)^2}{\Delta {\mr v}_b^2}\right),\end{equation} we find ${\partial f}/{\partial \rv}$ and evaluate the equation for a velocity of $\rv_b-\Delta \rv_b$. The result is \begin{equation}\tau_{ql}\simeq \frac{n_e}{\omega_{pe} n_b}.\label{eqn:tql}\end{equation}

\section{A Diffusion Treatment in 1-Dimension}

The second term on the left-hand side of Equation \ref{eqn:ql2} gives the effects of density gradients on the Langmuir waves. In the case of a constant density gradient \citep[e.g.][]{2001SoPh..202..131K} this term becomes \begin{equation}\frac{\pm\delta n}{L} \frac{\partial W(k,t)}{\partial k},\end{equation} where $\delta n$ is the change in density over the scale length $L$, and the sign is positive for a decreasing density and negative for an increasing one. We immediately see that an increasing gradient will shift the Langmuir waves to smaller wavenumbers, and vice versa, while alternating gradients will cause alternating shifts in wavenumber. In the next section we will see that this leads to a diffusion of the Langmuir waves in wavenumber space.

\subsection{The Diffusion Equation}\label{sec:diffEq}

We write the plasma density as \begin{equation}\label{Eq:nTilde}n_e[1+\tilde{n}(x, t)]\end{equation} with $n_e$ the constant background density and $\tilde{n}(x,t)$ the relative density fluctuation. This relative fluctuation is assumed to be weak, i.e. $\tilde{n}(x,t)\ll1$, and long wavelength, $\lambda \gg \lambda_{De}$, conditions often satisfied by the density perturbations the solar corona and the solar wind (see Section \ref{sec:flucs}).

The conditions of weak, long-wavelength fluctuations mean the fractional change in Langmuir wavenumber is small, \begin{equation}\left|\frac{1}{k}\frac{\partial k}{\partial x}\right| \ll k\end{equation} and so we can make the WKB, or geometric optics, approximation. The long-wavelength condition also allows us to treat the Langmuir waves as quasi-particles, averaging over their spatial and time scales, which are far shorter than those for the density fluctuations. In this case, we can describe the wave motion using standard equations of motion describing their propagation and their momentum (equivalent to wavenumber) change due to the action of a density gradient. These equations are \citep[e.g.][]{1965JFM....22..273W,1967PlPh....9..719V,1974R&QE...17..326Z}
\begin{equation}\label{eqn:WKB1} \frac{d k}{d t} = F(x,t)
\end{equation}
\begin{equation}
\frac {d x}{d t} ={\mr v}_g
\end{equation}
where $k$ is the Langmuir wavenumber, $\rv_g$ their group velocity, given by $\rv_g=3k\rv_{Te}^2/\omega_{pe}$ and $F(x,t)$ is the ``force'' acting on the waves due to the density gradients, which for Langmuir waves is given by \begin{equation}F(x,t)=  \frac{\partial \omega_{pe}(x)}{\partial x}=-\frac{\omega_{pe}}{2}  \frac{\partial \tilde{n}}{\partial x}.\end{equation} These equations are equivalent to the conservation equation describing the evolution of the spectral energy density $W(x, k, t)$
\begin{equation}
\label{eqn:liou}
\frac {\partial W(x, k, t)}{\partial t} +F(x,t)\frac{\partial W(x, k, t)}{\partial k}=0.
\end{equation} As noted above, we neglect the effects of spatial transport, i.e. a term $\partial W/\partial x$. For convenience we also omit the source terms due to electrons on the right-hand side (see Equation \ref{eqn:ql2}) during this derivation. 

From Equation \ref{eqn:WKB1} we can see that a random force, as arises due to random density fluctuations, will result in random changes in the Langmuir wavenumber and subsequently a diffusion of the Langmuir waves in wavenumber space. We can derive the equation describing this process via standard procedures \citep[e.g.][]{1963JETP...16..682V,1966PhRv..141..186S}, the outline of which follows. 

We decompose the spectral energy density of Langmuir waves $W(x,k,t)$ into the sum of its average and fluctuating parts $ W=\langle W\rangle  +\widetilde{W}$ and substitute this expression into Equation \ref{eqn:liou} to obtain \begin{equation}\label{eqn:liouDecmp}\frac {\partial \langle W\rangle }{\partial t}+ \frac {\partial \widetilde{W}}{\partial t} +F(x,t)\frac{\partial \langle W\rangle }{\partial k}+F(x,t)\frac{\partial \widetilde{W}}{\partial k}=0.\end{equation} We then average this, and use the facts that, by assumption $\langle F(x,t)\rangle=0$ and by definition $\langle\widetilde{W}\rangle=0$. The result is
\begin{equation}\label{eqn:mean}
\frac{\partial \langle W\rangle}{\partial t}
=-\langle F(x,t)\frac{\partial \widetilde{W}}{\partial k}\rangle.
\end{equation} Subtracting this from Equation \ref{eqn:liouDecmp} gives 
\begin{equation}
\frac {\partial \widetilde{W}}{\partial t} =- F(x,t)\frac{\partial \langle W\rangle }{\partial k}-\widetilde{\left(F(x,t)\frac{\partial \widetilde{W}}{\partial k}\right)}.
\end{equation}
This becomes 
\begin{equation}\label{eqn:fluct}
\frac {\partial \widetilde{W}}{\partial t}=-F(x,t)\frac{\partial \langle W \rangle}{\partial k},
\end{equation} by neglecting the term containing the product of the fluctuation $\widetilde{W}$ and the force $F(x,t)$ as these are both small. Neglecting these products is the key feature of the quasilinear approximation.

Integrating the equation for the fluctuations, Equation \ref{eqn:fluct}, gives
\begin{equation}\label{eqn:pert_sol}
\widetilde{W}(x,k,t)=-\int_{0}^t  F(x-{\mr v}_g\tau, t-\tau)\frac{\partial \langle W\rangle(x-{\mr v}_g\tau, k, t-\tau)}{\partial k} \rd\tau,
\end{equation} which is substituted into Equation \ref{eqn:mean} to give the equation describing the diffusion of wave energy in $k$-space:
\begin{equation}
\frac{\partial \langle W\rangle}{\partial t}
=\frac{\partial }{\partial k}D(k)\frac{\partial \langle W\rangle}{\partial k}
\end{equation} where $D(k)$ is the diffusion coefficient. Now we may drop the notation $\langle \rangle$ for the average, and instead write simply \begin{equation}\label{eqn:diff1d}
\frac{\partial W}{\partial t}
=\frac{\partial }{\partial k}D(k)\frac{\partial W}{\partial k}.
\end{equation}

\subsection{The Diffusion Coefficient}
The diffusion coefficient $D(k)$ in Equation \ref{eqn:diff1d} is given by
\begin{equation}\label{eqn:diff2}
D(k)=\int_0^\infty  \langle F(x, t)F(x-{\mr v}_g\tau, t-\tau)\rangle \rd\tau
\end{equation} where $\langle F(x, t)F(x-{\mr v}_g\tau, t-\tau)\rangle$ denotes the auto-correlation function of the force $F(x, t)$. In practice it is often useful to express this in terms of the spectrum of the density fluctuations that are producing the force.

To do this, we Fourier transform $F(x,t)$ in the the auto-correlation function, \\\hbox{$\langle F(x, t)F(x-{\mr v}_g\tau, t-\tau)\rangle$}, from space and time, $(x, t)$, to wavenumber and frequency, denoted $(q, \Omega)$, using the definition \begin{equation} F(x, t)=\int_{-\infty}^\infty \rd q\int_{-\infty}^\infty \rd\Omega F(q, \Omega) \exp{[2\pi i (qx -\Omega t)]},\end{equation} 
and find 
\begin{align}\label{eqn:Fxt1}\langle F(x, t)\,F(x', t')\rangle =&\int\int\int\int \langle F(q, \Omega)\,F(q',\Omega')\rangle \times \notag\\& \exp{(2\pi i(qx+q'x'-\Omega t-\Omega' t'))} \rd q \rd q' \rd\Omega \rd\Omega'
\end{align}
where $x'=x-\rv_g \tau\,,\: t'=t-\tau$. We assume the force is stationary (auto-correlation does not vary in time) and homogeneous (auto-correlation does not vary in space), and so the auto-correlation can depend only on the differences $x-x'\,,\: t-t'$. We therefore require $q=-q'\,,\: \Omega=-\Omega'$ and so the spectrum of the force, $S_F$, must be given by \begin{equation} \label{eqn:spectrum} \langle F(q, \Omega)\,,\: F(q', \Omega')\rangle=: S_F(q, \Omega)\delta (q+q') \delta (\Omega+\Omega'),\end{equation} and the diffusion coefficient is
\begin{equation}D(k)=\int_0^\infty \rd\tau \int_{-\infty}^\infty \rd q \int_{\infty}^\infty \rd\Omega S_F(q, \Omega) \exp{[2\pi i\tau(\Omega -\rv_g q)]}. \end{equation} Then using the definition \begin{equation} \int_{0}^{\infty} \rd\tau \exp[2\pi i(\Omega-q \rv_g)\tau]=\frac{1}{2}\delta(\Omega-q \rv_g) \end{equation} this becomes \begin{equation}\label{Dv2}D(k)=\frac{1}{2}\int_{-\infty}^\infty \rd q\int_{-\infty}^\infty \rd\Omega S_F(q, \Omega)\delta(\Omega-q{\mr v}_g).\end{equation}

Finally, we relate the spectrum of the force to the spectrum of density fluctuations, $S_n(q, \Omega)$. The force on Langmuir waves is $F(x, t)=-\frac{1}{2}\omega_{pe}\partial_x\tilde{n}(x, t)$ and so its correlation function, assuming $x'=0, t'=0$ from the conditions of stationarity and homogeneity, is $\langle F(x, t)F(0,0)\rangle=(\omega_{pe}^2/4) \partial_{xx}\langle \tilde{n}(x, t)\tilde{n}(0,0)\rangle$. Then using the Wiener Khintchine theorem (a standard result in Fourier theory), which says that 
\begin{equation}\label{Eqn:Wiener} S_F(q,\Omega)=\mathcal{FT} \langle F(x, t)\,F(x', t')\rangle, \end{equation}
twice we find that \begin{equation} S_F(q, \Omega)=\mathcal{FT}\left[ \langle F(x, t) F(0, 0)\rangle\right] = \frac{\omega_{pe}^2}{4}(2\pi i q)^2 S_n(q, \Omega).
\end{equation}
The diffusion coefficient is therefore finally \begin{equation}\label{eqn:diff_1d} D(k)= \frac{\omega_{pe}^2 \pi^2}{2}
\int_{-\infty}^\infty \rd q \int_{-\infty}^\infty \rd\Omega\; q^2 S_n(q,
\Omega)\delta(\Omega-q{\mr v}_g),
\end{equation}
where by definition \begin{equation}\langle\tilde{n}^2\rangle=  \int_{-\infty}^\infty \rd q \int_{-\infty}^\infty \rd\Omega\; S_n(q,\Omega).\label{Eq:norm1D}\end{equation} The delta function, $\delta(\Omega-q{\mr v}_g)$ implies that this is a resonant interaction, while the factor of $q^2$ means that fluctuations at large wavenumbers, i.e. small spatial scales, will have a more significant effect than those at longer wavelengths.

\subsection{Wave Modes}\label{Sec:1ddispRely}

In the previous section we outlined a standard mathematical derivation of a diffusion equation describing the effects of a fluctuating force with arbitrary wavenumber and frequency, $q$ and  $\Omega$ respectively. Now we develop this theory, firstly to consider fluctuations due to a compressive wave mode, for which these must be related by the wave dispersion relation,  $\Omega=\Omega(q)$, and in the next section to find the coefficient for specific fluctuation spectra. To incorporate the restriction $\Omega=\Omega(q)$ we define the spectrum as $S_n(q,\Omega)= S_n(q)\delta(\Omega-\Omega(q))$ so that \begin{equation}\int_{-\infty}^\infty \rd q \int_{-\infty}^\infty \rd\Omega\, S_n(q, \Omega)=\int_{-\infty}^\infty \rd q S_n(q)=\langle\tilde{n}^2\rangle.\end{equation} Substituting this into Equation \ref{eqn:diff_1d} then gives 
\begin{equation}\label{eqn:diff_1dDisp} D(k)= \frac{\omega_{pe}^2 \pi^2}{2}
\int_{-\infty}^\infty \rd q  q^2 S_n(q)\delta(\Omega(q)-q{\mr v}_g).
\end{equation}

Because of the delta function $\delta(\Omega(q)-q{\mr v}_g)$, we have some constraints on the wave modes we can consider in this model. For example, ion-sound waves have a dispersion relation which is approximately $\Omega(q)=q\rv_s$, which would imply that the diffusion coefficient is non-zero only where $\rv_s=\rv_g$, i.e at a single value of the Langmuir wavenumber. However, other wave modes may be of interest, as are fluctuations with arbitrary spectrum. 


\subsection{The Diffusion Coefficient for Specific Spectra}

Using Equation \ref{eqn:diff_1d} we may evaluate the diffusion coefficient for some common density fluctuation spectra, and examine how this affects its functional form. 

\subsubsection{Random Fluctuations} \label{sec:DiffGauss}
If we assume fluctuations that are random in space and time, their correlation function is Gaussian, and therefore from Equation \ref{Eqn:Wiener} so is the fluctuation spectrum. Taking a spectrum given by \begin{equation}  S_n(q, \Omega)=\frac{\langle\tilde{n}^2\rangle }{\pi q_0\Omega_0} \exp\left(-\frac{q^2}{q_0^2}-\frac{\Omega^2}{\Omega_0^2}\right)
\end{equation} where $q_0$ and $\Omega _0$  are characteristic wavenumber and frequency, normalised so that \begin{equation}\label{Eq:RMS}\int_{-\infty}^\infty \rd q \int_{-\infty}^\infty \rd\Omega S_n(q, \Omega)=\langle\tilde{n}^2\rangle \end{equation} the diffusion coefficient reads \begin{equation}\label{eqn:diffcoeffGauss}
D(k)= \omega_{pe}^2{\pi}^{3/2}\frac{q_0}{{\mr v}_0}\langle\tilde{n}^2\rangle \left(1+\frac{{\mr v}_g^2}{{\mr v}_0^2}\right)^{-3/2},
\end{equation}
where ${\mr v}_0=\Omega_0/q_0$ is the characteristic velocity of the fluctuations.

We see immediately that for these random fluctuations there will be two distinct regimes of diffusion, depending on whether the density fluctuation characteristic velocity ${\mr v}_{0}$ is much larger than the Langmuir wave group velocity $\rv_g$ or vice versa.

In the former case, i.e ${\mr v}_{0}\gg {\mr v}_{g}$, the diffusion coefficient becomes \begin{equation} \label{eqn:Dlim1} D(k)= \omega_{pe}^2{\pi}^{3/2}
\left(\frac{q_{0}}{{\mr v}_{0}}\right)\langle\tilde{n}^2\rangle,
\end{equation} and is independent of $\rv_g$ and therefore of the Langmuir wavenumber, $k$. This case of a constant diffusion coefficient drives the system towards the steady state of a flat Langmuir wave spectrum.

In the other extreme, i.e. ${\mr v}_{0}\ll {\mr v}_{g}$, the coefficient is \begin{equation}\label{eqn:Dlim2}
D(k)= \omega_{pe}^2{\pi}^{3/2} \left(\frac{q_0}{{\mr v}_{0}}\right)
\left(\frac{{\mr v}_{0}}{{\mr v}_{g}}\right)^{3}\langle\tilde{n}^2\rangle,
\end{equation} and the diffusion is strongly dependent on the wavenumber $k$. Assuming that initially $D(k) \partial W(k)/\partial k$ is not everywhere zero then the steady state solution, $\partial W(k)/\partial t=0$ is proportional to $k^4$, with more wave energy at large wavenumbers than the thermal case. 

\subsubsection{A Power-law Fluctuation Spectrum} \label{sec:DiffTurb}
Also of interest are fluctuations with a turbulent power-law spectrum in wavenumber at a single characteristic frequency, $\Omega_0$ with spectrum \begin{equation} S_n(q,\Omega)=
\langle\tilde{n}^2\rangle\left(\frac{\zeta-1}{q_0}\right)\left(\frac{q}{q_0}\right)^{-\zeta}\delta(\Omega-\Omega_0)\end{equation} for $q >q_0$ and zero elsewhere, with spectral index $\zeta>1$. The lower limit on $q$ is necessary for normalisation. Fluctuations of this type have been observed in the solar wind \citep[e.g.][]{1972ApJ...171L.101C,1983A&A...126..293C,1983PASAu...5..208R}, with spectral index around $5/3$, and were included in simulations of beam-wave interactions by \citet{2010ApJ...721..864R} (see Section \ref{sec:Inhom}).

The diffusion coefficient is now 
\begin{equation}\label{eqn:DiffPwrLaw}
D(k)=\frac{\omega_{pe}^2\pi^2}{2}(\zeta-1)\left(\frac{q_0}{{\mr v}_0}\right)
\left(\frac{{\mr v}_0}{{\mr v}_g}\right)^{3-\zeta}\langle\tilde{n}^2\rangle.
\end{equation} A power law index of $\zeta=3$ will again lead to a constant diffusion coefficient.

\section{Diffusion in 3-Dimensions}\label{ref:3dDiff}

In many cases we wish to consider density fluctuations which are isotropic or angularly varying, and therefore we must treat the situation in 3-D. In the 3-dimensional treatment the Langmuir wavenumber becomes a wave\emph{vector}, and the density fluctuation induced diffusion is now able to change both its magnitude and orientation. The restrictions on wave dispersion relation due to the condition $\Omega=q\rv_g$ found in 1-D are relaxed in 3-D, allowing us to explicitly consider wave modes such as ion-sound waves.

Previously, treatments of Langmuir wavenumber diffusion \citep[e.g.][]{1976JPSJ...41.1757N} were restricted to elastic scattering, where the fluctuation frequency is much less than the plasma frequency $\omega_{pe}$, and so is neglected. The Langmuir wave energy is conserved by the scattering, and so the Langmuir wavevector can be modified only in angle. 

Here we derive the diffusion equation and coefficients for the general case, and briefly discuss the differences between this and the elastic scattering case. For very low frequency waves the modifications are negligible, but they are important in some cases. For ion-sound waves the corrections due to inelasticity are small but non-zero, while for fluctuations with arbitrary frequency (rather than an explicit dispersion relation, $\Omega=\Omega(q)$) and high frequency components the corrections may be large.

\subsection{The Diffusion Equation}
The conservation equation for the 3-dimensional wave spectral energy density $W(\vec{x},\vec{k},t)$ [erg], analogous to Equation \ref{eqn:liou} is \begin{equation} \frac {\partial W(\vec{x}, \vec{k}, t)}{\partial t}
+ \vec{v}_g .\nabla W(\vec{x},\vec{k}, t) -\frac{1}{2}
\omega_{pe} \nabla \tilde{n}(\vec{x},t)\cdot\frac{\partial
W(\vec{x}, \vec{k}, t)}{\partial \vec{k}}=0,
\end{equation}
where $\vec{v}_g=(3 {\mr v}_{Te}^2/\omega_{pe})\vec{k}$  is the group velocity, and the total energy density of the waves is given by \begin{equation}E_L(\vec{x},t)=\int \rd \vec{k} W(\vec{x}, \vec{k},t).\end{equation} We proceed as above, again neglecting the spatial transport term, to find the 3-dimensional diffusion equation, \begin{equation}\label{eq:diff_3d}
\frac{\partial W(\vec{k},t)}{\partial t}=\frac{\partial}{\partial
k_i} D_{ij}   \frac{\partial W(\vec{k},t)}{\partial {k_j}}.
\end{equation}
The coefficient is now a tensor, and is given in the general case by
\begin{equation}\label{eqn:genDiff}
D_{ij}(\vec{k})=2\pi \omega_{pe}^2 \int \rd\Omega\int
\frac{\rd^3q}{(2\pi)^3}\; q_iq_j S_n(\vec{q}, \Omega)
\delta\left(\Omega-\vec{q}\cdot \vec{v}_g\right)
\end{equation}
where $S_n(\vec{q}, \Omega)$ is the spectrum of the density fluctuations. This spectrum is normalised so that \begin{equation} \label{Eq:norm3D} \int \rd^3q \int \rd\Omega S_n(q, \Omega)=\langle\tilde{n}^2\rangle. \end{equation}

For wave modes, with dispersion relation $\Omega=\Omega(\vec{q})$, we write $S_n(\vec{q},\Omega)=S_n(\vec{q})\delta(\Omega-\Omega(\vec{q}))$ and find
\begin{equation}\label{eqn:genDiffw}
D_{ij}(\vec{k})=2\pi \omega_{pe}^2\int \rd^3q\; q_iq_j S_n(\vec{q})
\delta\left(\Omega(\vec{q})-\vec{q}\cdot \vec{v}_g\right).
\end{equation} Here we have a dot product $\vec{q}\cdot \vec{v}_g$, and therefore additional freedom to specify a dispersion relation and still satisfy the resonant condition $\left(\Omega(\vec{q})-\vec{q}\cdot \vec{v}_g\right)=0$ . So for example we may put $\Omega(\vec{q})=|\vec{q}| \rv_s$ as is the case for ion-sound waves and still satisfy the condition $|\vec{q}| \rv_s =\vec{q}\cdot \vec{v}_g$ for more than one value of the group velocity. This may be contrasted with Section \ref{Sec:1ddispRely} where we had $\Omega=q\rv_g$ which strongly restricted the possible wave dispersion relations. 

\subsection{Diffusion in Spherical Coordinates}
In the elastic scattering approximation previously considered in the literature \citep[e.g.][]{1976JPSJ...41.1757NF,1982PhFl...25.1062G, 1985SoPh...96..181M}, diffusion occurs in angle only, and the problem is very easily treated using spherical polar coordinates. In our more general case the diffusion tensor will have additional angular and magnitude components. However, for the problem of beam-wave interactions, and many other cases of interest, we may assume azimuthal symmetry, so there is no diffusion in azimuth and the spherical coordinate expression remains a useful simplification. 

\begin{figure}
 \centering
\includegraphics[width=8cm]{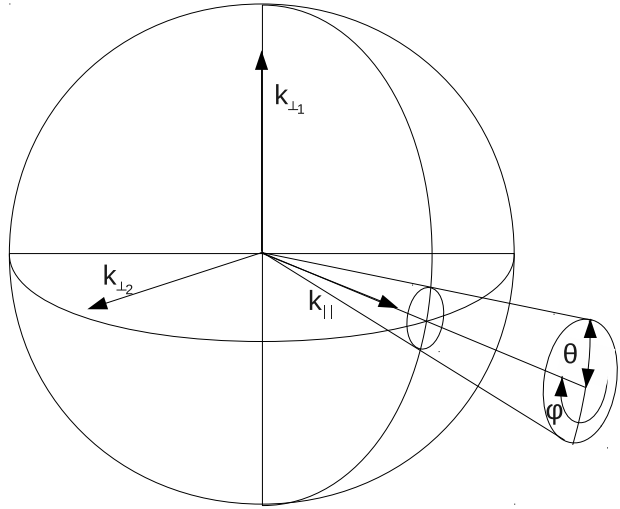}
\caption[Our Cartesian and Spherical coordinate systems.]{The Cartesian and spherical coordinate systems used in our derivations. In the spherical system, $\theta$ is the angle to the parallel direction, and for the Langmuir wavevector $\vec{k}$ under consideration $\phi=0$.}\label{fig:coordsPic}
\end{figure}

We begin by defining two sets of coordinates, one Cartesian and one spherical, as illustrated in Figure \ref{fig:coordsPic}. We can then transform the diffusion equation and its coefficients. In Cartesian coordinates, we define one axis to be parallel to the beam direction, labelled as $\parallel$, and two mutually perpendicular axes labelled, $\perp_1, \perp_2$, giving a standard right-handed Cartesian coordinate system. In our spherical coordinate system, $\theta$ is the angle to the beam direction, and $\phi$ the azimuth, measured clockwise around the beam direction. We now have the two equivalent representations of the Langmuir wavevector, namely $\vec{k}=(k_\parallel, k_{\perp_1}, k_{\perp_2})$ and $\vec{k}=(k, \theta, \phi)$. 

\subsubsection{The Diffusion Equation in Spherical Coordinates}

Equation \ref{eq:diff_3d} can also be written as 
\begin{equation}\frac{\partial W(\vec{k},t)}{\partial t}= \nabla_k \cdot \left(\bar{\bar{D}} \, \nabla_k W(\vec{k},t)\right)\end{equation} where $\bar{\bar{D}}$ denotes the diffusion tensor and $\nabla_k$ is the gradient operator in $\vec{k}$ space. Then using the definitions of $\nabla$ and  $\nabla \cdot$ in spherical coordinates, and the assumption of azimuthal symmetry which implies there can be no diffusion in $\phi$ we find 
\begin{align}\label{eqn:sphdiffnophi}
\frac{\partial W(\vec{k},t)}{\partial t}=&\left[ \frac{1}{k^2}\frac{\partial}{\partial k}\left(k^2D_{kk}\frac{\partial}{\partial k}+kD_{k\theta}\frac{\partial}{\partial\theta}\right) +  \right. \notag \\
 & \left. \frac{1}{\sin\theta}\frac{\partial}{\partial \theta} \sin\theta\left(\frac{D_{\theta\theta}}{k^2}\frac{\partial}{\partial\theta}+\frac{D_{\theta k}}{k}\frac{\partial}{\partial k}\right) \right] W(\vec{k},t).
\end{align}

\subsubsection{The Components of $\bar{\bar{D}}$}

The diffusion coefficients $D_{\theta\theta}, D_{kk}, D_{\theta k},D_{k\theta}$ in Equation \ref{eqn:sphdiffnophi} are derived from the Cartesian components (Equation \ref{eqn:genDiff} or \ref{eqn:genDiffw}) using standard tensor coordinate transforms. For a given Langmuir wavevector $\vec{k}$, using the assumption of azimuthal symmetry, we may define coordinates such that the azimuth of $\vec{k}$ is zero. Then taking $D_{\theta\theta} $ as an example, we find 
\begin{equation}D_{\theta\theta}= D_{\perp_1\,\perp_1}\cos^2\theta- 2D_{\perp_1\,\parallel}\sin\theta\cos\theta +D_{\parallel\,\parallel}\sin^2\theta.\end{equation}

Comparing this with the Cartesian expression in Equation \ref{eqn:genDiff}, we see that by defining the new quantities \begin{align}\label{eqn:dthetatheta}
q_{\theta\theta}&= q_{\perp_1}q_{\perp_1}\cos^2\theta- 2q_{\perp_1}q_\parallel\sin\theta\cos\theta +q_{\parallel}q_{\parallel}\sin^2\theta \\
\label{eqn:dkk}
q_{kk}&=q_{\perp_1}q_{\perp_1}\sin^2\theta + 2q_{\perp_1}q_{\parallel}\sin\theta\cos\theta +q_{\parallel}q_{\parallel}\cos^2\theta \\
\label{eqn:dktheta}
q_{k\theta}&=q_{\theta k}=\sin\theta\cos\theta (q_{\perp_1}q_{\perp_1}-q_{\parallel}q_{\parallel})+(\cos^2\theta-\sin^2\theta)q_{\perp_1}q_{\parallel}
\end{align} 
we may write simply
\begin{equation}\label{eqn:dSph}
D_{ij}(\vec{k})=2\pi \omega_{pe}^2 \int \rd\Omega\int \rd q\int \rd\cos\bar{\theta}\int \rd\bar{\phi}\; \frac{q^2}{(2\pi)^3} q_{ij} S_n(\vec{q}, \Omega) \delta\left(\Omega-\vec{q}\cdot \vec{v}_g\right)
\end{equation}
where $\vec{q}=(q, \bar{\theta}, \bar{\phi})$ and $q_{ij}$ stands for $q_{\theta\theta}, q_{kk}$ and $q_{k\theta}$.

\subsubsection{Evaluating the Diffusion Coefficients}

Equation \ref{eqn:dSph} is entirely general, but in many cases cannot be evaluated. One significant simplification is to introduce an isotropic wave dispersion relation $\Omega=\Omega(q)=\Omega(|\vec{q}|)$. The spectrum is then $S_n(\vec{q},\Omega)=S_n(\vec{q})\delta(\Omega-\Omega(q))$ and Equation \ref{eqn:dSph} becomes
\begin{equation}
D_{ij}(\vec{k})=2\pi \omega_{pe}^2 \int\rd q\int \rd\cos\bar{\theta}\int \rd\bar{\phi}\; \frac{q^2}{(2\pi)^3} q_{ij} S(\vec{q}) \delta\left(\Omega(q)-\vec{q}\cdot \vec{v}_g\right).
\end{equation}

In what follows, we introduce the notation $\cos\theta=\mu, \cos\bar{\theta}=\bar{\mu}$ to shorten the equations. Again, we note that the azimuth of the Langmuir wavevector, $\phi$, is assumed to be zero. We evaluate the dot product, \begin{equation}\label{eq:vgdotq} \vec{{v}}_g \cdot \vec{q}=\frac{3{\mr v}_{Te}^2}{\omega_{pe}}\vec{k}\cdot\vec{q}=\frac{3{\mr v}_{Te}^2}{\omega_{pe}}kq \left((1-\mu^2)^{1/2}(1-\bar{\mu}^2)^{1/2}\cos\bar{\phi}+\mu\bar{\mu}\right)
\end{equation} and use the delta function to integrate over $d\bar{\phi}$. This gives us 
\begin{equation}
D_{ij}(\vec{k})=2\pi \omega_{pe}^2 \int\rd q\int \rd\bar{\mu} \frac{q^2}{(2\pi)^3}q_{ij} S_n(\vec{q}) 
\end{equation} subject to the condition
\begin{equation}\left(\Omega(q)-\vec{q}\cdot \vec{v}_g\right) =0,\end{equation} or using Equation \ref{eq:vgdotq}
\begin{equation}
 \cos \bar{\phi}=\frac{\Omega'-\mu\bar{\mu}}{(1-\mu^2)^{1/2}(1-\bar{\mu}^2)^{1/2}}
 \end{equation} where $\Omega'=\Omega \omega_{pe}/(3\rv_{Te}^2 k q)$, and $\Omega'\le 1$.

Using this we evaluate the components $q_{ij}$ in Equations \ref{eqn:dthetatheta}-\ref{eqn:dktheta} finding
\begin{align}q_{k\theta}&=0 \\q_{kk}&=q^2\Omega'(q)^2 \\q_{\theta\theta}&=q^2 \left(\bar{\mu}^2+\mu^2\Omega'(q)^2-2\mu\bar{\mu}\Omega'(q)\right)/(1-\mu^2) \end{align}
and so the non-zero diffusion coefficients are \begin{align}
D_{kk}=&\frac{\omega_{pe}^3}{12 \pi^2 \rv_{Te}^2 k}  \int\, \rd q q^3 \int \rd\bar{\mu}\Omega^{'2} S_n(\vec{q})\times \notag \\ & \;\;\left( (1-\mu^2)(1-\bar{\mu}^2) -\Omega^{'2} + 2\mu\bar{\mu}\Omega' -\mu^2\bar{\mu}^2\right)^{-1/2}\label{Eqn:DkkFin}
\\ & &\notag \\D_{\theta\theta}=&\frac{\omega_{pe}^3}{12 \pi^2 \rv_{Te}^2 k}  \int \rd q q^3 \int \rd\bar{\mu} \frac{\bar{\mu}^2+\mu^2\Omega^{'2}-2\mu\bar{\mu}\Omega'}{1-\mu^2} S_n(\vec{q})\times \notag\\ &\;\; \left( (1-\mu^2)(1-\bar{\mu}^2) -\Omega^{'2} + 2\mu\bar{\mu}\Omega' -\mu^2\bar{\mu}^2\right)^{-1/2} \label{Eqn:DthThFin}
\end{align}
where the limits on the $\bar{\mu}$ integrals are the solutions of $(1-\mu^2)(1-\bar{\mu}^2) -\Omega^{'2} + 2\mu\bar{\mu}\Omega' -\mu^2\bar{\mu}^2=0$ i.e $\mu_\pm=\mu\Omega' \pm (1-\mu^2)^{1/2}(1-\Omega^{'2})^{1/2}$ (noting that $\Omega'\le 1$).

\subsubsection{Isotropic Fluctuations}
If we assume the fluctuations are isotropic, i.e. $S_n(q)=S_n(|\vec{q}|)$, then we can evaluate the $\bar{\mu}$ angular integral and obtain
\begin{equation}\label{eqn:diffkk}D_{kk} =\frac{\omega_{pe}^2}{216 \pi \rv_{Te}^3 (k\lambda_{De})^3} \int \rd q q\Omega(q)^{2} S_n(q)  \end{equation}  and \begin{equation}\label{eqn:diffthetatheta}D_{\theta\theta}=\frac{\omega_{pe}^2}{24 \pi \rv_{Te} k\lambda_{De}}\int \rd q q^3 \left( 1-\left( \frac{\Omega(q)}{3\rv_{Te} k\lambda_{De} q} \right)^2\right) S_n(q).\end{equation}
From the resonance condition we have $\left(\Omega(q)/(3\rv_{Te} k\lambda_{De} q)\right) \le 1 $ and so $D_{\theta\theta}$ is always positive.

\subsubsection{Elastic Scattering}
We may recover the result for elastic scattering by setting $\Omega=0$, meaning there is no change in energy due to scattering, in Equations \ref{eqn:diffkk} and \ref{eqn:diffthetatheta}, finding \begin{equation}D_{kk}=0\end{equation} and \begin{equation}
D_{\theta\theta}=\frac{\omega_{pe}^2}{24 \pi {\mr v}_{Te}
k\lambda_{De}}\int \rd q q^3 S_n(q),
\end{equation} as given by \citet{1985SoPh...96..181M}. Diffusion occurs in angle only, at a rate independent of the angle $\theta$. Therefore it will tend to isotropise the Langmuir wave spectrum.

\section{Effects of Diffusion on Waves and Electrons}
The detailed evolution of the beam-wave system may only be explored using simulations, but we may infer the general behaviour by considering the form of the diffusion coefficients, and looking at the limiting cases. We give here a brief theoretical discussion of the simplest cases.

\subsection{1-Dimensional Diffusion}

\begin{figure}
 \centering
\includegraphics[width=10cm]{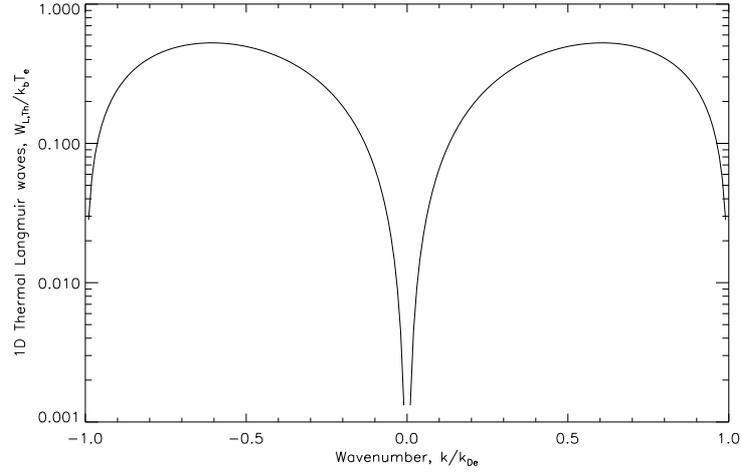}
\caption[The thermal Langmuir wave spectrum.]{The thermal level of Langmuir waves projected onto 1D and normalised to $k_B T_e$, as a function of wavenumber $k$.}\label{fig:LTh}
\end{figure}

In 1-D the wave spectrum evolves according to Equation \ref{eqn:ql2}, 
\begin{equation}
\frac{\partial W(k,t)}{\partial t}
= \frac{\partial }{\partial k}D(k)\frac{\partial W}{\partial k}+ \frac{\omega_{pe}^3 m_e}{4\pi n_e}{\mr v} \ln\left(\frac{{\mr v}}{{\mr v}_{Te}}\right) f(\rv,t)
+\frac{\pi\omega_{pe}^3}{n_ek^2}W(k,t)\frac{\partial f(\rv,t)}{\partial {\mr v}}. \notag
\end{equation} Considering only the diffusion term and assuming a constant diffusion coefficient $D(k)=const$, it is clear that the steady state solution is a flat Langmuir wave spectrum. The thermal level of Langmuir waves in collisionless plasma (given by Equation \ref{eq:LThNoColl}) is plotted in Figure \ref{fig:LTh}. Diffusion due to density fluctuations will then tend to increase the Langmuir wave level at small wavenumbers. The analogous effect at large wavenumbers will not be seen, as above $0.3 k_{De}$ the waves are very rapidly Landau damped by the background electrons.

When we include an electron beam generating Langmuir waves, we expect to see spreading of their spectrum in wavenumber and consequently spreading of the beam in velocity space. The fraction of the waves which shift to smaller wavenumbers will cause electron acceleration, which we investigate in the next chapter using numerical simulations. 

\subsection{Beam Aligned Fluctuations}\label{sec:pllelFluc}
The 1-dimensional case discussed previously can be derived from the 3-dimensional equations by a suitable choice of the density fluctuation spectrum. We define \begin{equation}S_n(|\vec{q}|,\Omega)=S_n(q,\Omega)\delta(|1-\bar{\mu}|),\end{equation} with arbitrary frequency,
and the components in Equations \ref{eqn:dkk} and \ref{eqn:dthetatheta} become $q_{\theta\theta}=0$ and $q_{kk}= q^2$ and so the diffusion coefficients are, using Equation \ref{eqn:dSph}, $D_{\theta\theta}=0$ and 
\begin{equation}
D_{kk}= 4 \pi^2\omega_{pe}^2 \int \rd\Omega \int q^4 \rd q S_n(q,\Omega) \delta(\Omega-q \rv_g).
\end{equation} 

We may relate the 3-D spectrum to its 1-D counterpart via their respective normalisations, Equations \ref{Eq:norm1D} and \ref{Eq:norm3D}, finding $S_n^{1D}(q,\Omega)=2\pi q^2 S_n(q,\Omega)$ and thus we recover the diffusion coefficient as given by Equation \ref{eqn:diff_1d} and the diffusion equation as in Equation \ref{eqn:diff1d}. This confirms that the 1-dimensional model discussed above exactly describes the case of beam parallel density fluctuations.

\subsection{Angular Diffusion}\label{Sec:angdiff}
For a well collimated electron beam, the Langmuir waves generated will be confined to the region in wavenumber space given by $\cos\theta \sim 1$. The effects of inelastic scattering in 3-D can then be considered in 1-D by rewriting Equation \ref{eqn:sphdiffnophi} in terms of the independent variables $k$ and $k_\parallel$ where $k_\parallel$ is the component of the wavevector parallel to the beam direction, given by $k_\parallel=k\cos\theta$, and taking the limit $\cos\theta\sim 1$, or equivalently $k\sim k_\parallel$, to obtain 
\begin{equation} \label{eqn:diffkpar} \frac{\partial}{\partial t} W(k_\parallel,t)=\frac{1}{k_\parallel^2}\frac{\partial}{\partial k_\parallel}\left(k_\parallel^2D_{kk}\frac{\partial}{\partial
k_\parallel}\right)W(k_\parallel,t) -2\frac{D_{\theta\theta}}{k_\parallel}  \frac{\partial}{\partial
k_\parallel} W(k_\parallel,t).
\end{equation}

The first term here is a diffusion of waves in the parallel direction due to the diffusion in magnitude of the wavenumber, while the second is advection of the waves towards smaller parallel wavenumber caused by the angular diffusion. This latter term will be advective only in the limit $\cos\theta\sim1$, which becomes less applicable as angular diffusion proceeds. However, while the Langmuir wave angular spread remains small, both terms will lead to the transfer of energy from waves at larger parallel wavenumbers to those at smaller parallel wavenumbers, and therefore cause energy transfer from slower to faster electrons, leading to an acceleration effect. 

Angular diffusion due to elastic scattering \citep[e.g.][]{1976JPSJ...41.1757NF,1982PhFl...25.1062G, 1985SoPh...96..181M} has been shown to lead to suppression of the beam plasma instability by moving waves out of the resonant region in wavenumber space. In most treatments however, the electron distribution is assumed to be fixed. The combination of angular diffusion and wave absorption is treated as simple absorption with a constant coefficient, and the energy transferred to electrons due to this absorption is ignored. 

From Equation \ref{eqn:diffkpar} we may infer that in fact this energy reabsorption can lead to electron acceleration and the formation of high energy tails in the electron distribution. The 3-D PIC simulations by \citet{2012A&A...544A.148K} consider the effects of Langmuir wave scattering due to wave-wave interactions, and confirm that in this case, the reabsorption by the electrons of Langmuir waves at smaller wavenumbers indeed leads to an acceleration effect. 

We can compare the effects of magnitude and angular diffusion by evaluating the diffusion coefficients in Equations \ref{eqn:diffkk} and \ref{eqn:diffthetatheta} using a simple example spectrum. We take \begin{align} S_n(q,\Omega)&= C\langle \tilde{n}^2 \rangle \; \mathrm{for} \;q_1<q<q_2 \\ &= 0 \;\mathrm{else} \notag \end{align} with $C$ a constant giving proper normalisation, and a dispersion relation $\Omega(q)=q \rv_0$. We find the ratio of the coefficients to be \begin{equation}\frac{D_{kk}}{D_{\theta\theta}} = \frac{3}{16}\frac{\rv_s^2/\rv_{Te}^2}{(k\lambda_{De})^2} \left(1-\left(\frac{\rv_0}{3\rv_{Te}k\lambda_{De}}\right)^2\right)^{-1}\end{equation}

Then for plasma with equal ion and electron temperatures, $T_i=T_e=1MK$ and fluctuations with a velocity $\rv_0\simeq \rv_s$ the sound speed, this ratio is approximately $D_{kk}/D_{\theta\theta} \simeq 0.04$ for beam generated Langmuir waves at wavenumbers around $0.1 k_{De}$, and $0.2$ at $0.05 k_{De}$. We therefore expect that contributions from both effects may be visible even for relatively low frequency fluctuations. 

\subsection{Timescales}
Finally, we may give some general predictions regarding when the diffusion effects are important, and which effects may dominate. In all cases, we have two competing processes in operation: the quasilinear interaction between the electrons and waves, and the diffusive process causing wave evolution. Clearly, the relative importance of these processes will depend on their timescales. These are, for the former, the quasilinear time (Equation \ref{eqn:tql}), \begin{equation}\tau_{ql}=\frac{n_e}{\omega_{pe} n_b}\end{equation} and for the latter, the diffusive time which may be approximated as \begin{equation}\tau_{D}=\frac{k_0^2}{D(k_0)}\end{equation} with $k_0$ the characteristic Langmuir wavenumber and $D$ the appropriate diffusion coefficient, either for 1-D diffusion (see Equation \ref{eqn:diff1d}), magnitude diffusion in 3-D (see Equation \ref{eqn:diffkpar}) or convection due to angular diffusion in 3-D (see Equation \ref{eqn:diffkpar}) respectively. 

We expect the diffusive process to be significant for $\tau_D \sim \tau_{ql}$ in each case, and suppression of the beam-plasma instability to occur if $\tau_D \gg \tau_{ql}$ for a process which causes the transfer of wave energy out of the resonant region in $k$ space. We confirm this numerically for the 1-D diffusion in \citet{RBK}, and the next chapter.

\chapter{Quasilinear Simulations of Langmuir Wave Evolution}
\label{ref:Chapter3}
\renewcommand{\vec}[1]{\mathbf{#1}}

An accelerated electron population which propagates down to the solar chromosphere can produce HXR emission via bremsstrahlung (see Section \ref{sec:brems}). However, as they propagate though the dense plasma, the electrons can lose significant energy due to collisional effects and so to obtain a given electron distribution in the emission region, far more electrons may have to be accelerated. Previously, wave generation was considered to be a pure energy loss process, and therefore expected to only increase the initially required number of fast electrons. In general however, observations of the HXR spectra can only reach down to around $20$~keV and the spectrum below this is unrecoverable from HXR. Therefore if we can transfer energy from energies below this into the observable part of the spectrum, we will appear to have more energy in the electrons. 

In Section \ref{sec:pllelFluc} we discussed the effects of beam parallel density fluctuations, and suggested that these could lead to the acceleration of beam electrons. In addition we concluded that 3-D diffusion due to, for example, isotropic fluctuations would include a similar parallel diffusion component, and thus give a similar effect. In this chapter we investigate electron acceleration using the mathematical treatment developed in Chapter \ref{ref:Chapter2}. We consider plasma similar to the solar corona, with density fluctuations similar to those which have been observed, and treat both a sample case of a Maxwellian beam, in order to investigate the detailed effects, and the collisional relaxation of a power law beam, to see the effects this acceleration may have in a physical situations, and how it may be observed. A large part of this work has been previously published in \citet{KRB, RBK}.

\section{The Simulation Equations}

In Section \ref{sec:qleqs} we gave the quasilinear equations in homogeneous collisionless plasma, Equations \ref{eqn:ql1} and \ref{eqn:ql2}, as originally derived by \citet{1963JETP...16..682V,1967PlPh....9..719V}. However, in the dense plasma of the solar corona, collisional effects are important for both electrons and waves, and so we must add terms describing these. The collisional operator for electrons is given by \citep[e.g.][]{1981phki.book.....L}
\begin{equation}\label{eqn:ql3}
{\mr St}_{col}(f)=\Gamma\frac{\partial}{\partial {\mr v}}\left(\frac{f}{{\mr v}^2}+\frac{{\mr v}_{Te}^2}{{\mr v}^3}\frac{\partial f}{\partial {\mr v}}\right),
\end{equation} where $\Gamma=4 \pi e^4 n_e \ln \Lambda/m_e^2$, with $\ln\Lambda$ the Coulomb logarithm. For the temperatures and densities of the corona and solar wind, the empirical formulae of \citet{1966RvPP....4...93S} give values for $\ln \Lambda$ of approximately 15-20. The collisional damping rate for Langmuir waves is $\gamma_{col} \simeq \Gamma/4{\mr v}_{Te}^3$. For the parameters we use below, the collisional time is $\tau_{coll}\simeq 3.0\times10^{-4}$~s for electrons at $\rv_{Te}$.

We include the diffusion operator describing the effects of beam-aligned density fluctuations on the Langmuir waves from Equation \ref{eqn:diff1d}, and obtain \begin{equation}\label{eqn:3ql1}
\frac{\partial f}{\partial t}= \frac{4\pi^2
e^2}{m_e^2}\frac{\partial}{\partial {\mr v}}\left( \frac{W}{{\mr
v}}\frac{\partial f}{\partial {\mr v}}\right) +
\Gamma\frac{\partial}{\partial \mr v}\left(\frac{f}{\mr
v^2}+\frac{\mr v_{Te}^2}{\mr v^3}\frac{\partial f}{\partial \mr
v}\right),
\end{equation}
\begin{equation}\label{eqn:3ql2}
\frac{\partial W}{\partial t} =\frac{\omega_{pe}^3 m_e}{4\pi n_e}\mr
v \ln\left(\frac{\mr v}{\mr v_{Te}}\right)f
+\frac{\pi\omega_{pe}^3}{n_ek^2}W\frac{\partial f}{\partial {\mr
v}}-\frac{\Gamma}{4 {\mr v}_{Te}^3}W+\frac{\partial}{\partial
k}\left(D\frac{\partial W}{\partial k}\right).
\end{equation}
The coefficient of diffusion, $D$, is given by Equations \ref{eqn:diffcoeffGauss} or \ref{eqn:DiffPwrLaw}, depending on the density fluctuation spectrum assumed.

The initial value problem, which is sufficient to reproduce the physics important here, is solved using finite difference methods, discussed in general for partial differential equations in the books by e.g. \citet{smith1985numerical, thomas1995numerical}, and for the specific case of beam-plasma interaction in \citet{1982SoPh...78..141T, 2001CoPhC.138..222K}. We use explicit finite difference schemes, first order in time and second order in velocity, or equivalently wavenumber, space. 

The code is written in FORTRAN 90, and based on that used in \citet{2001SoPh..202..131K} with additional modifications from \citet{2002PhRvE..65f6408K,2009ApJ...707L..45H} describing collisional effects and the interaction of Langmuir waves with ion-sound waves. In Chapter \ref{ref:Chapter4} we add additional subroutines for the generation of electromagnetic emission. We use a variable timestep according to the stability criterion for the quasilinear equations. The accuracy of the numerical schemes used is discussed in \citet{2001CoPhC.138..222K}. 

\section{Simulations of a Maxwellian Beam}

In order to illustrate the effects of Langmuir wave evolution it is useful to first consider the simplest possible case of beam-wave interaction, by assuming a Maxwellian electron beam. This case offers three significant simplifications. Firstly, the timescale for beam-wave interaction is $\tau_{ql} \simeq 2\times10^{-5}$~s and so collisional effects are negligible for electrons of velocity $\rv\gtrsim 5 \rv_{Te}$. Secondly, the condition $\partial f/\partial \rv >0$ for Langmuir wave growth is immediately satisfied by such a beam, so waves are rapidly produced. Finally, the plateau in the electron distribution produced due to the energy transfer from electrons to waves \citep{1980MelroseBothVols} does not significantly evolve in time, in contrast to the case of a beam formed either due to transport effects \citep[e.g.][]{1958SvA.....2..653G,1975SoPh...43..211M} or collisional relaxation \citep[e.g.][or Section \ref{sec:collRelax}]{1984ApJ...279..882E}.

\subsection{Initial Conditions}\label{sec:MaxInit}
The initial electron distribution function $f({\mr v},t)$ [electrons cm$^{-3}$ (cm/s)$^{-1}$] is the superposition of a
Maxwellian background and a Maxwellian beam:
\begin{equation}
f({\mr v}, t=0)= \frac{n_e}{\sqrt{2\pi} {\mr v}_{Te}}
\exp\left(-\frac{{\mr v}^2}{2 {\mr v}_{Te}^2}\right) +\frac{n_{b} }{\sqrt{\pi} \Delta {\mr v}_b} \exp\left(-\frac{({\mr v}-{\mr v}_b)^2}{\Delta {\mr v}_b^2}\right)
\end{equation} where $n_b$ is the beam density, $\rv_b$ its average velocity and $\Delta \rv_b$ its velocity space width. 

The initial spectral energy density of Langmuir waves $W(k,t)$ [ergs cm$^{-2}$] is set to the thermal level. This is found from the simultaneous steady state solution of Equations \ref{eqn:3ql1} and \ref{eqn:3ql2}, ignoring the collisional terms and is
\begin{equation}\label{eq:LThNoColl}
W(k, t=0)= \frac{k_b T_e}{4 \pi^2} k^2\ln\left(\frac{1}{k\lambda_{De}}\right),
\end{equation}
which agrees with the thermal level of plasma waves in a collisionless Maxwellian plasma as given by \citep{1973plas.book.....K,1995lnlp.book.....T}. 

In our first set of simulations, we fix the parameters of the beam and background plasma, and vary only those of the density fluctuations. We consider plasma similar to that in dense coronal loops. The electron and ion temperatures are approximately equal here, in contrast to solar wind regions where generally the ion temperature is lower \citep[e.g.][]{1998JGR...103.9553N, 1979JGR....84.2029G}, and we take $T_e=T_i=1$~MK. The background density is relatively high, $n_e=1.2\times 10^{10}$~cm$^{-3}$, which corresponds to a local plasma frequency of $\omega_{pe}/2\pi=1$~GHz. We take a beam of density $n_b=10^5\simeq 10^{-5}n_e$, which as will be seen leads to high levels of Langmuir waves, $10^5$ over the thermal level or more. The beam velocity is set to ${\mr v}_b=5\times10^9$~cm~s$^{-1}$ and $\Delta{\mr v}_b=0.3{\mr v}_b$, in the mid range of observed velocities of solar fast electron beams.

We begin by considering random density fluctuations, with Langmuir wavenumber diffusion coefficient $D(k)$ given by Equation \ref{eqn:diffcoeffGauss}. This coefficient contains three free parameters, $\langle \tilde{n}^2\rangle$, $q_0$ and $\rv_0$. However $q_0$ appears only in the coefficient magnitude, arising due to the normalisation of the RMS average density fluctuation. Qualitatively, the effects of varying $q_0$ are therefore identical to those of changing the RMS fluctuation level $\sqrt{\langle \tilde{n}^2\rangle}$, so we may fix $q_0$ without loss of generality. We set $q_0=10^{-4} k_{De}$.

The magnitude of relative density fluctuations $\sqrt{\langle \tilde{n}^2\rangle}$ ranges between $10^{-4}$ and $10^{-2}$ covering the range of values commonly observed in the corona \citep{1972ApJ...171L.101C, 1979ApJ...233..998S}. If we assume the fluctuations are due to ion-sound waves, which may exist (subject to $T_i < T_e$) in the corona at appropriate frequencies, the characteristic velocity will be the sound speed, $\rv_s=1.8\times10^7$~cm~s$^{-1}$ for the plasma parameters as stated, or approximately $\rv_{Te}/20$. We consider a range of velocities around this, varying $\rv_0$ between $0.01{\mr v}_{Te}$ and ${\mr v}_{Te}$.

\subsection{Electron and Wave Distributions}
\setlength{\belowcaptionskip}{0.cm}
\begin{figure}
\center
\includegraphics[width=5.5cm]{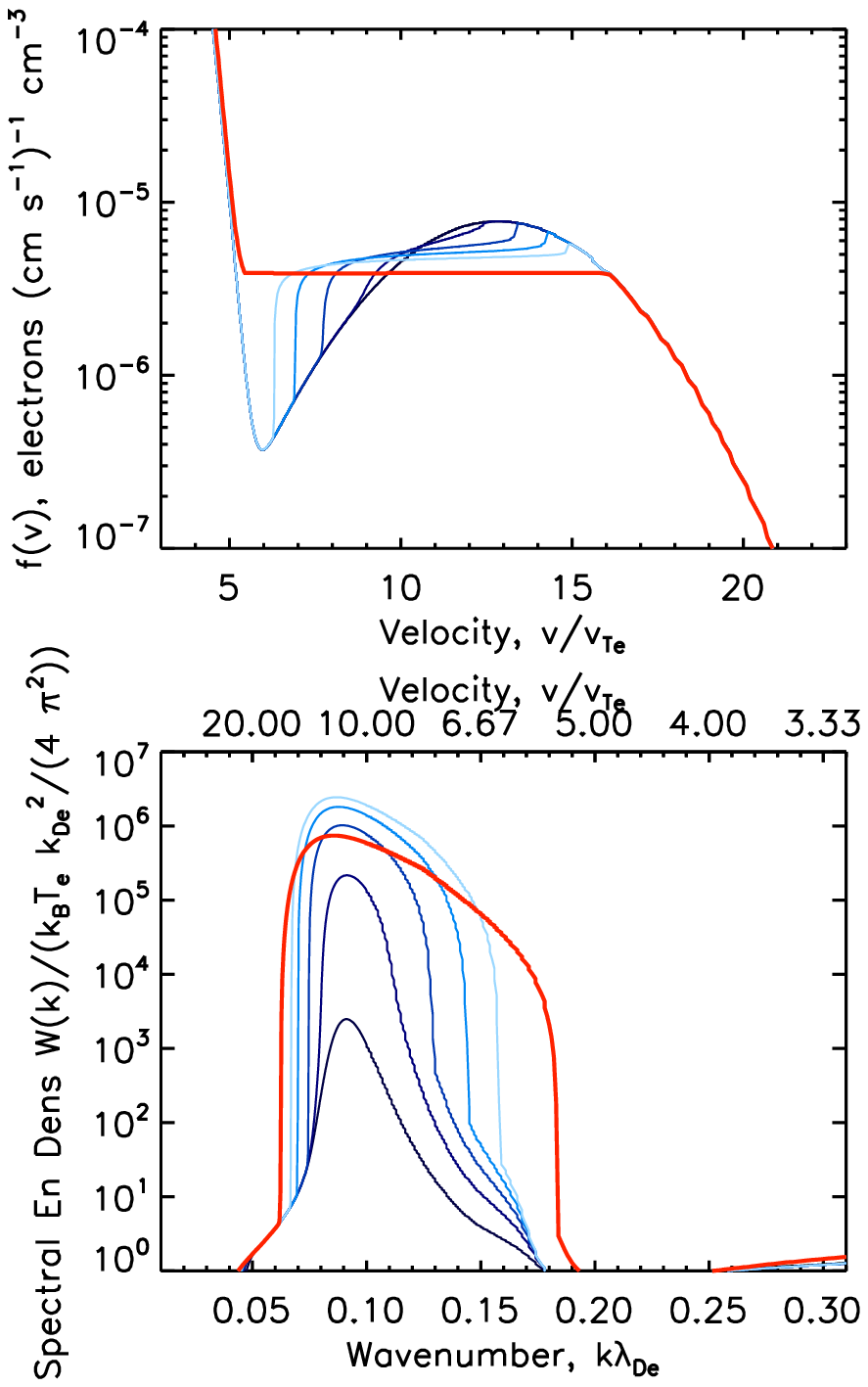}\includegraphics[width=5.5cm]{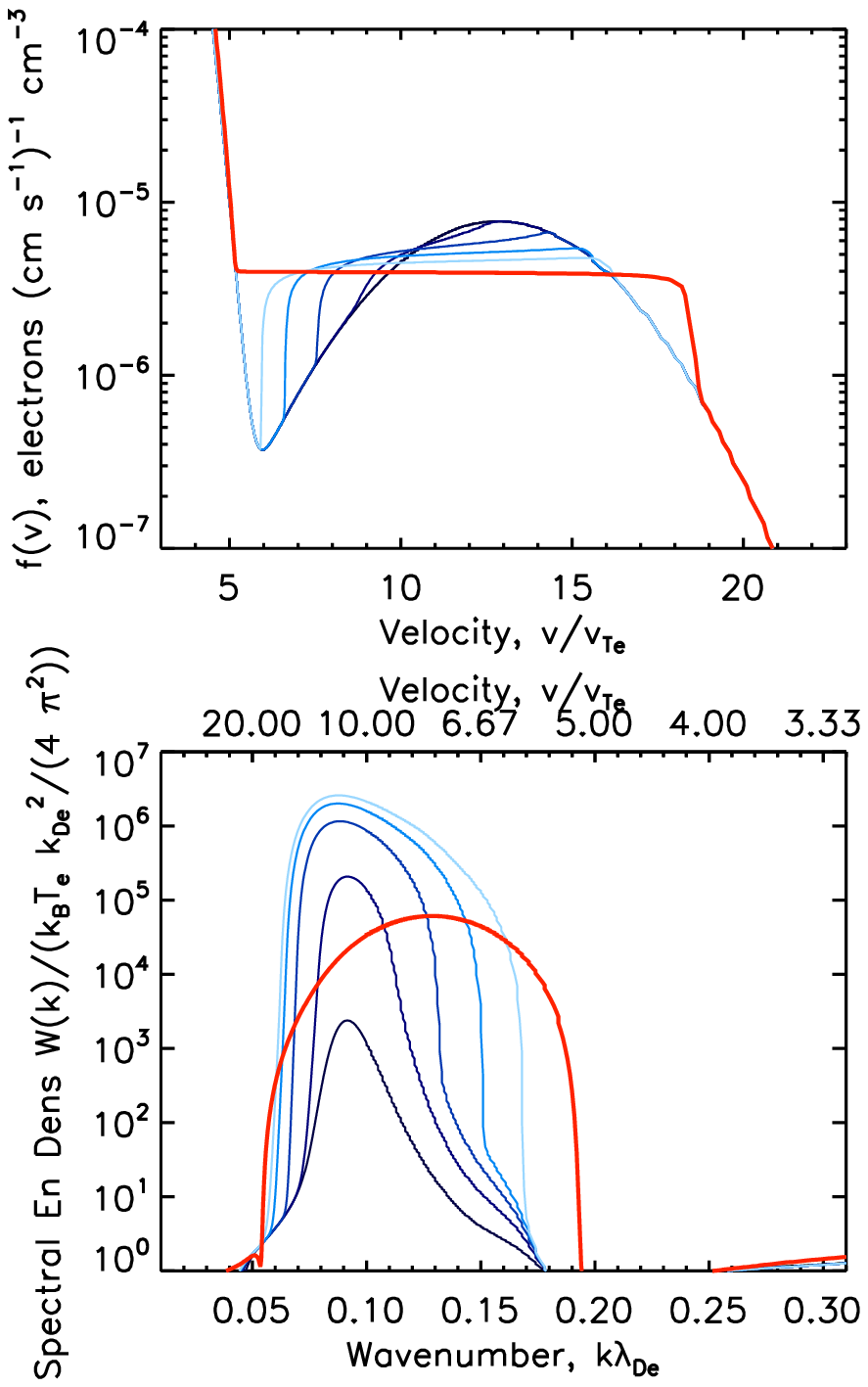}\vspace{0.4cm}
\includegraphics[width=5.5cm]{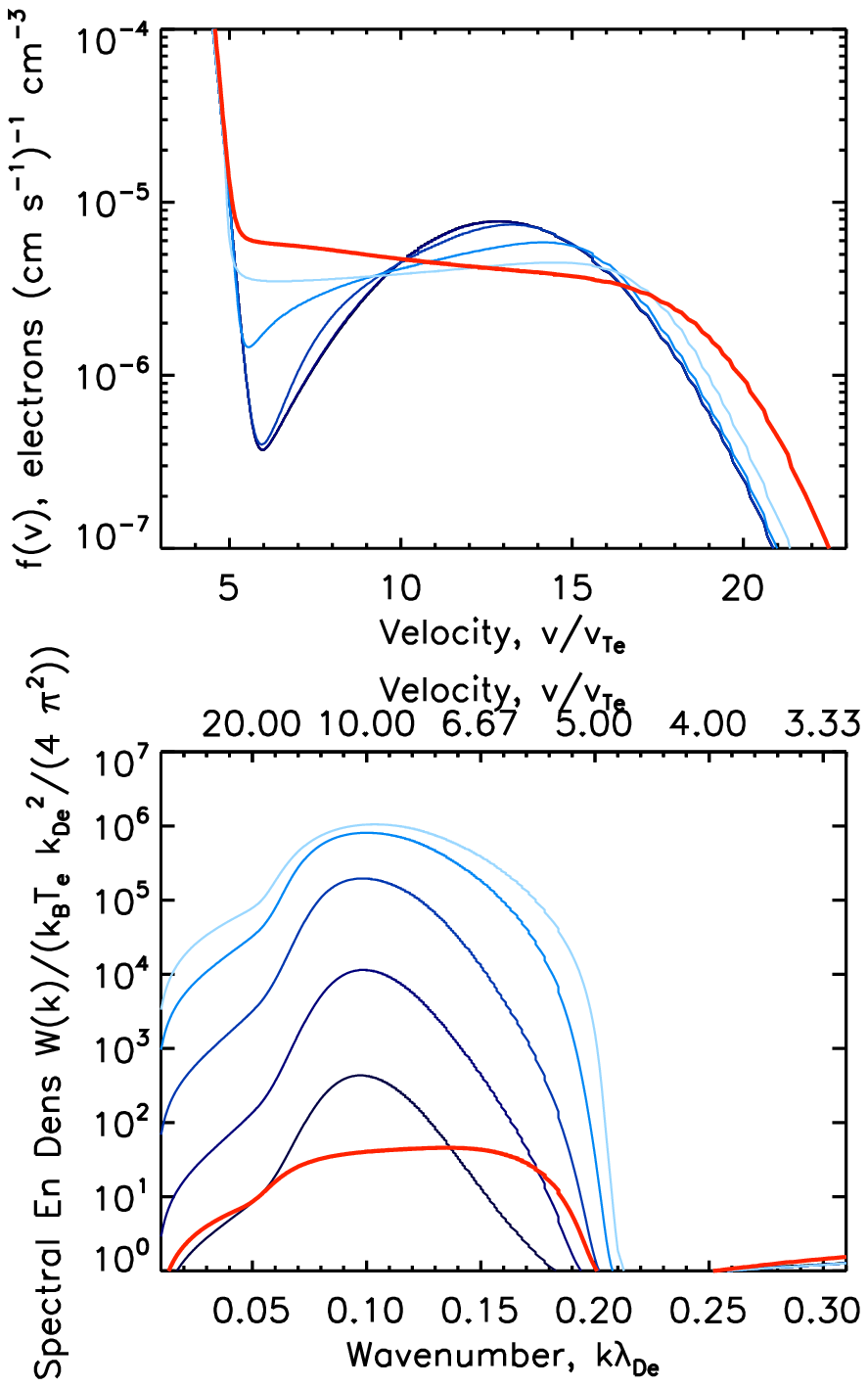}\includegraphics[width=5.5cm]{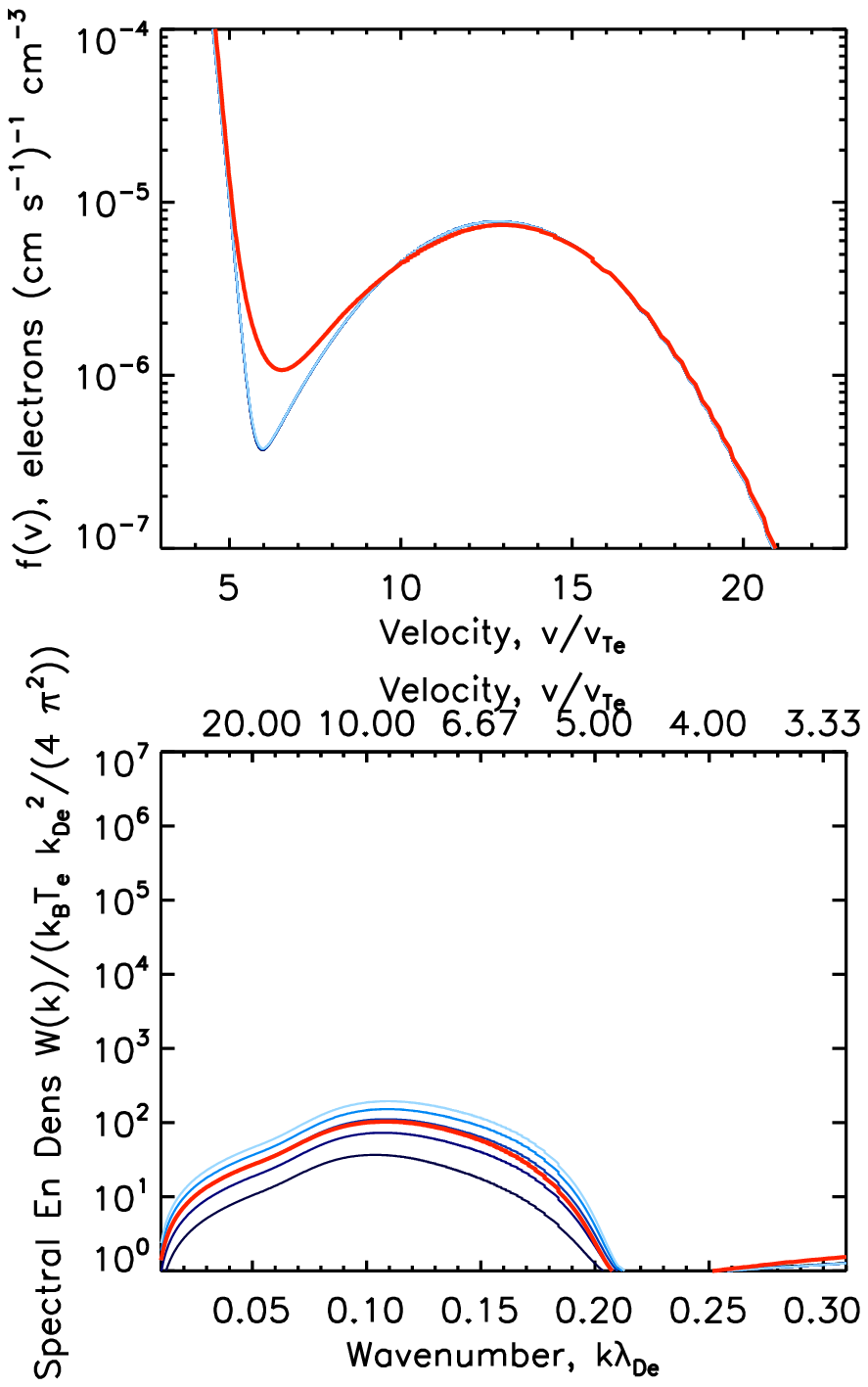}
\caption[Simulated electron and wave distributions for a Maxwellian initial beam.]{Electron distribution $f(\mr v)$ (top), and Langmuir wave spectral energy density $W(k)$ (bottom) for homogeneous (top left pair) and inhomogeneous plasma, with $q_0=10^{-4}k_{De}$, ${\mr v}_0/{\mr v}_{Te}=0.3$ and $\sqrt{\langle \tilde{n}^2\rangle}$ of $3.7\times10^{-4}$ (top right pair), $3.7\times10^{-3}$ (bottom left pair) and $1.2\times10^{-2}$ (bottom right pair). Blue lines from dark to light show the beam relaxation during the first ten quasilinear times, while the red shows the relaxed state reached at $t=100\tau_{ql}$.}\label{fig:w_wo}
\end{figure}
\setlength{\belowcaptionskip}{0.4cm}
Figure \ref{fig:w_wo} shows the evolution of the electron and the wave distributions during the initial quasilinear relaxation of the beam over the first few quasilinear times, and the relaxed state reached at $t=100\tau_{ql}$. The top left panel shows the results in homogeneous density plasma. Langmuir waves grow rapidly over the first few quasilinear times, forming a flat plateau after around $10 \tau_{ql}$.  For this beam velocity and density, the plateau in the electron distribution spans from $6 {\mr v}_{Te}$ ($E =1.5$~keV) to $16 {\mr v}_{Te}$ ($11$~keV).  This plateau remains at $100\tau_{ql}$, as shown by the red line, and will persist until collisional effects begin to be important. Plateau destruction occurs on a timescale given by $\Gamma/\rv^3$ where $\Gamma$ is the collisional rate. This is approximately 1 second at ${\mr v}=15 {\mr v}_{Te}$, of the order of $10^4$ or $10^5$ quasilinear times, and is far longer than the duration of the simulations, so these effects will not be seen.

\subsubsection{Density Fluctuations}

The other panels in Figure \ref{fig:w_wo} show the electron and wave spectra in plasma with fluctuating density. The bottom left panel, showing a moderate level of fluctuations, $\sqrt{\langle \tilde{n}^2\rangle}=3.7\times 10^{-3}$, gives the best illustration of these effects. The waves generated by the beam spread rapidly in wavenumber space, with subsequent slight decrease of their peak intensity. This spreading is mirrored in the electron distribution, and formation of the plateau is slowed. 

This may also be expressed as broadening the resonance between the waves and electrons, since the process of generation of a wave at wavenumber $k$, its shift via diffusion to a new wavenumber $k\pm \Delta k$, and subsequent absorption at velocity $\rv=\omega_{pe}/(k\pm \Delta k)$ can also be thought of as a resonant interaction at $k$ and $\rv$, but with finite width, as opposed to the delta function resonance seen in the quasilinear equations. \citet{1966PhFl....9.1773D} originally showed this for the case of particle diffusion in velocity, but Langmuir wavenumber diffusion affects the resonance in a similar manner.

At large velocities we see an increase in the electron distribution $f({\mr v})$ from the upper edge of the plateau to the highest velocities in the simulation. In other words, electrons have been accelerated. Because waves diffuse to both larger and smaller wavenumbers, we also see changes in the electron distribution down to $\sim 6 \rv_{Te}$. Below this, the energy in background electrons is significantly more than that in the Langmuir waves (the weak turbulence assumption, defined in Equation \ref{eqn:weakTurb}) and their effect is not visible. 

The other two panels in Figure \ref{fig:w_wo} show the effects of stronger and weaker fluctuations. For very weak fluctuations, $\sqrt{\langle \tilde{n}^2\rangle}=3.7\times 10^{-4}$, there is slight spreading of the Langmuir wave spectrum, giving a wider electron plateau and therefore slight acceleration. Otherwise the relaxation proceeds as in the homogeneous case.

When the inhomogeneity becomes very strong, as illustrated by the bottom right panel where $\sqrt{\langle
\tilde{n}^2\rangle}=1.2\times 10^{-2}$, diffusion transports the waves out of their region of excitation in $k$-space on a
timescale which is much smaller than their growth timescale, so the wave level is barely increased above the thermal level. As we can see, by time $t=100\tau_{ql}$ the electron distribution remains essentially unchanged from the initial distribution. Waves are produced and reabsorbed, but the broadening of their spectrum due to diffusion occurs so rapidly that this absorption is mainly by the thermal electrons at a few $\rv_{Te}$, corresponding to wavenumbers near $k_{De}$, where there is far more energy than is in the waves (see Equation \ref{eqn:weakTurb} giving the limit of weak turbulence theory). Thus we get suppression of the beam-plasma instability. 

Elsewhere, the effects of suppression have been considered mainly in the context of elastic angular scattering of Langmuir waves, as discussed in Section \ref{Sec:angdiff}. Our treatment confirms a related suppression effect due to diffusion in the Langmuir wavenumber magnitude, rather than a direct shift $k\rightarrow k \pm \Delta k$. Moreover, although we have assumed here density fluctuations aligned with the electron beam, as noted in Section \ref{Sec:angdiff} isotropic fluctuations, or indeed any angular distribution with component along the beam propagation direction will lead to diffusion in both angle and magnitude of the wavenumber. Our treatment of inelastic scattering considers both possible effects simultaneously, allowing their relative importance to be easily assessed.

\subsection{Parametrising the Effects of Diffusion}

To find the timescale associated with wavenumber diffusion, we approximate the diffusion equation, Equation \ref{eqn:diff1d}, as 
\begin{equation}\frac{\partial W}{\partial t}
=\frac{\partial }{\partial k}D\frac{\partial W}{\partial k} \simeq \frac{1}{\Delta k} D(k_b) \frac{\Delta W}{\Delta k}
\end{equation}
for $\Delta k$ the shift of the Langmuir waves from the generation region out of resonance with the beam, and $D$ evaluated at a characteristic Langmuir wavenumber $k_b$. For a Maxwellian electron beam centred at $\rv_b$, these can be evaluated using the resonance condition, $\omega_{pe}=k \rv$ and are \begin{equation}\label{eq:kb}k_{b}=\omega_{pe}/{\mr v}_{b}\; \mathrm{and} \; \Delta k_{b}=\omega_{pe} \left(\frac{\Delta {\mr v}_{b}}{{\mr v}_{b}^{2}}\right).\end{equation} The parameters $\omega_{pe}, {\mr v}_b$ and $\Delta {\mr v}_b$ are given above, and the timescale is therefore \begin{equation}
 \tau_D = \frac{(\Delta k_{b})^2}{D(k_b)}.
\end{equation}

The other important timescale in the system is that for beam wave interaction, namely the quasilinear time, $\tau_{ql}$. These two timescales indicate the rates of energy transfer by the two processes and thus the relative strength of wave generation and wave diffusion. Therefore it is natural to consider their ratio \begin{equation}
R=\frac{\tau_{D}}{\tau_{ql}}
\end{equation} when discussing the strength of the wave diffusion rather than the absolute magnitude of the diffusion coefficient.

\begin{figure}
\center
\includegraphics[width=12cm]{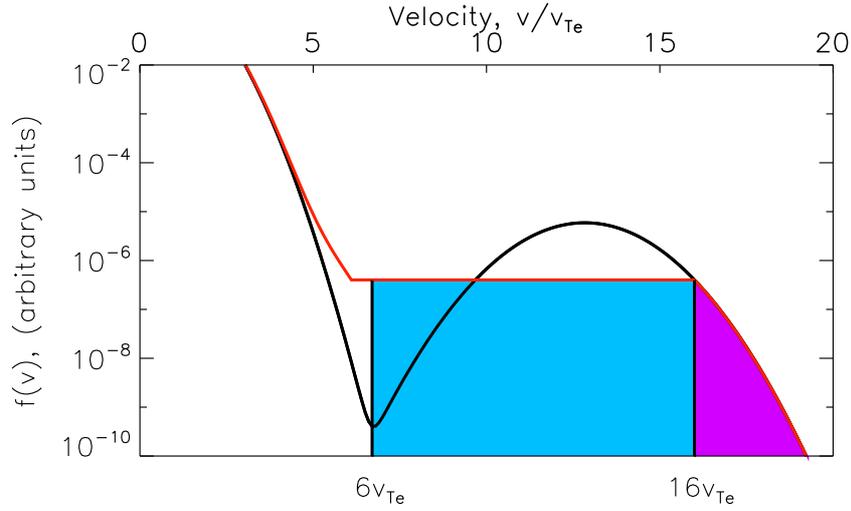}
\caption[Regions in the electron distribution.]{The initial electron distribution (black) and the relaxed state after $100 \tau_{ql}$ (red), showing the definitions of the various regions: the thermal core below $6\rv_{Te}$, the beam above $6\rv_{Te}$ (shaded blue), and the tail above $16\rv_{Te}$ (shaded pink).}\label{fig:regions}

\end{figure}

In the electron distributions in Figure \ref{fig:w_wo} we see three distinct regions in velocity space, illustrated in Figure \ref{fig:regions}. Below around $6 \rv_{Te}$ is the core region, where little or no effect from diffusion is seen. From $6\rv_{Te}$ and up is the ``beam'' region, containing the beam electrons. Above $16\rv_{Te}$ is the ``tail'' region, where we see the accelerated electrons appearing. We set the upper limit in velocity at $30 \rv_{Te}$. Thus the energy of the initial Maxwellian beam is \begin{equation}E_0=\int_{6\rv_{Te}}^{30\rv_{Te}} m_e \rv^2 f(\rv,t=0) \rd\rv \end{equation} and we define the total energy in beam electrons at time $100\tau_{ql}$ as \begin{equation}E_{beam}=\int_{6\rv_{Te}}^{30\rv_{Te}} m_e \rv^2f(\rv, t=100\tau_{ql}) \rd\rv \end{equation} and the tail electrons as \begin{equation}E_{tail}=\int_{16\rv_{Te}}^{30\rv_{Te}} m_e \rv^2f(\rv, t=100\tau_{ql}) \rd\rv.\end{equation}

In the homogeneous case at time $t=100\tau_{ql}$, the beam energy is $E_{beam}=0.76E_0$ and the tail energy $E_{tail}=0.2E_0$ respectively. Generally, although not always, a beam energy greater than this implies slowed relaxation of the beam, while an increased tail energy implies electron acceleration, and so the ratios $E_{beam}/E_{0}$, $E_{tail}/E_{0}$ can be used as a measure of the extent of suppression and/or acceleration. Also of interest is the quantity $E_{tail}/n_{tail}$, the average energy of a single electron in the tail region, giving a measure of the average energy gain of an individual tail electron. In the homogeneous case this is $26$~keV.

\begin{figure}
\includegraphics[width=15cm]{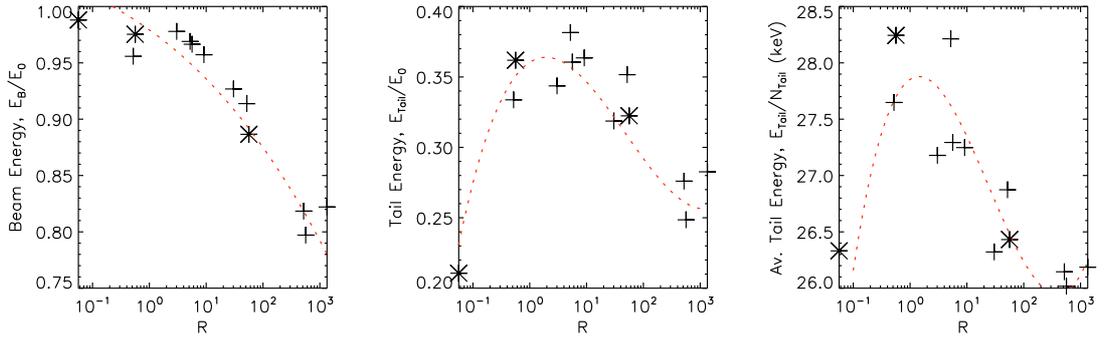}
  \caption[Beam and tail electron energies for a Maxwellian initial beam.]{Plots of total beam (left) and tail (middle) electron energy and the average energy of a tail electron (right) against the parameter $R=\tau_D/\tau_{ql}$, for multiple simulation runs. The three levels of inhomogeneity shown in Figure \ref{fig:w_wo} are marked by asterisks. The red dashed lines are third order polynomial fits to the data.}\label{fig:parst}
\end{figure}

In Figure \ref{fig:parst}, we plot these ratios as functions of the parameter $R$ for a range of values of $\sqrt{\langle \tilde{n}^2\rangle}$ and ${\mr v}_0$. The three cases of diffusion in Figure \ref{fig:w_wo} are marked by asterisks, and represent three distinct regions in $R$. In the strong inhomogeneity case, with $\sqrt{\langle \tilde{n}^2\rangle}=1.2\times 10^{-2}$ and $R\ll1$, the beam-plasma instability is suppressed, so the beam remains close to its initial Maxwellian form and we find
$E_{beam}=0.99E_0$ and $E_{tail}=0.21E_0$, while the average tail electron energy is unchanged.

Very weak inhomogeneity, with $\sqrt{\langle \tilde{n}^2\rangle}=3.7\times 10^{-4}$ and $R\gg1$, has little effect on the beam relaxation, so the beam energy is close to the homogeneous value, but the tail energy is slightly increased due to the broadened plateau, giving $E_{beam}=0.78E_0$ and $E_{tail}=0.23E_0$. Comparing the total tail region energy to the average tail electron energy suggests in this case we have accelerated a reasonable number of electrons, but not to very large velocities, as seen in Figure \ref{fig:w_wo}. The intermediate case, with $\sqrt{\langle \tilde{n}^2\rangle}=3.7\times 10^{-3}$ and $R\sim 1$, gives $E_{beam}=0.95E_0$ and $E_{tail}=0.35E_0$, and a significantly increased average tail electron energy. 

To summarise these findings, for $R\gg 1$ the density fluctuations are weak and relaxation proceeds as in homogeneous plasma with no change to the beam energy and no electron acceleration. As the diffusion coefficient is increased and $R$ approaches 1, we begin to see substantial electron acceleration, quantified by the tail energy $E_{tail}$, due to the energy transfer from slower to faster electrons through Langmuir wave diffusion. Finally, when $R\ll1$ we have strong diffusive broadening of the wave spectrum, and thus suppression of the beam-plasma instability as wave energy is lost from the resonant region on a much shorter timescale than $\tau_{ql}$. Moreover, in the region of strong acceleration, the total tail electron energy is almost doubled, and is increased by more than half over around 2 orders of magnitude in $R$ around this. 

\subsection{Effects of Fluctuation Characteristic Velocity}

\begin{figure} \centering 
\includegraphics[width=6cm]{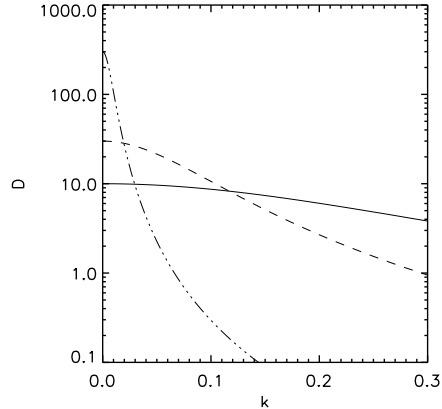}
\caption[The Langmuir wave diffusion coefficient as function of wavenumber.]{The diffusion coefficient (in normalised units) for a fixed level of fluctuations, $\sqrt{\langle\tilde{n}^2\rangle}=2\times10^{-3}$, and three values of the characteristic velocity, $\rv_0/\rv_{Te}=0.95$ (solid line), 0.3 (dashed line) and 0.03 (dot-dashed line).}\label{fig:DiffCoeff}  
\end{figure}

For random density fluctuations we saw in Section \ref{sec:DiffGauss} that there are two simple regimes of diffusion when $\rv_0\ll \rv_g$ or vice versa. However, the values of $\rv_0$ of interest here lie primarily in the transitional region where $\rv_0\sim \rv_g$ and thus the diffusion coefficient shape is strongly dependent on this value. In Figure \ref{fig:DiffCoeff}  we plot the diffusion coefficient as a function of wavenumber $k$ for three values of the fluctuation characteristic velocity, $\rv_0$. Significant dependence is seen around the main region of Langmuir wave excitation, $k_b \sim0.1k_{De}$ (Equation \ref{eq:kb}), which is largely accounted for in our definition of the diffusive timescale, $\tau_D$. 

\begin{figure}\centering
\includegraphics[width=10cm]{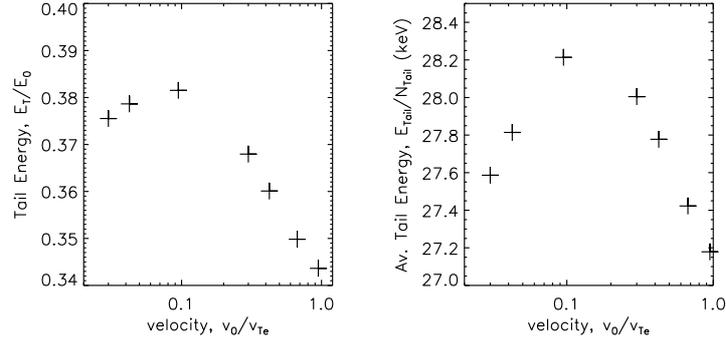}
\caption[Beam and tail electron energies: variations with characteristic velocity.]{Total and average tail electron energy against $\rv_0/\rv_{Te}$ for a fixed level of fluctuations, $\sqrt{\langle\tilde{n}^2\rangle} =2\times10^{-3}$.} \label{fig:parsVel} \end{figure}

To confirm that this is so, we can minimise the diffusive timescale with respect to $\rv_0$, finding that this occurs at $\rv_0/\rv_{Te}= 3 \sqrt{3} k_b/k_{De}$. Substituting $k_b$ for our parameters, we find the maximum acceleration should occur for $\rv_0/\rv_t \simeq 0.2$. In Figure \ref{fig:parsVel} we plot the total and average tail electron energies as a function of $\rv_0$ for a fixed level of fluctuations, $\sqrt{\langle\tilde{n}^2\rangle}=2\times10^{-3}$, and see a clear peak at exactly this value of $\rv_0$. The dependence of the acceleration effect on characteristic velocity therefore accounts for the vertical scatter of the points in Figure \ref{fig:parst}, which is confirmed by comparing the range in \ref{fig:parsVel}, approximately $\Delta(E_{beam}/E_0)\simeq 0.04$ and $\Delta(E_{tail}/n_{tail}) \simeq 1$~keV, to that in Figure \ref{fig:parst}.

\subsection{Power-law Fluctuations}

\begin{figure}
\centering
\includegraphics[width=15cm]{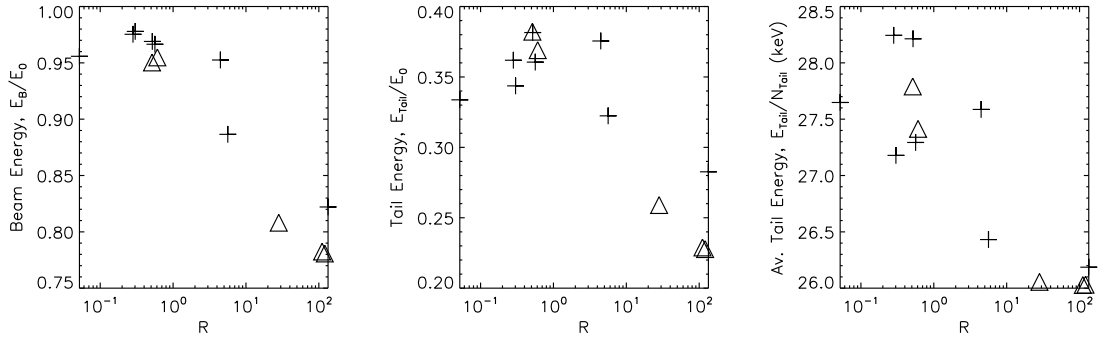}
\caption[Beam and tail electron energies: variations with fluctuation spectrum.]{Plots of total beam (left) and tail (middle) electron energy and the average energy of a tail electron (right) against the parameter $R=\tau_D/\tau_{ql}$ for random (crosses) and power-law density fluctuations (triangles). Power law indices are $5/3$ and $7/3$ and the other fluctuation parameters vary within the limits described in the text.} \label{fig:parsPwr} \end{figure}

In Section \ref{sec:DiffTurb} we calculated the diffusion coefficient for a turbulent power-law spectrum of density fluctuations. This coefficient is given by Equation \ref{eqn:DiffPwrLaw}, and was seen to be similar in some respects to the coefficient for random fluctuations given by Equation \ref{eqn:diffcoeffGauss}. In Figure \ref{fig:parsPwr} we compare the acceleration effects due to these two coefficients for a range of fluctuation parameters. We use power-law indices of $5/3$ and $7/3$, similar to those observed in the solar corona and wind \citep[e.g.][]{1972ApJ...171L.101C,1983A&A...126..293C,1983PASAu...5..208R}. The effects of acceleration for the two fluctuation spectra are seen to be similar for the parameters considered here, but it must be noted that only a few test cases have been run. 

\subsection{The Effects of Beam and Plasma Parameters}

\begin{figure}
\includegraphics[width=15cm]{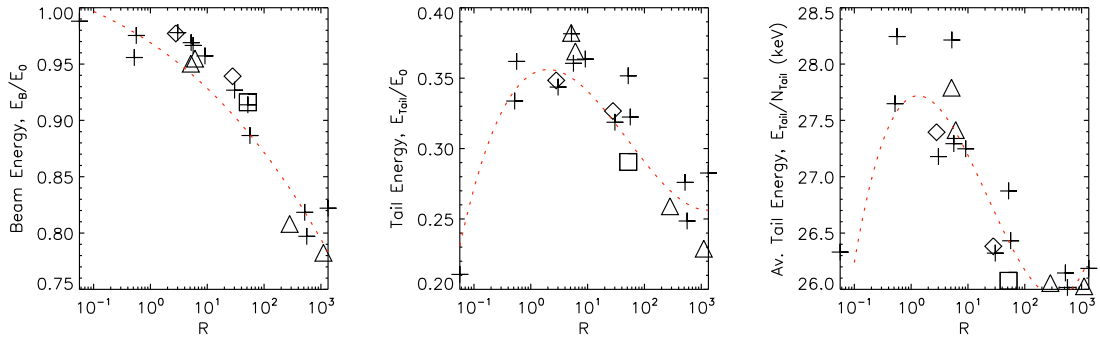}
\caption[Beam and tail electron energies: variations with beam and plasma parameters.]{Plots of total beam (left) and tail (middle) electron energy and the average energy of a tail electron (right) against the parameter $R=\tau_D/\tau_{ql}$. Various values of $\sqrt{\langle\tilde{n}^2\rangle}, \rv_0$ are shown. At $\omega_{pe}=1$GHz we show: random fluctuations with $n_b=10^5$~cm$^{-3}$ (crosses) and $n_b=10^4$~cm$^{-3}$ (squares), powerlaw fluctuations with indices $\frac{5}{3}$ and $\frac{7}{3}$ (triangles). Finally for $\omega_{pe}=200$MHz we show random fluctuations with $n_b=10^3$~cm$^{-3}$ (diamonds) }\label{fig:ParsAll}
\end{figure}

Thus far we have considered only the effects of changing the fluctuation parameters, but we can expect the trends established for the chosen beam and plasma parameters to hold true when these quantities are also varied. The simulations presented in this section show that this is indeed the case, but only within certain ranges.

Firstly, both of the relevant timescales, $\tau_{ql}$ and $\tau_D$, obey the same relation with respect to the local plasma density, and thus varying this has no effect on the parameter $R$. We show two cases of diffusion with $\omega_{pe}=200$~MHz in Figure \ref{fig:ParsAll}, and confirm that the same relationships with $R$ hold in this case. 

The beam density, $n_b$, controls the level of Langmuir waves generated, and therefore could be expected to affect the acceleration process. However, as shown in Figure \ref{fig:ParsAll}, the effect is almost entirely accounted for by the dependence of $R$ on the timescale $\tau_{ql}$, and the same regimes of acceleration and suppression are seen. On the other hand, the results for a beam density of $n_b=10^3$~cm$^{-3}$ at 1~GHz (not shown in the figure) do not behave as expected, showing neither acceleration nor suppression. This is due partly to the influence of collisional effects on the beam and waves over the longer timescale for relaxation of such a weak beam, and also to the small level of Langmuir waves such a beam can generate. The acceleration relies on the redistribution of energy between different Langmuir wavenumbers, and is thus limited by the amount of energy available in waves. 

\subsection{Conclusions}

To summarise this section, for the test case of a Maxwellian beam, we have found that plasma density fluctuations can lead to a significant electron acceleration or suppression of the beam-plasma instability. These effects occur for broad range of beam, plasma and fluctuation parameters. The details of the density fluctuations change the effects in relatively small ways, but their extent is controlled almost entirely by the single quantity $R$, the ratio of the quasilinear and wave diffusion timescales. When $R$ is small, the diffusion process dominates, and the beam-plasma instability is suppressed. When $R$ is close to 1, there is significant acceleration. When $R$ is much larger than 1, there is no effect as the beam-plasma interaction is far more rapid than the diffusion process. However, as the effect is due to energy transfer between different velocities by means of Langmuir wave evolution, a sufficiently large level of Langmuir waves must be generated, and so we require a sufficiently dense beam.

\section{Simulations of the Collisional Relaxation of a Power-Law Beam}\label{sec:collRelax}

In the previous section we looked in detail at the effects of Langmuir wavenumber diffusion for a simple test case of a Maxwellian beam. This served as a useful ``shortcut'' for the production of Langmuir waves, as the initial beam distribution was already unstable to their generation. However, in physical situations the accelerated electrons often obey a power law distribution, with $\partial f/\partial \rv <0$ at all velocities, which only becomes unstable to wave generation due to subsequent evolution. 

In some situations, this instability arises due to ``time-of-flight'' effects. The faster electrons outpace the slower ones, producing a distribution with more electrons at high velocities than at lower ones and therefore a reverse slope. Coronal electron beams typically have very high velocities, between 0.1 and 0.6 $c$. An electron beam moving upwards from the acceleration region may be estimated to become unstable after a minimum distance of approximately $10^9$ or $10^{10}$~cm \citep{1982ApJ...263..423K, 2011A&A...529A..66R}. The timescale to become unstable is then between $0.1$~s and a few seconds. 

A reverse slope distribution can also be generated due to collisional effects. The collisional term in Equation \ref{eqn:3ql1} is \begin{equation}
\frac{\partial f}{\partial t}= \Gamma\frac{\partial}{\partial \mr v}\left(\frac{f}{\mr
v^2}+\frac{\mr v_{Te}^2}{\mr v^3}\frac{\partial f}{\partial \mr
v}\right),
\end{equation} which is roughly proportional to $\rv^{-3}$, and so slower electrons will lose energy more rapidly than faster ones. The result is called a ``gap distribution'' \citep{1975SoPh...43..211M, 1985ApJ...296..278W}. At a plasma frequency of 2~GHz the collisional timescale for electrons around $\rv_{Te}$, which is given by $\rv_{Te}^3/\Gamma$, is approximately $10^{-4}$~s. A reverse slope can therefore be generated at velocities around $8\rv_{Te}$ after at time of $(8\rv_{Te})^3/\Gamma$ or approximately $10^{-2}$~s.  

In dense plasma it can therefore be interesting to consider the collisional relaxation of the beam, as the timescale for this is shorter than that for transport effects to become important, especially for the slightly slower beams. We use a 1-D model and consider the evolution at a single spatial location. This provides a good model of, for example, a very dense coronal loop. The results in this section have been published in \citet{KRB}: the discussion as presented here has been written specifically for this thesis. 

\subsection{Initial Conditions}

We use the quasilinear equations, Equations \ref{eqn:3ql1} and \ref{eqn:3ql2}, as in the previous section and a diffusion coefficient given by Equation \ref{eqn:diffcoeffGauss}. The initial electron distribution is now a Maxwellian background plasma plus a power law beam smoothly fitted to this:
\begin{equation}
f({\mr v}, t=0)= \frac{n_e}{\sqrt{2\pi} {\mr v}_{Te}} \exp\left(-\frac{{\mr v}^2}{2 {\mr v}_{Te}^2}\right) +\frac{2 n_{b}}{\sqrt{\pi}\, {\mr v}_b}\frac{\Gamma(\delta)}{ \Gamma(\delta-\frac{1}{2})}
\left[1+({\mr v}/{\mr v}_{b})^2\right]^{-\delta}.
\label{eq:f_t0}
\end{equation} Here, $\Gamma(\delta)$ denotes the gamma function and appears due to the normalisation of the distribution, while $\delta$ is the power law index for the energetic particles in energy space and $n_{b}$ the number density of non-thermal electrons, $n_{b}\ll n_e$. 

This distribution is a power law $f({\mr v},t=0)\sim {\mr v}^{-2\delta}$ at high velocities ${\mr v}> {\mr v}_b=10{\mr v}_{Te}$ and flattens below this. This helps to fit the beam smoothly to the Maxwellian core, and also means that collisional relaxation can more quickly lead to Langmuir wave generation, as there are less electrons at low velocities than for a pure power law. 

The initial electron distribution is normalised to the electron number density [electrons cm$^{-3}$],
so that \begin{equation}
 \int_{0}^\infty \frac{2 n_{b}}{\sqrt{\pi}\, {\mr v}_b}\frac{\Gamma(\delta)}{ \Gamma(\delta-\frac{1}{2})}
\left[1+({\mr v}/{\mr v}_{b})^2\right]^{-\delta} \rd{\mr v}=n_{b}.
\end{equation}

The thermal level of Langmuir waves when collisions are included is calculated as in Section \ref{sec:MaxInit} by finding the steady state solution of Equation \ref{eqn:3ql2}. We ignore the beam electrons and consider only the Maxwellian core to obtain
\begin{equation}
W(k, t=0)= \frac{k_b T_e}{4 \pi^2}\frac{k^2\ln\left(\frac{1}{k\lambda_{De}}\right)}
{1+\frac{\ln \Lambda}{16\pi n_e}\sqrt{\frac{2}{\pi}}k^3 \exp\left(\frac{1}{2k^2\lambda_{De}^2}\right)} \simeq \frac{k_b T_e}{4 \pi^2} k^2\ln\left(\frac{1}{k\lambda_{De}}\right)
\label{eq:W_t0}
\end{equation} as used in the previous simulations.

We consider again dense coronal regions, in this case a slightly larger density of $n_e=5\times 10^{10}$~cm$^{-3}$, corresponding to a local plasma frequency of $\omega_{pe}/2\pi=2$~GHz. The electron and ion temperature are again equal, $T_e=T_i=1$~MK. The power law index of the initial beam is $\delta=4$, corresponding to an index of 8 in velocity space.

During the collisional relaxation of the beam, a large fraction of the initial energy will be lost. To obtain a sufficiently large reverse slope $\partial f/\partial \rv$, and therefore high levels of Langmuir waves, we must take an initial beam density far higher than in the case of a Maxwellian beam. We therefore take $n_b \simeq 10^9$ cm$^{-3}$, which leads to Langmuir wave levels of $10^5$ over thermal at their peak, sufficient for their modification to have a strong effect on the beam evolution.

The simulations here require much longer run times than those in the previous section as the evolution occurs on large multiples, several thousand or so, of the collisional timescale. We therefore consider the evolution over $\sim1$~s which can require a simulation run-time of several days, depending on which effects are considered. Thus we cannot explore the parameter space in the same way as for the Maxwellian beam, and instead present only a few sample cases. We begin with homogeneous plasma, then add a constant plasma density gradient, then show a few examples of the effects of Langmuir diffusion. Finally we directly consider the effects of wave-wave interactions for which the participating wavenumbers are approximately equal, in addition to the diffusion caused by small wavenumber fluctuations.

\subsection{Homogeneous Plasma}

\begin{figure} \centering
\includegraphics[width=6cm]{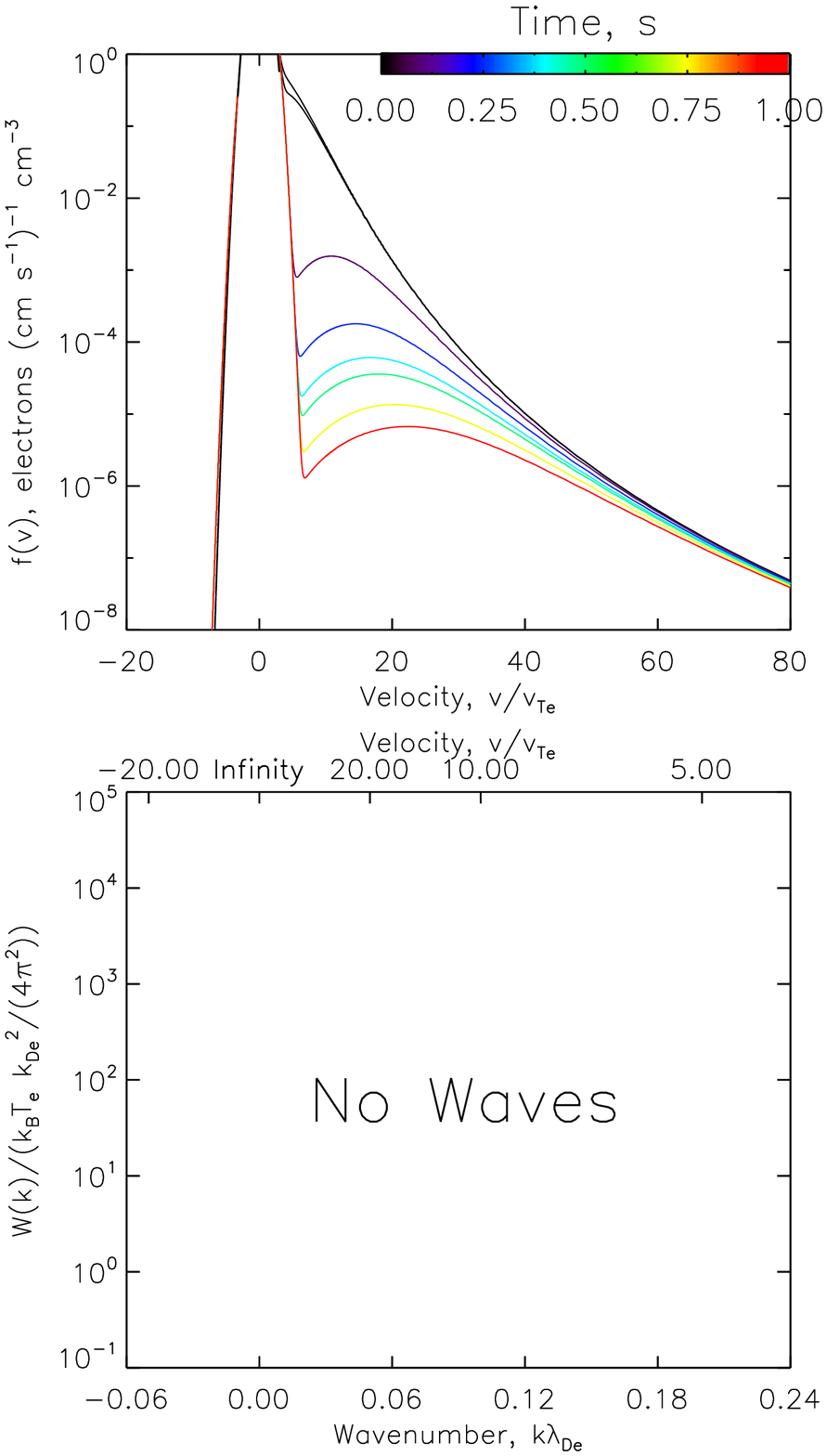}
\includegraphics[width=6cm]{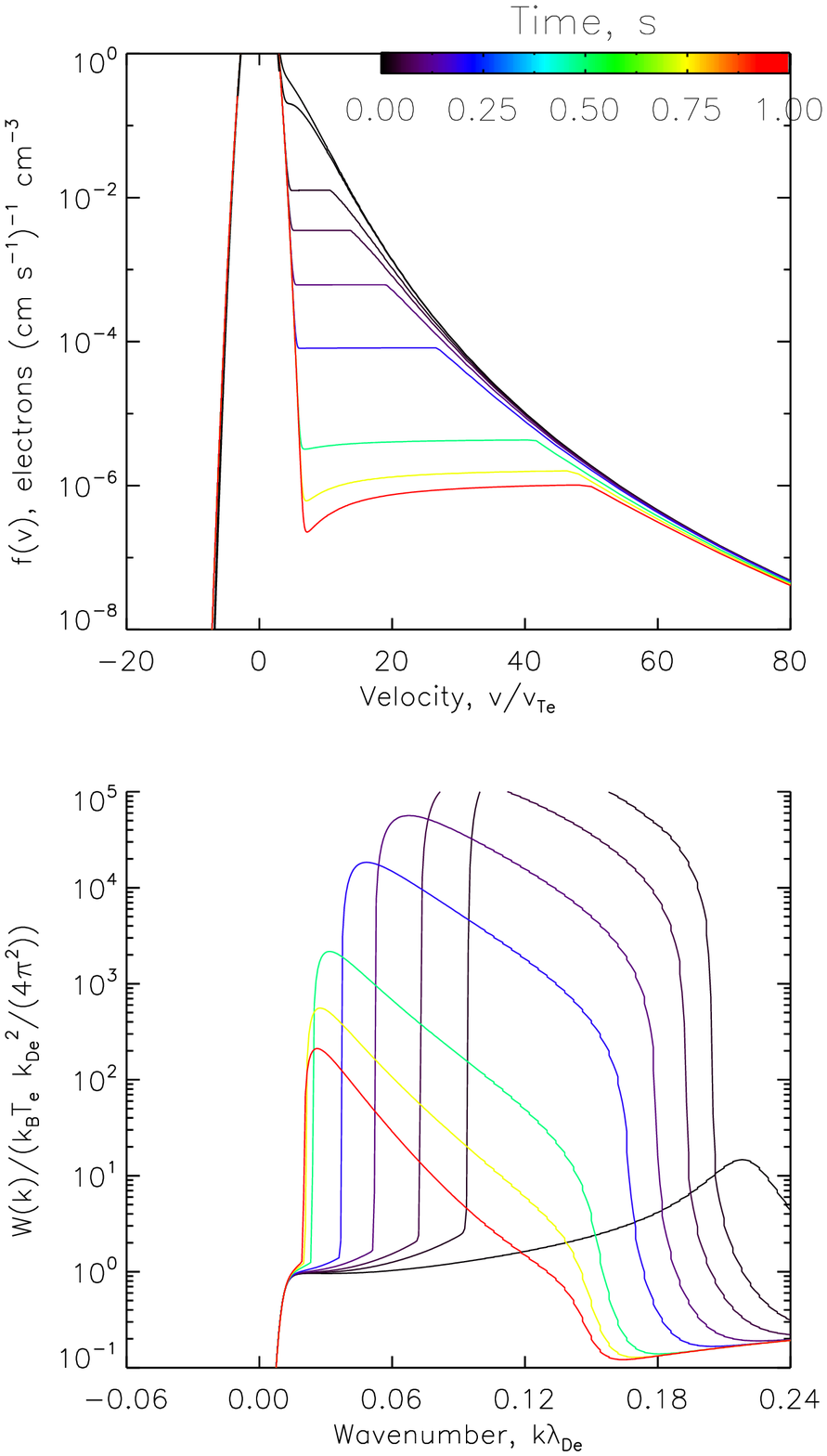}\\

\caption[Electron and wave distributions for collisionally relaxing electron beam: homogeneous plasma.]{Collisional relaxation of an electron beam in a plasma. Top: the electron distribution function $f({\mr v})$. Bottom: the spectral energy density of Langmuir waves $W(k)$. The left pair show relaxation without Langmuir wave generation, the right pair include this. Each coloured line shows the distribution at a different time, as shown in the colour bar.}
\label{fig:collisions}
\end{figure}

When wave generation is ignored, the initial electron distribution of Equation \ref{eq:f_t0} will evolve due to collisions as shown in the left panel of Figure \ref{fig:collisions}. The electrons lose energy, the beam distribution decreases, and a reverse slope is generated over the first $\sim 0.1$~s. The region of reverse slope gradually widens and moves to higher velocities as time progresses, but at the same time the number density of electrons in the reverse slope region decreases. 

When we include the effects of wave generation, this reverse slope is quickly flattened, as seen in the right panel of Figure \ref{fig:collisions}. The collisional effects continually reform the reverse slope, generating more Langmuir waves, and the combination gives a flat plateau and high Langmuir wave level, with the plateau height gradually decreasing as energy is lost due to collisions. At long times, the beam density becomes so low that few waves are generated and eventually, on timescales of several seconds, the wave level becomes so low that the distribution returns towards that of the collisions only case. On even longer time scales all energy is lost from the beam electrons, and the distribution returns to a Maxwellian at a slightly higher temperature than the initial distribution.

\subsubsection{Time Integrated Electron Distribution}

Because the slope in the electron distribution changes over time, and the reverse slope region moves to higher velocity, we cannot define a simple beam-wave interaction time like the quasilinear time in Equation \ref{eqn:tql}, as this will be time dependent. This makes it difficult to quantify when the diffusion of Langmuir waves in wavenumber will lead to electron acceleration, and how the density fluctuation parameters affect this.
\begin{figure} \centering
\includegraphics[width=6cm]{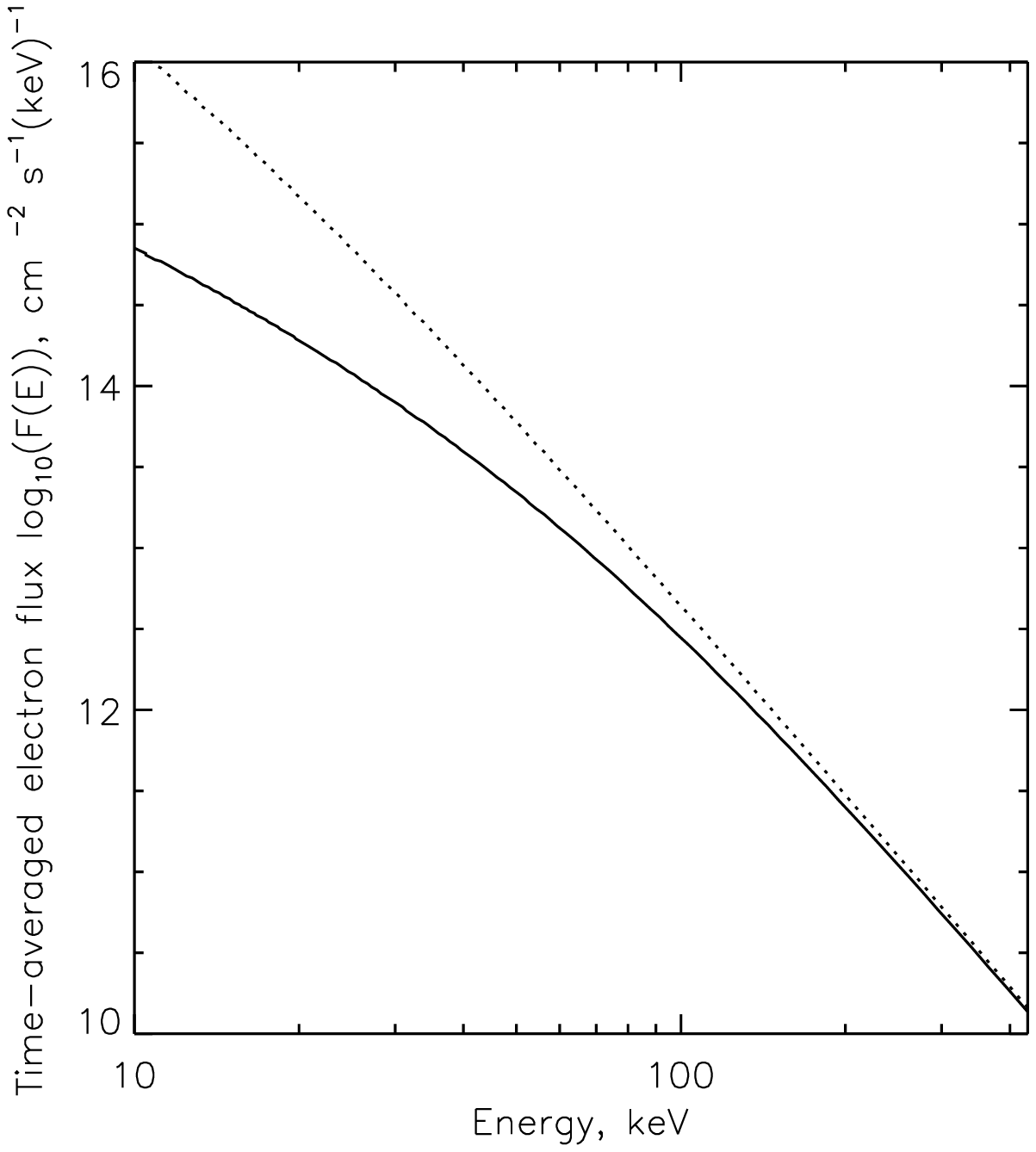}
\includegraphics[width=6cm]{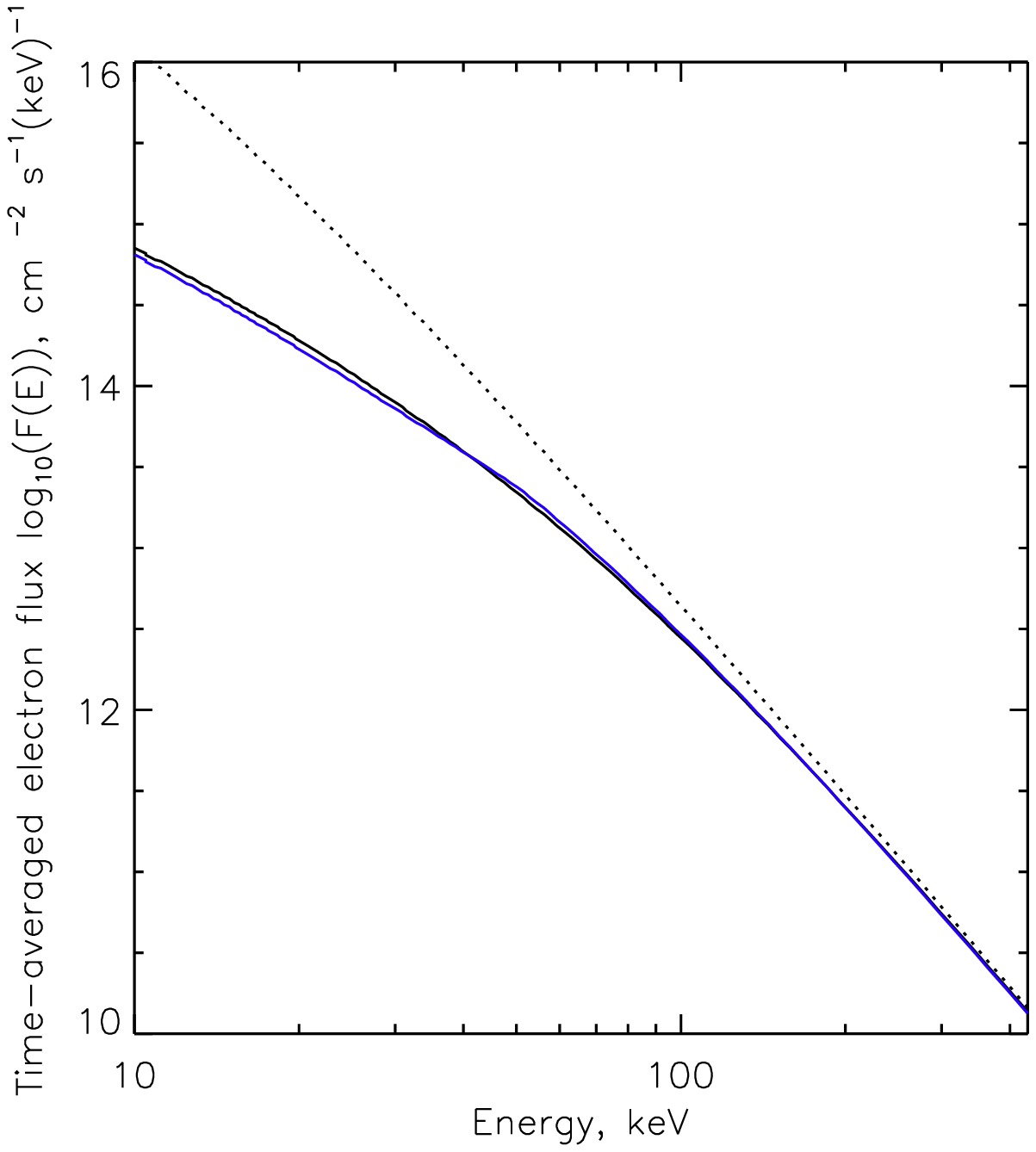}

\caption[Time-averaged electron flux: homogeneous plasma.]{The time averaged electron flux spectrum [electrons keV$^{-1}$ cm$^{-2}$ s$^{-1}$] as a function of electron energy. The dashed line shows the flux if the beam remained the initial power-law. The left panel shows the flux in the no-waves case, while the right panel shows both this (black line) and the case including Langmuir wave generation (blue line).}
\label{fig:collisionsFlux}
\end{figure}

Instead we consider the time-averaged electron distribution as function of energy, defined by \begin{equation} F(E)=\frac{1}{T}\int_0^T f(E,t) \rd t,\end{equation} where \begin{equation}\label{eqn:f_E}f(E,t)=f(\rv(E),t) m_e \rv(E)\end{equation} is the electron distribution as a function of energy. Here we consider the average over $T=1$~s. This time-averaged flux can be directly related to the HXR emission from the electrons, as discussed later in Section \ref{sec:XRayFlx}. Here we simply note that changes in this flux due to Langmuir wave evolution will be visible observationally via this HXR emission. 

In Figure \ref{fig:collisionsFlux} we show this time-averaged flux for the initial beam, and for the collisional evolution with and without wave generation. The latter are almost identical, confirming the results in \citet{1987ApJ...321..721H,1987A&A...175..255M, 2009ApJ...707L..45H,2011A&A...529A.109H} that Langmuir wave generation only very weakly affects the time integrated electron flux.

\subsection{A Constant Density Gradient}

The role of a density gradient in electron self-acceleration was originally considered by e.g. \citet{1969JETP...30..759B, 1969JETP...30..131R} in order to explain experimental observations of electron acceleration during beam relaxation. In the context of solar electron beams, the recent 1-D simulations by \citet{2002PhRvE..65f6408K} found such an electron acceleration effect due to an increasing plasma density. We reproduce this case here as it is a useful test for our simulations, and also calculate the time-averaged electron flux to compare to the homogeneous case. It is also interesting to compare this density gradient case to the fluctuating density considered in the next section.

Consider the Liouville equation, Equation \ref{eqn:liou}: \begin{equation}
\frac {\partial W(x, k, t)}{\partial t} -\frac{\partial \omega_{pe}}{\partial x}\frac{\partial W(x, k, t)}{\partial k}=\mathrm{Source\,\, terms}. 
\end{equation} In the case of an increasing density gradient we have
\begin{equation}\label{eq:omega_L}
\frac{\partial \omega_{pe}}{\partial x}\simeq\frac{\omega_{pe}}{L}
\end{equation}
where \begin{equation}\label{eq:L}
L\equiv \omega_{pe}(x)\left(\frac{\partial \omega _{pe}(x)}{\partial x} \right)^{-1}
=\frac{n_{e}(x)}{2}\left(\frac{\partial n_{e}(x)}{\partial x} \right)^{-1}
\end{equation}
is the characteristic scale of density inhomogeneity. The fractional change in Langmuir wavenumber $\Delta k$ due to the inhomogeneity must be small, $|\Delta k / k|\ll 1$ in order that we remain in the geometric optics (WKB) approximation \citep[e.g.][]{1967PlPh....9..719V}. 

The shift in wavenumber experienced by the Langmuir waves is \begin{equation}\Delta k \simeq \frac{\pm\omega_{pe} \Delta t}{|L|}. \end{equation} From this we can estimate the required density gradient for a significant shift in $k$. For a timescale of the order of the collisional time, $10^{-4}$~s, and a plasma frequency of 2~GHz, taking $\Delta k= 0.05 k_{De}$ we find $L\sim 10^6$~cm. 

\begin{figure}
\centering
\includegraphics[width=6cm]{fig2a.eps}
\includegraphics[width=6cm]{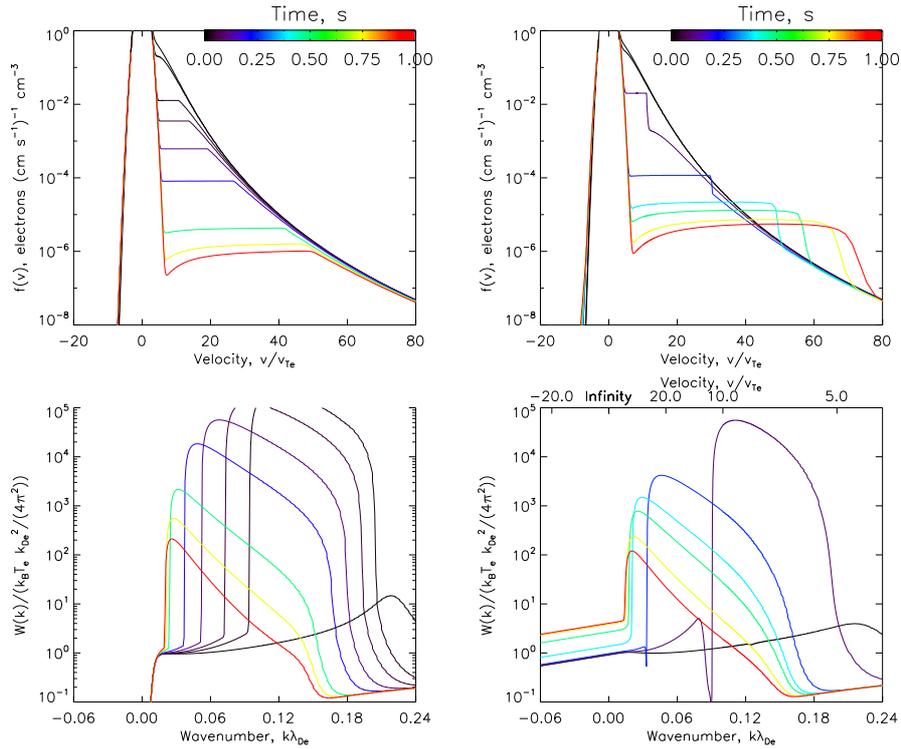}\\
\caption[Electron and wave distributions for collisionally relaxing electron beam: plasma with a constant density gradient.]{Electron and wave distributions as as in Figure \ref{fig:collisions}, for homogeneous plasma (left) and plasma with a constant density gradient as in Equation \ref{eq:omega_L}, with $L=10^6$~cm.} \label{fig:Grad}
\end{figure}

In Figure \ref{fig:Grad} we show the electron and wave spectra in plasma with a constant density gradient of length scale $L=10^6$~cm, and also the homogeneous case for comparison. Several interesting features may be seen. Firstly, as expected we see the advection of Langmuir waves towards smaller wavenumbers, and the consequent transfer of energy from small to large velocities. The plateau in the electron distribution is broadened. The rate of energy loss from the beam is also decreased, with the widened plateau decreasing in height significantly slower than in the homogeneous or collisions-only cases. Also of interest is the fact that the increased electron number density seen between $\sim 40 \rv_{Te}$ and $\sim 80 \rv_{Te}$ not only exceeds the number in the homogeneous case but also the number in the initial power law beam. After $1$~s, at $60 \rv_{Te}$ we have a tenfold or more increase in the electron distribution above this initial value. 

In the left panel of Figure \ref{fig:FluxGradFlucs} we show the time averaged electron flux for the homogeneous case and the case of a constant density gradient. Again there are several features of interest. We see a clear increase in electron flux between around $20$ and $200$~keV, reaching perhaps an order of magnitude over the homogeneous case at its peak. The flux also significantly exceeds that from the initial beam distribution. In addition, there is a distinctive kink in the flux spectrum, which shows a relatively linear distribution from 10 to 200~keV then drops sharply back to be the same as in the homogeneous case. 

We may conclude that the presence of a density gradient in the plasma is able to produce electron self-acceleration in our simulation model, and moreover that this change in the electron distribution has significant effects on the time-averaged electron flux, and therefore the HXR emission.    

\subsection{Density Fluctuations}
\begin{figure}
\centering
\includegraphics[width=6cm]{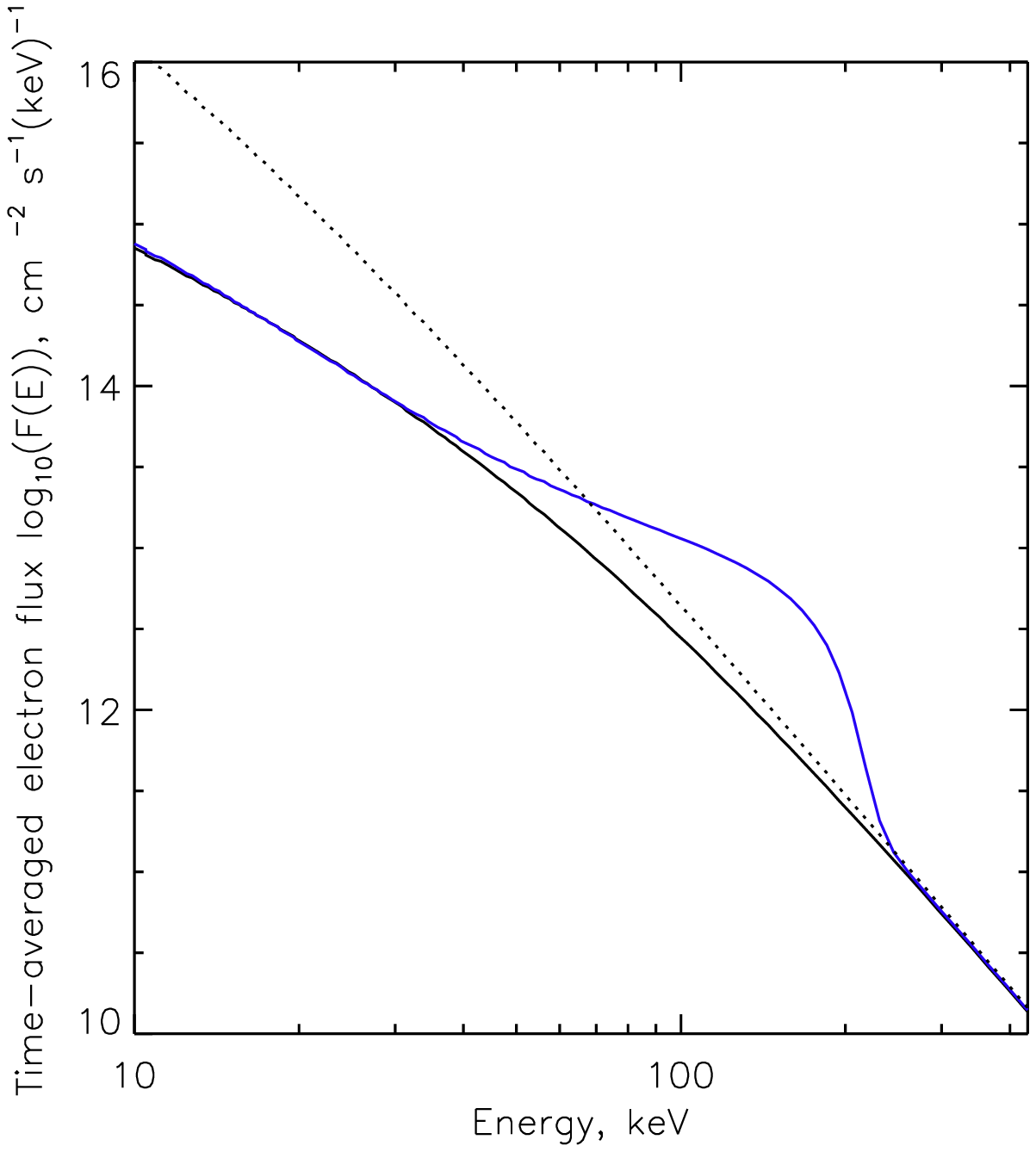} \includegraphics[width=6cm]{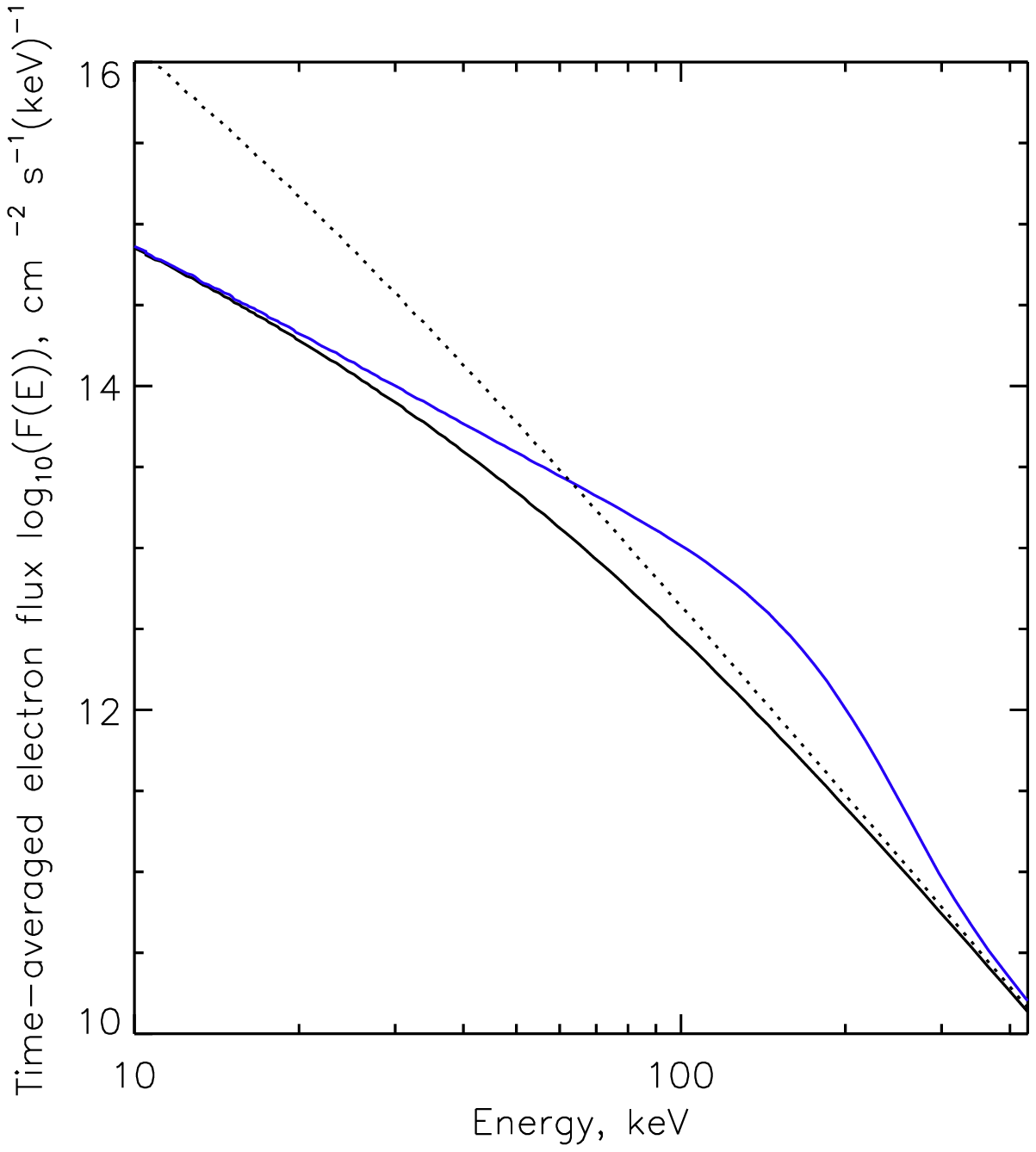}

\caption[Time-averaged electron flux: plasma with a constant density gradient, or density fluctuations.]{The time averaged electron fluxes. Dashed lines: the initial power-law beam. Black lines: homogeneous plasma. Blue lines: constant density gradient (left panel); fluctuating density (right panel).} \label{fig:FluxGradFlucs}
\end{figure}

In the previous section we considered the effects of plasma density fluctuations on a Maxwellian electron beam in detail, and saw a significant electron acceleration effect, controlled primarily by the parameter $R=\tau_{ql}/\tau_D$, the ratio of the beam-plasma interaction and Langmuir wavenumber diffusion timescales. In the current case of a power-law beam relaxing due to collisions, we cannot define a simple analog of the beam-plasma interaction timescale found in Equation \ref{eqn:tql}, as the actual Langmuir wave growth rate is the convolution of wave growth and the growth of the reverse slope $\partial f/\partial \rv$. However, we can still infer that the maximum acceleration effect will occur when the two processes operate at approximately equal rates. 

In addition, we saw that the details of the density fluctuation spectrum and parameters were comparatively unimportant. Therefore we consider here only random fluctuations, with a diffusion coefficient given by Equation \ref{eqn:diffcoeffGauss}, assume as before that $q_0=10^4 k_{De}$ and take their characteristic velocity to be $\rv_0 =10^7$~cm~s$^{-1}=0.3 \rv_{Te}$ and the density fluctuation magnitude to be $\sqrt{\langle \tilde{n}^2\rangle} =10^{-3}$. These parameters were found by a few trials to lead to the most significant acceleration for these beam parameters.

\begin{figure}
\centering
\includegraphics[width=6cm]{fig2a.eps}
\includegraphics[width=6cm]{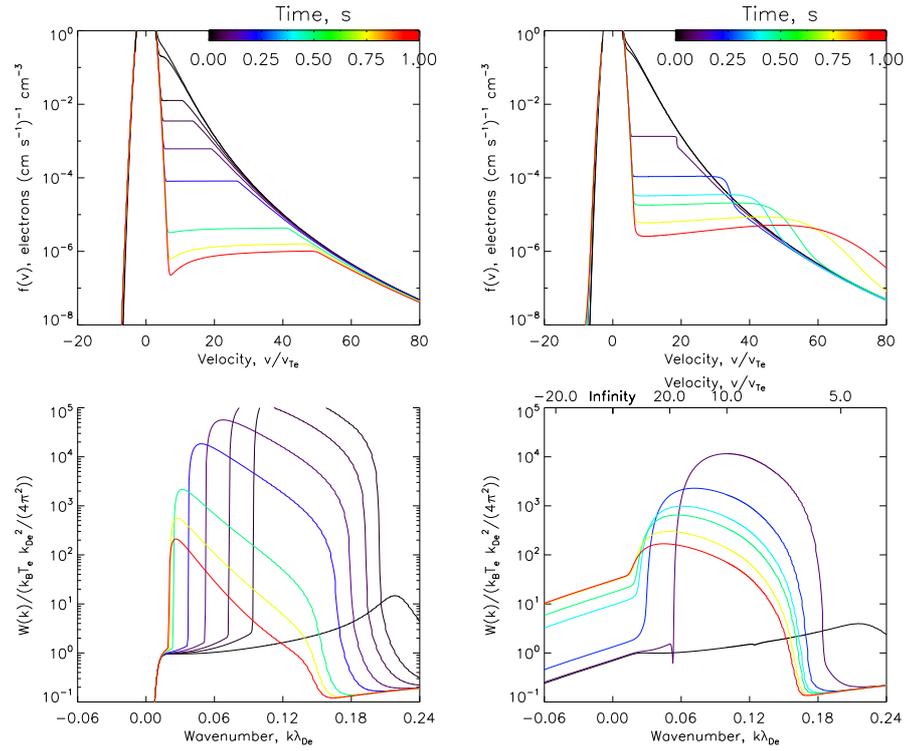}
\caption[Electron and wave distributions for collisionally relaxing electron beam: plasma with random  density fluctuations.]{Electron and wave distributions as as in Figure \ref{fig:collisions}, for homogeneous plasma (left) and plasma with random density
fluctuations, given by Equation \ref{eqn:diffcoeffGauss}, with parameters as stated in the text.}
\label{fig:diffusion}
\end{figure}

The resulting electron and wave distributions are shown in Figure \ref{fig:diffusion}. In comparison to the homogeneous or density gradient cases, we see that a fluctuating density leads to spreading of the Langmuir waves to both larger and smaller wavenumbers, with a consequent decrease in their peak value. However, compared to the density gradient case, as expected, the acceleration effect is neither as significant, nor as sharply cut off. The time-averaged electron flux shown in the right panel of Figure \ref{fig:FluxGradFlucs} again shows a significant increase from around 20~keV up to 300~keV in this example, with a kink around 200~keV, similar to the case of a constant density gradient shown in the left panel. 

One final point may be mentioned here. Because in this case the condition for strong acceleration, $R=1$, cannot be satisfied for the entire beam relaxation process, it may be of interest to consider a time-varying level of fluctuations. The density fluctuations will be damped by their interaction with Langmuir waves, and therefore decay. A self consistent treatment of the problem is possible in the framework in the previous chapter, by including the evolution of the density fluctuations. This has been done in the past by, for example, \citet{1967PlPh....9..719V, vedenov1968theory}, and would be interesting to include in future simulations. 

\subsection{Wave-wave Interactions}\label{sec:wave_wave}

Langmuir waves are subject to various scattering and decay processes and can themselves excite other wave modes. For instance, scattering off thermal ions can backscatter a Langmuir wave, or transform it into an electromagnetic wave, as will be discussed in the next chapter. Three-wave interactions of the form $X+Y \rightleftarrows Z$ are the primary source of ion-sound waves in the plasma, as their large damping rates mean they cannot persist and must be rapidly generated. In general there are two kinematically allowed processes which can generate an ion-sound wave (denoted $s$) from a Langmuir wave (denoted $L$), decay to a scattered Langmuir wave $L'$: $L \rightleftarrows L'+ s$ and decay or coalescence giving an electromagnetic wave (transverse, denoted $t$) $L \pm s \rightleftarrows t$. 

The latter process will be considered in the following chapter as it is an essential component in the plasma emission mechanism, but for the parameters considered here may be shown to have negligible effect on the Langmuir waves. For the former process, energy and momentum conservation state that the parent Langmuir wave at wavenumber $k_L$ will scatter to approximately the opposite wavenumber $k_{L'}\simeq -k_L$ and produce an ion-sound wave at wavenumber $q \simeq 2 k_L$. More precisely, the scattered wavenumber is decreased by a small increment $\Delta k$, \begin{equation}\label{eq:deltaK} \frac{\Delta k}{k_{De}}=\frac{2}{3}\sqrt{\frac{m_e}{m_i}}\sqrt{1+\frac{3T_i}{T_e}} \sim \frac{1}{30}\end{equation} and so successive scatterings will lead to the generation of Langmuir waves at smaller and smaller wavenumbers. Eventually this forms a so-called Langmuir wave condensate \citep[e.g.][]{2001PhPl....8.3982Z,2002PhRvE..65f6408K,2011ApJ...727...16Z} at very small wavenumbers. 

For waves with comparable wavenumbers, the quasi-particle treatment of Section \ref{sec:diffEq} is no longer valid, and so we must consider these decays individually, using the three-wave equations \citep[e.g.][]{1980MelroseBothVols, 1995lnlp.book.....T}. In practise, these equations may be reduced to 1-D in a similar manner to the quasilinear equations. Here we use an implementation and simulation code as in \citet{2002PhRvE..65f6408K}. We add a source term ${\rm St}_{decay}(W,W_S)$ due to the decay $L \rightleftarrows L'+s$ to the right hand side of Equation \ref{eqn:3ql2} and a new equation to the set describing the evolution of the ion-sound waves, described by their spectral energy density $W_S$: 

\begin{align}\label{eqn:3ql_s}
\frac{\partial W_S(k)}{\partial t}=&-2\gamma_S(k)
W_S(k)\notag \\&-\alpha ({\omega^S_k})^2\int
\left(
\frac{W(k_L)}{\omega^L_{k_L}}\frac{W_S(k)}{\omega^S_{k}}-
\frac{W(k_{L'})}{\omega^L_{k_{L'}}}\left(\frac{W(k_L)}{\omega^L_{k_L}}+
\frac{W_S(k)}{\omega^S_{k}}\right)\right)\times\notag \\&
\delta(\omega^L_{k_{L\prime}}-\omega^L_{k_L}-\omega^S_k)dk_{L\prime}
\end{align} where $k_L$ is the initial Langmuir wavenumber and  $k_L'$ is the scattered wave.

The source term for the Langmuir waves is
\begin{align}\label{eqn:3ql_sSrc}
&{\rm St}_{decay}(W(k),W_S(k_S))=\alpha\omega_{k} \times \notag \\ &
\int\omega_{k_S}^S\left[ \left(
\frac{W(k_L)}{\omega^L_{k_L}}\frac{W_S(k_S)}{\omega^S_{k_S}}-
\frac{W(k)}{\omega^L_k}\left(\frac{W(k_L)}{\omega^L_{k_L}}+
\frac{W_S(k_S)}{\omega^S_{k_s}}\right)\right)\delta (\omega^L_{k}-\omega^L_{k_L}-\omega^S_{k_S})\right.
\notag \\&
-\left.
\left(
\frac{W(k_{L'})}{\omega^L_{k_{L'}}}\frac{W_S(k_S)}{\omega^S_{k_S}}-
\frac{W(k)}{\omega^L_k}\left(\frac{W(k_{L'})}{\omega^L_{k_{L'}}}-
\frac{W_S(k_S)}{\omega^S_{k_s}}\right)\right)
\delta(\omega^L_{k}-\omega^L_{k_{L'}}+\omega^S_{k_S}) \right.\biggr] \,dk_S
\end{align} where we distinguish between $k_L= k-k_S$ and $k_{L'}=k+k_S$, and the constants $\alpha, \beta$ and the sound wave damping rate are 
\begin{equation}\label{alphaL}\alpha=\frac{\pi \omega^2_{pe}(1+3T_i/T_e)}{4n_ek_b T_e} \;\;
\beta=\frac{\sqrt{2\pi}\omega^2_{pe}}{4n_ek_b T_i(1+T_e/T_i)^2},
\end{equation}
\begin{equation}\label{gam_sk}\gamma_S(k)=\sqrt{\frac{\pi}{8}}\omega^S_k\left[\frac{\rv_s}{\rv_{Te}}+\left(\frac{\omega
^S_k}{k\rv_{Ti}}\right)^3\exp\left[- \left(\frac{ \omega ^S_k}{
k\rv_{Ti} }\right)^2 \right]\right].
\end{equation}

\subsubsection{Prescribed Ion-sound Wave Level}

We begin with the simple situation of a fixed level of ion-sound waves, generated by some external source. We set this to the thermal level,
\begin{equation}W_S(k)= k_B T_e k_{De}^2 \frac{k_{De}^2}{k_{De}^2+k^2} \end{equation} and neglect its evolution. 

\begin{figure}
\centering
\includegraphics[width=6cm]{fig4a.eps}
\includegraphics[width=6cm]{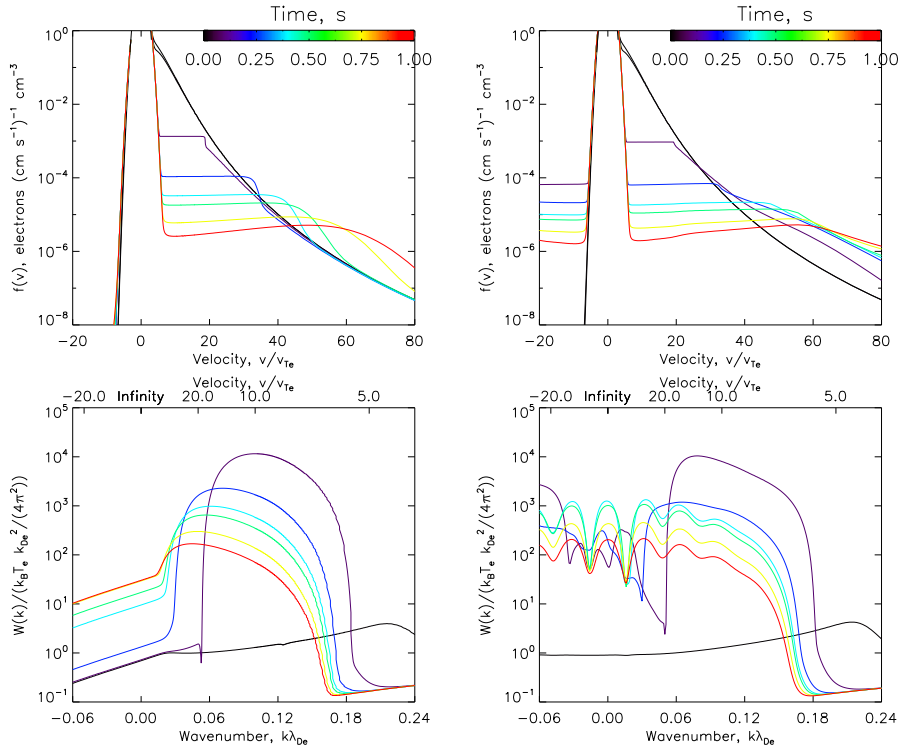}
\caption[Electron and wave distributions for collisionally relaxing electron beam: effects of wave-wave interactions.]{Electron and wave distributions as in Figure \ref{fig:collisions}, for plasma with random density
fluctuations as in Figure \ref{fig:diffusion}, without (left) and with (right) scattering due to a fixed level of ion-sound waves.}
\label{fig:random_nonlin}
\end{figure}

\begin{figure}
\centering
\includegraphics[width=6cm]{fig4b.eps}
\includegraphics[width=6cm]{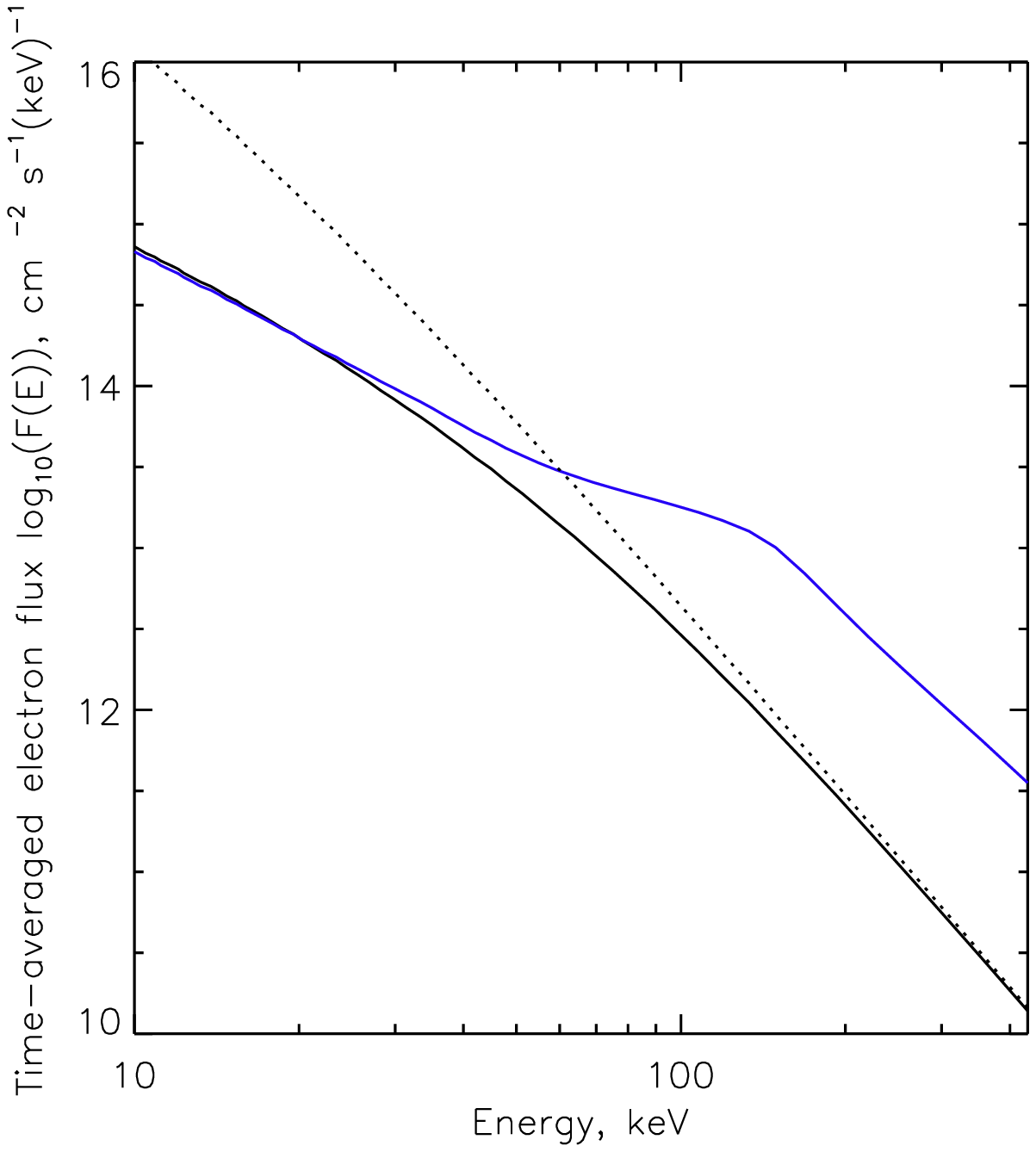}
\includegraphics[width=6cm]{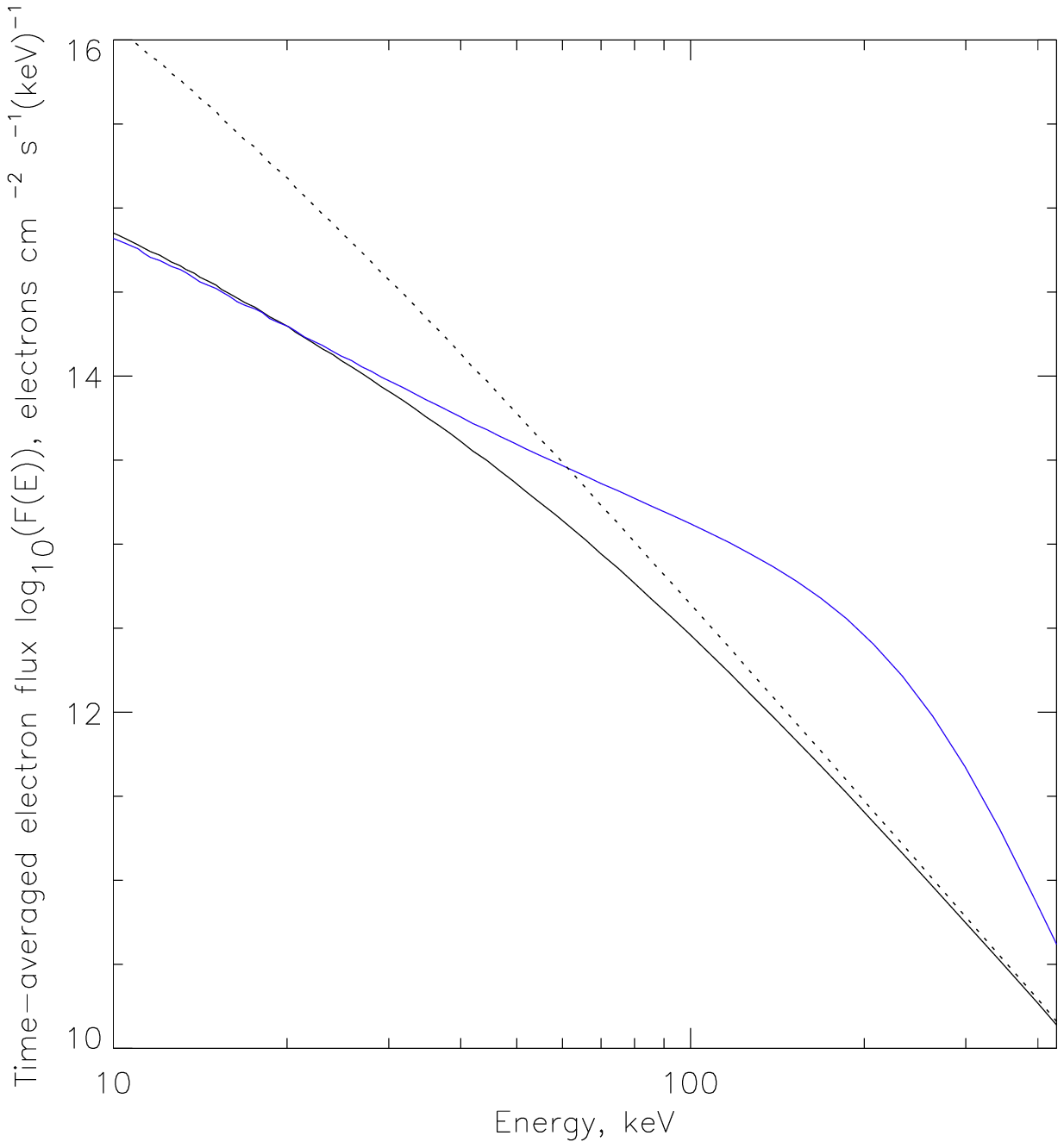}
\caption[Time-averaged electron flux: effects of wave-wave interactions.]{Time-averaged electron fluxes as as in Figure \ref{fig:collisionsFlux}, for plasma with random density
fluctuations as in Figure \ref{fig:FluxGradFlucs}. Top left: no wave-wave interactions. Top right: fixed level of ion-sound waves. Bottom: fully self-consistent ion-sound wave interactions.}
\label{fig:Fluxrandom_nonlin}
\end{figure}

The results are shown in Figures \ref{fig:random_nonlin} and \ref{fig:Fluxrandom_nonlin}. The strong variations in Langmuir wave level at small wavenumbers are due to repeated scatterings by the ion-sound waves, the thermal level of which is relatively large at small wavenumbers. We see from Equation \ref{eq:deltaK} that the Langmuir wavenumber $k/k_{De}$ decreases by $0.03$ on each scattering. As will be seen in the next subsection, when the ion-sound waves are allowed to evolve self-consistently these strong variations are not seen.

It should be noted that at the extreme high energies a relativistic treatment is required, and so the effect above around 300~keV is not meaningful. The most significant differences in the two cases are seen at large energies, 100 to 300~keV, where we see significantly more acceleration.  The Langmuir wave scattering is particularly effective at the small wavenumbers corresponding to these high velocities. The other interesting feature of Figure \ref{fig:random_nonlin} is the appearance of electrons at negative velocities. These correspond to electrons accelerated by the back-scattered Langmuir waves. 

\subsubsection{Self-Consistent 3-Wave Scattering}

Fixing the level of ion-sound waves can lead to energy being gained by the Langmuir waves and not lost from the sound waves, but as the energy in sound waves is far less than that in Langmuir waves, this effect is generally small. The self-consistent treatment does however lead to slightly slower scattering because of the strongly peaked sound wave spectrum, and the resulting Langmuir wave spectrum does not show as clearly the successive scattering peaks. On the other hand, the main features of the acceleration are preserved, although the exact spectral shape can be altered. 

\begin{figure}
\centering
\includegraphics[width=6cm]{fig5a.eps}
\includegraphics[width=6cm]{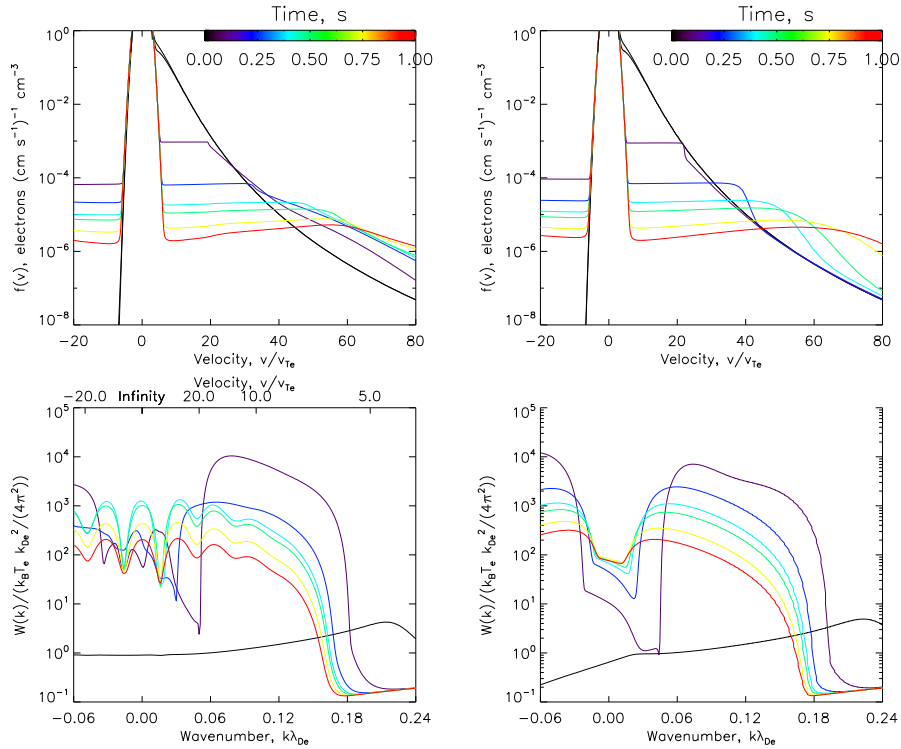}
\caption[Electron and wave distributions for collisionally relaxing electron beam: effects of fully self-consistent wave-wave interactions.]{Electron and wave distributions as in Figure \ref{fig:collisions}, for plasma with random density
fluctuations as in Figure \ref{fig:diffusion}, and scattering due to ion-sound waves for a fixed sound wave level (left) and in the self-consistent treatment (right).}
\label{fig:random_nonlin_2}
\end{figure}

An example of self-consistent ion-sound wave scattering with significant density fluctuation induced wavenumber diffusion included is shown in Figures \ref{fig:random_nonlin_2} and \ref{fig:Fluxrandom_nonlin}. The main differences between the fixed and self consistent cases are seen to occur around very small wavenumbers, or energies above 300~keV. In this regime, these simulations start to become invalid due to relativistic effects, so the significance of this is small. 

\subsection{The Effects of Beam Density}
\begin{figure}
\centering
\includegraphics[width=6cm]{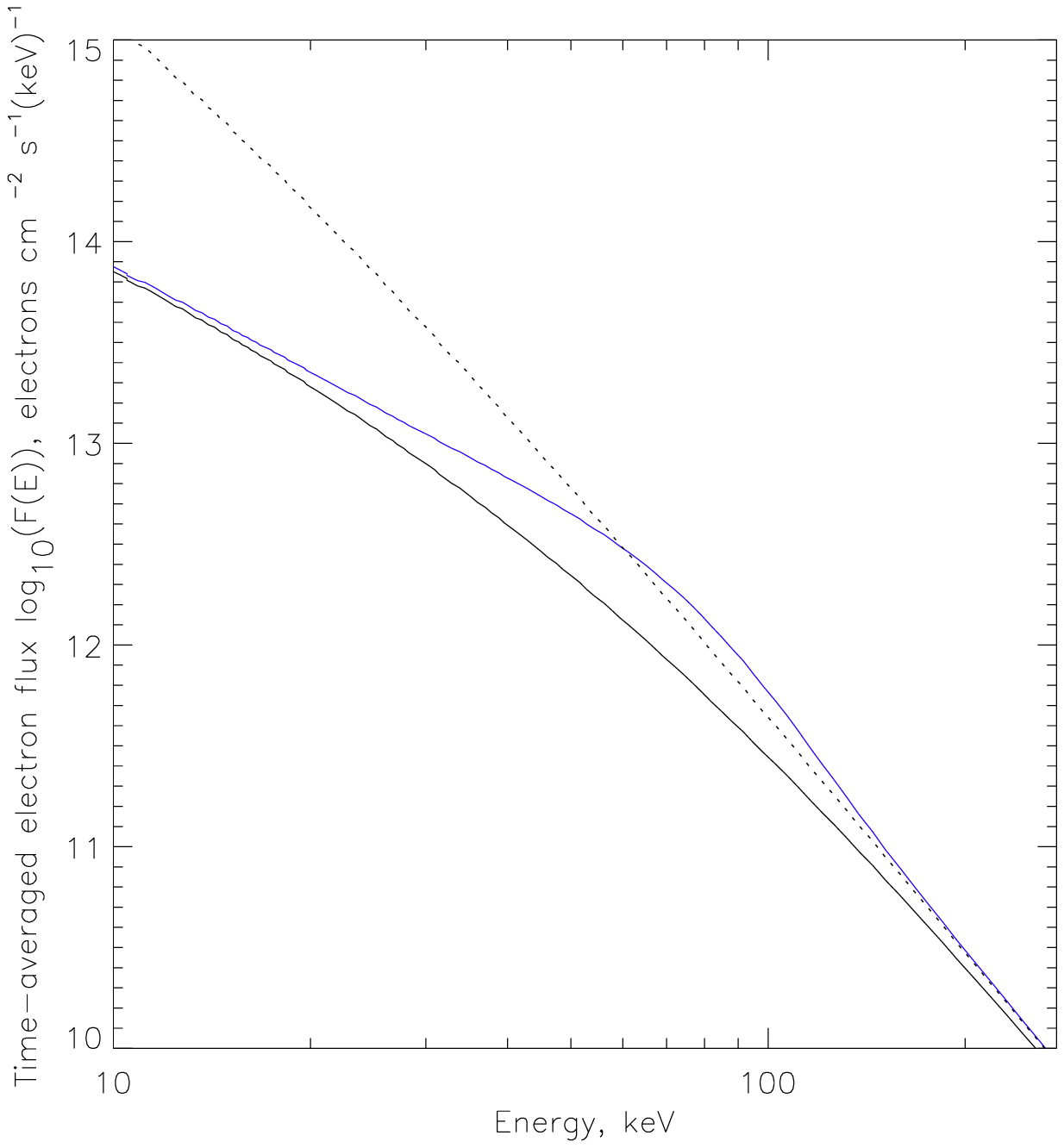}
\includegraphics[width=6cm]{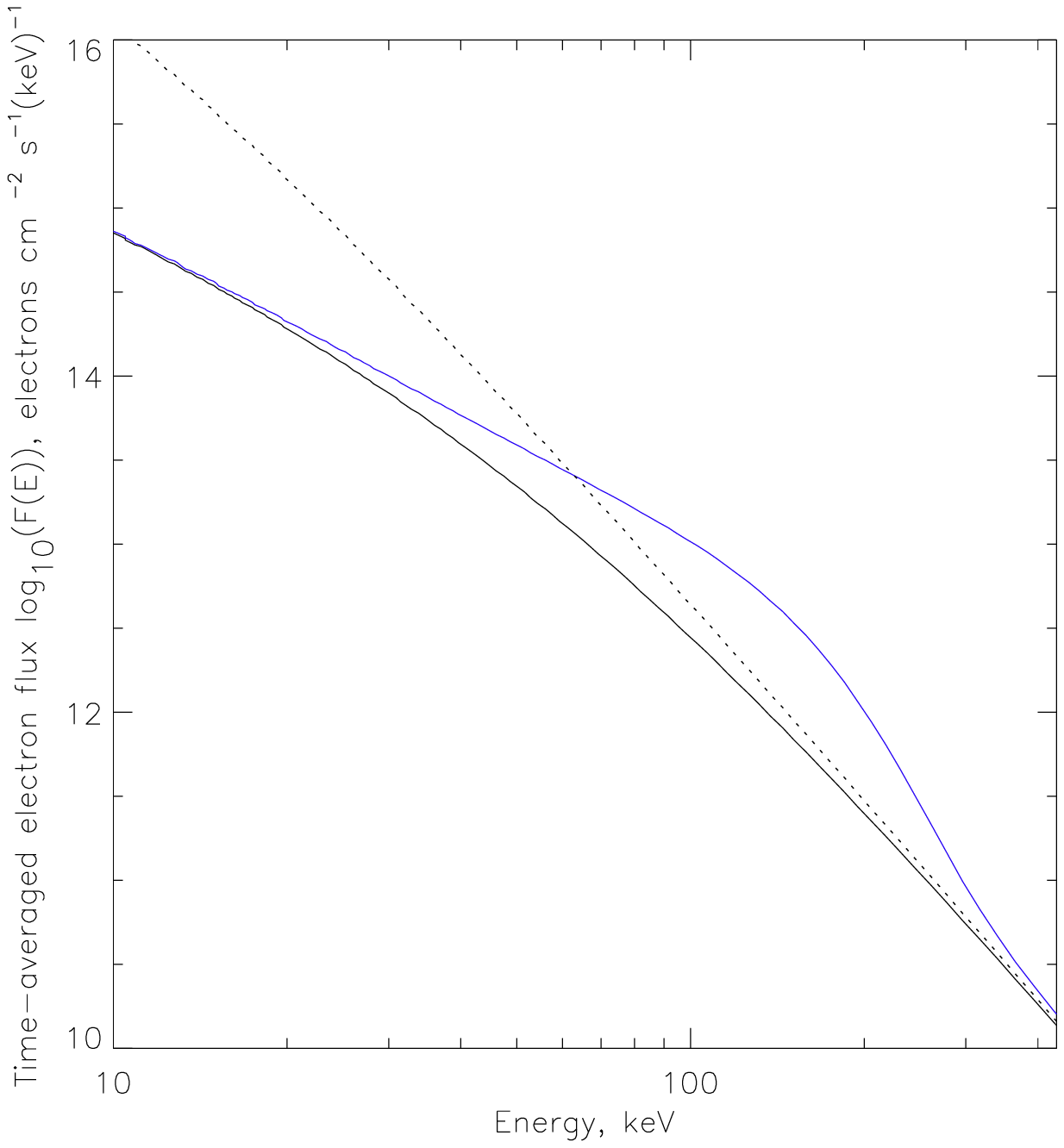}
\caption[Time-averaged electron flux: effects of beam density.]{Time-averaged electron fluxes for an initial beam with density $n_b=10^8$cm$^{-3}$ (left) and $n_b=10^9$cm$^{-3}$ (right) each showing the initial beam (dashed line), the collisions only case (black line) and the case with fluctuating density (blue line).}
\label{fig:beamDens}
\end{figure}

As the acceleration effect we are considering relies on the transfer of energy via Langmuir waves, we expect that a denser beam would allow more acceleration to occur. This was confirmed already for the case of a Maxwellian beam. In Figure \ref{fig:beamDens} we show examples of the acceleration for beam densities of $n_b=10^8$~cm$^{-3}$ and $n_b=10^9$~cm$^{-3}$. In both cases, the ``most effective'' acceleration was found by a few trials: as expected from the Maxwellian case the corresponding fluctuation parameters depend on beam density in order that the timescales for the processes are comparable. We see that the acceleration in both cases occurs in a similar region in energy: 20 to 200~keV in the weaker beam case and 20 to 300~keV in the stronger beam case, but the magnitude of the effect is strongly dependent on the beam density. 

\section{Energy Transfer Due to Diffusion}

\begin{figure}
\centering
\includegraphics[width=7cm]{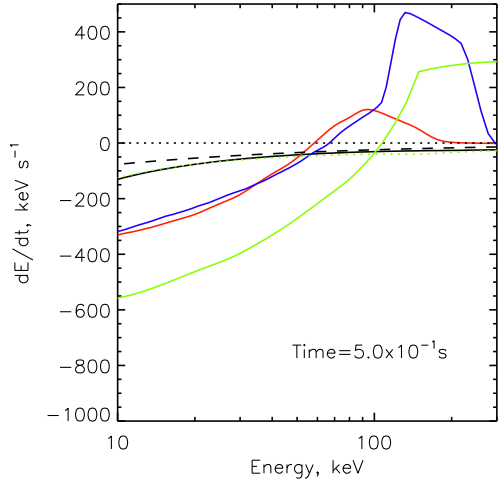}
\caption[Energy loss rates of collisionally relaxing electrons.]{The effective energy loss rate of an electron $\langle{\partial E}/{\partial t}\rangle_{eff}$ at 0.5s. Theoretical collisional losses, $\langle{\partial E}/{\partial t}\rangle_{eff} =-\Gamma\sqrt{m^3/(2E)}$ (Equation \ref{eq:collAn}) (black dashed line). Numerical loss rates from simulation results for: collisional losses as in Figure \ref{fig:collisions} (black line);
Constant density gradient as in Figure \ref{fig:Grad} (blue line); Random density
fluctuations as in Figure \ref{fig:diffusion} (red line); Wave-wave interactions
in inhomogeneous plasma as in Figure \ref{fig:random_nonlin} (green line).}
\label{fig:dE_dt}
\end{figure}

The acceleration process which we discuss is due to energy transfer between electrons at different velocities, with the total energy of the system remaining constant. Therefore, if we sum the energy loss and gain rate across electrons at all energies the result cannot be positive. 
In the purely collisional case, the effective energy loss rate, $\langle d E/d t\rangle_{eff}$ can be calculated analytically. This is done by assuming a single particle evolution \citep[e.g.][]{2009A&A...508..993B}, and writing a continuity equation for the energy distribution function in Equation \ref{eqn:f_E}. In \citet{KRB} this is given as 
 \begin{equation}\label{eq:continuity}
    \frac{\partial f(E,t)}{\partial t}+\frac{\partial }{\partial E}\left(\langle\frac{d E}{d t}\rangle_{eff}f(E,t)\right)=0
\end{equation} with solution in the purely collisional case \begin{equation}\label{eq:collAn}\langle{\partial E}/{\partial t}\rangle_{eff} =-\Gamma\sqrt{m^3/(2E)}.\end{equation} We can also calculate numerically the effective energy loss rates for the various simulation models used. These are plotted in Figure \ref{fig:dE_dt}. 

The changes in energy loss rate are consistent with the observed acceleration. For purely collisional evolution, this is small and negative, and varies slowly with energy. Once density variations are considered, we see that the losses at small energy increase, but those at larger energies decrease, and in fact become positive, i.e particles at large energies are gaining energy. Density fluctuations lead to increased energy in fast electrons, but net losses over the energy range shown in the figure. Ion-sound wave interactions similarly show significantly increased overall losses, partly due to energy transfer to backwards propagating electrons. 

\section{Hard X-ray Flux}\label{sec:XRayFlx}

The electron flux averaged over the collisional lifetime of the electrons, $F(E)$, is directly related to the observed HXR emission via the cross-section for bremsstrahlung \citep{2003ApJ...595L.115B}, and although recovering the electron flux from the photon distribution is difficult mathematically, it requires no assumptions about the physical processes which may cause evolution of the electron distribution. In the previous few sections we have seen that the evolution of Langmuir waves due to their interaction with plasma density fluctuations can significantly affect this time-averaged electron flux, in contrast to the situation in homogeneous plasma (Figure \ref{fig:collisions}) where the changes are negligible. We thus conclude that the Langmuir wave evolution will affect the observed HXR emission. 

Specifically, we see significant changes in the electron flux above 20~keV, up to around 200~keV. Below around 10-20~keV the HXR emission is dominated by the thermal spectrum, and so this evolution transfers energy from the unobservable part of the distribution to the observable energies. Thus the energy and number of electrons required in the beam as a whole is reduced when compared to the purely collisional case. Stated another way, we may obtain more intense HXR emission above 20~keV from a less dense initial beam. Therefore, the processes discussed here are an important consideration when interpreting such HXR spectra.

\section{Conclusions}

For a collisionally relaxing electron beam, the generation of Langmuir waves, and their subsequent evolution due to variations in the plasma density, has been shown to lead to a significant electron acceleration effect. This follows the conclusions for a Maxwellian beam, where acceleration was also seen. We may extrapolate from the result in the Maxwellian test case that the strongest acceleration occurs for $R\sim1$ that smaller scale and or more intense density fluctuations will lead to more acceleration. 

In the power law beam case, we may expect suppression of the beam plasma instability due to very rapid diffusion to lead to the return of the electron spectrum to the collisions only case, which is indistinguishable observationally from that of homogeneous plasma including Langmuir wave generation.

Small wavenumber ($q \sim k_{De}$) ion-sound waves are generated automatically due to the presence of high levels of Langmuir waves, and cause scattering of Langmuir waves to smaller wavenumbers and therefore a similar acceleration effect. In this case the acceleration can reach much larger velocities. However, in general, there is no significant distinction observationally between the acceleration due to a density gradient, a fluctuating density or ion-sound wave scattering, and the differences are only in degree. 

For the plasma density and beam parameters chosen, the acceleration increases the time-averaged electron flux in exactly the important range for observations, namely between 20 and 200~keV, and can increase the flux by a few or perhaps ten times in the most effective cases. Finally, we see that as the extent of acceleration is constrained by the level of Langmuir waves generated, the acceleration will be more effective for more intense beams. In other words, ignoring the effects of Langmuir wave generation and evolution will lead to a significant overestimate of the required number of initially accelerated electrons, and the magnitude of the overestimate will be larger the more electrons there are in the initial beam.

\chapter{Simulations of Radio Emission from Dense Coronal Loops}
\renewcommand{\vec}[1]{\mathbf{#1}}
\newcommand{\AAvL}{Av}
\newcommand{\AAv}{Av}
\newcommand{\AATxt}{angle-averaged }
\label{ref:Chapter4}

The high levels of Langmuir waves which can be generated by a fast electron beam can, as discussed in Section \ref{sec:plasEmm}, lead to intense radio emission. The classic example of emission via this ``plasma emission'' mechanism is given by the radio bursts known as Type IIIs. Their intense brightnesses indicate a coherent emission mechanism, and their wide frequency range, with bursts observed from the low kHz up to 500~MHz, and occasionally up to a few GHz, means that this mechanism must operate over a very broad range of parameters. For example, in a high frequency coronal burst we may have a density around $10^{10}$~cm$^{-3}$ and a temperature of 1~MK, while for an interplanetary burst at 1~MHz we have a density of $10^{4}$~cm$^{-3}$ and a temperature of perhaps 10000~K. 

\citet{1958SvA.....2..653G} first proposed a mechanism by which these bursts could be produced, and this has been subsequently discussed and modified by various authors \citep[e.g.][]{1964NASSP..50..357S,1970SvA....14...47Z, 1970SoPh...15..202S,1976SoPh...46..515S,1980SSRv...26....3M, 1983SoPh...89..403G,1985ARA&A..23..169D, 1987SoPh..111...89M}.  Yet in spite of the large amounts of work invested in the problem, the exact details of their production are still not fully understood. It is known that they occur due to an accelerated electron beam propagating in the decreasing density plasma of the corona and solar wind, which generates Langmuir waves and subsequently electromagnetic emission at the plasma frequency. 

There are thus three main factors which influence their production. Firstly, the details of electron beam acceleration and propagation; secondly, the production of the Langmuir waves and their spectral evolution; and finally, the process of converting some fraction of the Langmuir wave energy into electromagnetic emission. In addition to these factors, when considering remotely observed radiation we must account for the effects of propagation, such as time delay in the emission, scattering and absorption. 

In the solar wind it is possible to make in situ measurements of electric field, and so infer the levels of Langmuir waves, and also to measure the electron distribution function. Low frequency compressive electrostatic modes are also found in association with these Langmuir waves, which may be identified as ion-sound waves \citep[e.g.][]{1986ApJ...308..954L, 1977JGR....82..632G,1978JGR....83...58G, 1998ApJ...498..465T}, and provide further evidence regarding the evolution of the Langmuir waves, and the production of radio emission. However, deeper in the corona, the observed radio emission is the primary diagnostic of the Langmuir waves. Where such emission occurs, we can find a lower limit on the Langmuir wave levels required, but in general it is very difficult if not impossible to analytically relate the level of Langmuir waves to the level of radio emission. 

Simulations are therefore essential to address the details of the Langmuir wave evolution, and the inclusion of radio emission in these is necessary to relate this to observations. In this chapter, we first develop a model of plasma radio emission, and subroutines allowing this to be incorporated into our code from the previous chapter. We then consider effects of plasma density inhomogeneities on radio emission due to their effects on the Langmuir waves and electrons, for the model of a collisionally relaxing electron beam in dense plasma. 

\section{The Plasma Emission Mechanism}
The mechanism behind plasma emission was outlined in Section \ref{sec:plasEmm}. For emission at GHz frequencies which we consider here, only harmonic emission will be of interest. This is largely due to the very large optical depth for fundamental emission (see Section \ref{sec:collisDamp}), and also the low efficiency of the fundamental emission mechanisms. Thus, the processes of interest are the decay $L\rightleftarrows L' + s$ and scattering by ions $L + i\rightleftarrows L' + i'$ of Langmuir waves, and the coalescence of two Langmuir waves to produce emission at twice the local plasma frequency, $L + L' \rightleftarrows t$. The observational features of the emission at these frequencies were outlined in Section \ref{sec:CoronalIII}.

\subsection{Simulating Plasma Radio Emission}

We know that the production of plasma radio emission depends intimately on three related processes: the production and evolution of an electron beam, the subsequent Langmuir wave generation and evolution, and the final step of conversion into electromagnetic emission. Thus, simulations of Type III bursts may approach the problem from several perspectives. For example, considering only the beam transport allows us to find the frequency drift of the burst, and explore how its frequency-time profile depends on factors such as the beam velocity and the geometry of the magnetic field lines along which the beam propagates. This has been done by several authors and provides interesting results \citep[e.g.][]{1996A&A...314..303K, 2000SoPh..194..345R,2004SoPh..222..299L}. 

Considering Langmuir wave generation provides some additional information on the brightness of the radio emission. However, in general this depends on the exact spectrum of the Langmuir waves, and therefore to see the details of the time profile at even a single frequency, and explain the rise-decay profiles and the exact shape of the dynamic spectrum, we must address all three factors simultaneously. On the other hand treating any one possible effect in full is a significant undertaking, both theoretically and from a computational perspective. Therefore, there is a trade-off between the level of detail in any one step, and the ability to consider the whole process for a variety of parameters. Here we employ simple but well justified models of each step, in order to investigate specifically the effects of plasma inhomogeneity on the emission. 

The most significant simplification, which we have discussed in previous chapters, is to assume the dynamics of the electrons and Langmuir waves lie along a single spatial dimension. Detailed simulations of beam propagation and Langmuir wave production were performed by \citet{1998PlPhR..24..772K, 1999SoPh..184..353M} and have recently been extended \citep{2009ApJ...695L.140K, 2010ApJ...721..864R} to follow a beam from the sun all the way to 1AU. The evolution of the Langmuir waves due to density fuctuations was shown to have significant effects on the Langmuir wave spectrum. Similar effects for the downwards moving beams which generate HXR emission were found by \citet{2013A&A...550A..51H}. 

For a weak beam, with a small initial angular spread, the magnetic field of the corona and solar wind is sufficient to maintain a quasi-one-dimensional beam \citep{1990SoPh..130..201M}, as the electrons will propagate along the field lines. In this case, the ion-sound waves generated from Langmuir wave decays will also be restricted to this same direction. However, the electromagnetic emission occurs primarily at a significant angle to the parent Langmuir wave, specifically around $\pi/4$ for harmonic and $\pi/2$ for the fundamental. Moreover, the emission probabilities have strong angular dependence, as do the participating wavenumbers. Our first step is therefore to define a model of the emission geometry, and use this to derive a mathematical treatment of the emission. This is the topic of the next section. 
\section{3-D Equations for Scattering and Decay}

It is necessary to distinguish between processes described as ``scattering'' and those described as ``decay'' or ``coalescence'', the former involving the interaction of a wave with an individual plasma particle, either an electron or an ion, while the latter involve coherent motions of the plasma, i.e another wave mode. The equations describing the scattering and decay/fusion processes in 3-D are given in the books by \citet{1980MelroseBothVols, 1995lnlp.book.....T} and are reproduced in the next sections. 

A weak magnetic field, where $\Omega_{ce}\ll \omega_{pe}$, has little or no effect on the emission probabilities for plasma emission processes \citep[e.g.][]{1972AuJPh..25..387M} and so we use the equations for unmagnetised plasma. However, as we will discuss in Section \ref{sec:Polaris}, the field cannot be completely neglected in plasma emission, as it affects the polarisation states of the resulting waves. 

\subsection{Ion Scattering Processes}
Described by $\sigma+i\rightleftarrows \sigma'+i'$, where $\sigma, \sigma'$ denote wave modes, these describe the action of scattering by an individual plasma particle, for our purposes an ion, $i$, averaged over their distribution, which in this case is thermal and described by a Maxwellian. The change in momentum of the wave during scattering is absorbed by the ion, but this is small and so we may neglect the evolution of the ion distribution, instead assuming it very rapidly returns to thermal. The product wave can be of the same species, or a different species. 

Writing $W_\sigma, \vec{k}, \omega^\sigma(\vec{k})$ for the spectral energy density, wavevector and frequency of wave type $\sigma$, the equations in 3-D describing the process $\sigma+i\rightleftarrows \sigma'+i'$ are:
\begin{align}\frac{d W_\sigma(\vec{k}_\sigma)}{d t}=&\int \frac{d \vec{k}_{\sigma'}}{(2\pi)^3} w_i^{\sigma\sigma'}(\vec{k}_{\sigma'}, \vec{k}_\sigma)\times \notag \\ &\left[\frac{\omega_\sigma}{\omega_{\sigma'}}W_{\sigma'}(\vec{k}_{\sigma'})-W_\sigma(\vec{k}_\sigma)-\frac{(2\pi)^3}{T_i}\frac{\omega_\sigma-\omega_{\sigma'}}{\omega_{\sigma'}}W_\sigma(\vec{k}_\sigma)W_{\sigma'}(\vec{k}_{\sigma'})\right] \label{eqn:IS}\end{align}
and 
\begin{align}\frac{d W_{\sigma'}(\vec{k}_{\sigma'})}{d t}=&\int \frac{d \vec{k}_\sigma}{(2\pi)^3} w_i^{\sigma{\sigma'}}(\vec{k}_{\sigma'}, \vec{k}_\sigma)\times \notag \\&\left[\frac{\omega_{\sigma'}}{\omega_\sigma}W_\sigma(\vec{k}_\sigma)-W_{\sigma'}(\vec{k}_{\sigma'})-\frac{(2\pi)^3}{T_i}\frac{\omega_{\sigma'}-\omega_\sigma}{\omega_\sigma}W_{\sigma'}(\vec{k}_{\sigma'})W_\sigma(\vec{k}_\sigma)\right]. \end{align}
The probability $w_i^{\sigma\sigma'}$ depends on the wave modes involved, but for scattering of Langmuir waves it is \begin{equation} w_i^{\sigma\sigma'}= \frac{C_{ion} |\hat{\vec{e}}_\sigma \cdot \hat{\vec{e}}_{\sigma'}| }{|\vec{k}_{\sigma'}-\vec{k}_\sigma|} \exp{\left(-\frac{(\omega_{\sigma'}-\omega_\sigma)^2}{2|\vec{k}_{\sigma'}-\vec{k}_\sigma|^2 v_{Ti}^2}\right)}, \end{equation} with the
constant \begin{equation}C_{ion}=\frac{\sqrt{\pi} \omega_{pe}^2}{2 n_e v_{Ti} (1+T_e/T_i)^2},\end{equation} and $\hat{\vec{e}}_\sigma$ a unit vector in the direction of the electric field of the wave mode $\sigma$. Langmuir waves are longitudinal so that $\hat{\vec{e}} \parallel \vec{k}$, while electromagnetic waves are transverse so $\hat{\vec{e}} \perp \vec{k}$. 
      
\subsection{3-wave Decays}
Described by $\sigma \rightleftarrows \sigma' +\sigma''$, the three wave processes describe either decay due to the presence of another wave mode, or fusion of two existing wave modes. We write $W_\sigma, \vec{k}, \omega^\sigma(\vec{k})$ for the spectral energy density, wavevector and frequency of waves in mode $\sigma$. Then energy and momentum conservation are described by the equations $\vec{k}= \vec{k}'+ \vec{k}''$ and $\omega^\sigma(\vec{k})=\omega^{\sigma'}(\vec{k}')+\omega^{\sigma''}(\vec{k}'')$ and we have \begin{align}\label{eqn:3Wv3D1}\frac{d W_\sigma(\vec{k})}{dt}=&\omega^\sigma(\vec{k}) \int\int w_{\sigma\sigma'\sigma''}(\vec{k}, \vec{k}', \vec{k}'')\times \notag \\ &\left[\frac{W_{\sigma'}(\vec{k}')}{\omega'}\frac{W_{\sigma''}(\vec{k}'')}{\omega''}-\frac{W_{\sigma}(\vec{k})}{\omega}\left(\frac{W_{\sigma'}(\vec{k}')}{\omega'}+\frac{W_{\sigma''}(\vec{k}'')}{\omega''}\right)\right]d\vec{k}'d\vec{k}''\end{align}
\begin{align}\frac{d W_{\sigma'}(\vec{k'})}{dt}=&- \omega^{\sigma'}(\vec{k}') \int\int w_{\sigma\sigma'\sigma''}(\vec{k}, \vec{k}', \vec{k}'') \times \notag \\ & \left[\frac{W_{\sigma'}(\vec{k}')}{\omega'}\frac{W_{\sigma''}(\vec{k}'')}{\omega''}-\frac{W_{\sigma}(\vec{k})}{\omega}\left(\frac{W_{\sigma'}(\vec{k}')}{\omega'}+\frac{W_{\sigma''}(\vec{k}'')}{\omega''}\right)\right]d\vec{k}d\vec{k}''\end{align} and 
\begin{align}\label{eqn:3Wv3D3}\frac{d W_{\sigma''}(\vec{k''})}{dt}=&- \omega^{\sigma''}(\vec{k}'') \int\int w_{\sigma\sigma'\sigma''}(\vec{k}, \vec{k}', \vec{k}'') \times \notag \\ & \left[\frac{W_{\sigma'}(\vec{k}')}{\omega'}\frac{W_{\sigma''}(\vec{k}'')}{\omega''}-\frac{W_{\sigma}(\vec{k})}{\omega}\left(\frac{W_{\sigma'}(\vec{k}')}{\omega'}+\frac{W_{\sigma''}(\vec{k}'')}{\omega''}\right)\right]d\vec{k}d\vec{k}'\end{align}
where $w_{\sigma\sigma'\sigma''}(\vec{k}, \vec{k}', \vec{k}'')$ is the emission probability. The form of this depends strongly on the wave modes involved, and again is dependent on their electric field directions, $\hat{\vec{e}}$. The probabilities for relevant processes are given in the following derivations (Equations \ref{eqn:fundSprob} and \ref{eqn:LLTProb}).  

\subsection{Decay versus Scattering}

Coherent oscillations of the plasma particles can exist only when their damping rate is far less than their frequency. For ion-sound waves, which are oscillations of the electrons and ions, the requirement that $\gamma_s \ll \omega_s$ requires that we have $T_i\ll T_e$, and we cannot speak of a decay process involving ion-sound waves unless this condition is satisfied. On the other hand, scattering by individual ions, neglecting plasma collective effects, is also important, and this is particularly efficient when $T_i=T_e$. The two processes are not entirely distinct: both are scattering by plasma ions but the former involves collective motion of these, and is therefore a resonant interaction. It turns out that the effects of decay when $T_i\ll T_e$ and of scattering when $T_i=T_e$ are very similar, to within a numerical factor of 4 or so \citep{1995lnlp.book.....T}. 

In addition to this, when $T_i\ll T_e$ the wavenumber matching conditions for decay may mean that this cannot proceed, and scattering may be important, although its efficiency is small for this ratio. Thus, in order to consider the full range of temperature ratios observed in the solar corona and solar wind \citep[e.g.][]{1998JGR...103.9553N, 1979JGR....84.2029G}, which range from $T_i/T_e \lesssim 0.1$ to $1$ or even 2, we must consider both the decay and scattering formulations, and employ the correct one for the parameters chosen, to avoid overestimating the effects on Langmuir waves. 

\section{A Model for Plasma Emission}

\subsection{Langmuir Wave Evolution}
In the previous chapters we described a model of Langmuir wave generation from a collisionally relaxing beam including the effects of ion-sound wave scattering, and perturbations in the ambient plasma density. However, we focused on the effects on the beam-generated Langmuir waves with $k >0$ and the consequent effects on the electron distribution. For plasma emission, we are specifically interested both in the backscattered Langmuir waves with $k <0$ and the exact spectrum of ion-sound waves (when $T_i\ll T_e$ and these can exist). 

Therefore, here we combine Equations \ref{eqn:3ql1} and \ref{eqn:3ql2} for the beam-wave interactions including the effects of plasma density fluctuations, with Equations \ref{eqn:3ql_s} and \ref{eqn:3ql_sSrc} describing scattering by ion-sound waves, and in addition consider a term in the Langmuir wave equation describing scattering by individual ions. 
Beginning from Equation \ref{eqn:IS}, noting that as $\sigma$ and $\sigma'$ are the same species we sum the contributions from the two equations, we substitute a 1-dimensional Langmuir wave spectrum, which is non-zero only along the direction of the electron beam, given by \begin{equation}W^{3D}(\vec{k})= W^{\AAvL}(k) \delta(|1-\cos\theta|).\end{equation} Then we have $\cos \theta_{LL'}=1$ and the ion scattering is described by 
\begin{align}\label{eqn:LIon}\frac{d W_L^{\AAvL}(k_L)}{d t}=&\int \rd k_{L'} \frac{C_{ion}}{|k_L-k_{L'}|} \exp{\left(-\frac{(\omega_L-\omega_{L'})^2}{2|k_L-k_{L'}|^2 v_{Ti}^2}\right)} \times \notag\\&\left[\frac{\omega_{L'}}{\omega_L}W_L^{\AAvL}(k_L)-W_L^{\AAvL}(k_{L'})-\frac{(2\pi)^3}{T_i}\frac{\omega_{L'}-\omega_L}{\omega_L}W_L^{\AAvL}(k_{L'})W_L^{\AAvL}(k_L)\right].\end{align}  

From a computational perspective, this equation is cumbersome to simulate because of the remaining integral over $dk_{L'}$. However, the exponential factor means the integrand is sharply peaked around $\omega_L \simeq \omega_{L'}$, which allows us to consider the integral over only a small range in $k_L$ and thus speed this up considerably. 

\subsection{Fundamental Emission}\label{sec:C4Fund}
Fundamental emission at $\omega_{pe}$ generally does not occur at frequencies above a few hundred MHz. Our simulations consider the situations of equal or almost equal electron and ion temperatures, where ion-sound waves are strongly damped and thus contributions to the Langmuir wave spectrum from the processes  $L \pm s\rightleftarrows t$ can be ignored. Direct scattering of Langmuir waves into fundamental emission, $L + i\rightleftarrows t+i'$ also has negligible effect on the Langmuir waves. An \AATxt model for the radio emission is given in Appendix A, as this will be important for future work on lower frequency emission, but is omitted here for brevity.  

\subsection{Harmonic Emission}

We focus instead on the process of harmonic emission due to the coalescence of two Langmuir waves, i.e $L + L'\rightleftarrows t$. Writing the participating wave vectors as $\vec{k}_1, \vec{k}_2$ for the Langmuir waves and  $\vec{k}_T$ for the electromagnetic wave, and further writing $\vec{k}_1=(k_1, \theta, \phi)$, the general equation for the process is
\begin{align}\label{eqn:Harm3D}
\frac{dW_T(\vec{k}_T)}{dt}=&\omega_{k_T}^T \int\int\int w^{LLT}(\vec{k}_1, \vec{k}_2,\vec{k}_T) k_1^2 \sin\theta\times\notag \\&\left[\frac{W_L(\vec{k}_1)}{\omega_{k_1}^L}\frac{W_L(\vec{k}_2)}{\omega_{k_2}^L}-\frac{W_L(\vec{k})}{\omega_{k_1}^L}\frac{W_T(\vec{k}_T)}{\omega_{k_T}^T}-\frac{W_T(\vec{k}_T)}{\omega_{k_T}^T} \frac{W_L(\vec{k}_2)}{\omega_{k_2}^L}\right] \rd k_1\rd\theta_1 \rd\phi_1
\end{align}
with probability
\begin{equation}\label{eqn:LLTProb}
w^{LLT}(\vec{k}_1, \vec{k}_2,\vec{k}_T)=\pi \omega_{pe} \frac{(k_2^2-k_1^2)^2 (\vec{k}_T\times\vec{k}_1)^2}{16 m_e n_e k_T^{2}k_1^2k_2^2}\delta(\omega_{k_T}^{T}-\omega_{k_1}^L-\omega_{k_2}^L)
\end{equation}
and the condition $\vec{k}_1+\vec{k}_2=\vec{k}_T$. 

The ``head-on-approximation'' (HOA) where the coalescing Langmuir waves are almost parallel was proposed in early models of plasma emission \citep[e.g.][]{1979A&A....73..151M}. This implies that $k_T \ll k_L$, as $\vec{k}_T$ must be at a non-zero angle to the parent Langmuir wave or the emission cannot occur. However, if we consider an electromagnetic wave at approximately twice the local plasma frequency we have $\omega=(\omega_{pe}^2+k^2 c^2)^{1/2} \simeq 2\omega_{pe}$ and therefore the typical wavenumber is $k_T \simeq (\sqrt{3}\rv_{Te}/c) k_{De}$. For beam generated Langmuir waves we have typical wavenumbers of $k_L\simeq (\rv_{Te}/\rv_b) k_{De}$, so for a typical beam velocity of $\rv_b \simeq 0.3 c$ we have $k_T/k_L \sim \sqrt{3} \rv_b/c \simeq 0.5$ which is by no means small. Therefore we wish to find a better approximation than the HOA.

\subsubsection{An Approximate Emission Geometry}
\begin{figure}
 \center
\includegraphics[width=12cm]{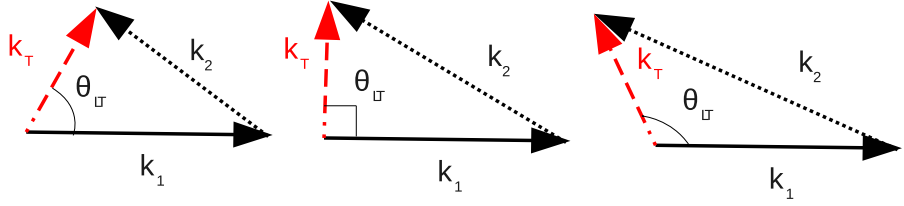}
\caption[Wavenumber matching diagrams for harmonic plasma emission.]{Vector diagrams of wavenumber matching for harmonic emission in several configurations. $k_1,k_2$ are the forwards and backwards Langmuir wavevectors, and $k_T$ is the resulting electromagnetic wavevector.}\label{fig:HamrVecs}
\end{figure}

Firstly, we define the angle $\theta_{LT}$ to be the angle between the forwards Langmuir wave with wavenumber $k_1$ and the electromagnetic wave with wavenumber $k_T$, as shown in the vector diagrams in Figure \ref{fig:HamrVecs}. The complications then become clear if we consider the emission equation. In the HOA we have to good accuracy that $(|\vec{k}_T-\vec{k}|^2-k^2)^2/k_2^2 \simeq k_T^2 \cos^2\theta$ and also that $k_1\simeq k_2$, so the entire angular dependence of the probability is via a term like $\cos^2\theta_{LT}\sin^2\theta_{LT}$. Relaxing the HOA introduces additional angular dependence into the probability because, as shown in Figure \ref{fig:HamrVecs}, the solution for $k_2$ from the wavenumber matching also depends on $\theta_{LT}$. 

\begin{figure}
 \center 
\includegraphics[width=7cm]{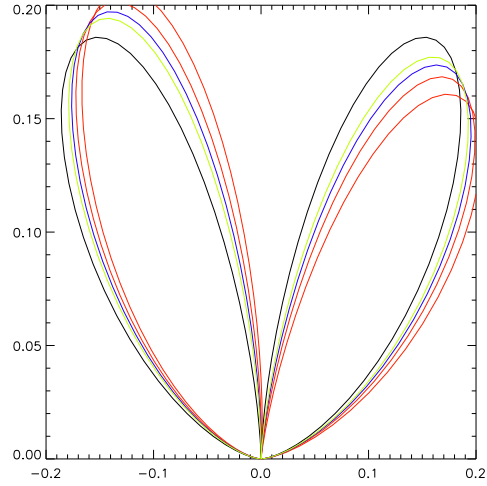}
\caption[Normalised emission probabilities for harmonic plasma emission.]{Polar plot of normalised emission probability from Equation \ref{eqn:LLTProb} against the angle $\theta_{LT}$. Lines from red to green are for increasing magnitudes of $k_T$, and the black line shows the probability in the HOA. \label{fig:probs}}
\end{figure}

Taking the conservation conditions $\vec{k}_1+\vec{k}_2=\vec{k}_T$ and $\omega_{k_1}+\omega_{k_2}=\omega^T_{k_T}$ we can find their solutions as a function of $k_T$. Then we can find the exact emission probability from Equation \ref{eqn:LLTProb}. In Figure \ref{fig:probs} we plot this for several values of the electromagnetic wavenumber $k_T$, along with the probability in the HOA approximation. In all cases the probability shows quadrupolar forms, and the differences from the simple HOA solution decrease as the electromagnetic wavenumber increases. Most importantly, both the angle of the peak, and the width of the probability remain similar. If we also consider the average of the two lobes, there is very little difference from the simple HOA.

Therefore, it seems reasonable to assume, when solving the wavenumber matching, that all emission occurs at $\theta_{LT}=\pi/4$, which is the angle of maximum emission probability in the HOA and very close to it in the exact solution. Then we independently calculate an angle-averaged probability using these solutions. From the width of the emission probability curves, it is reasonable to assume emission occurs only within the angular range $[\pi/8, 3\pi/8]$, which covers its full-width-half-maximum (FWHM). 

\begin{figure}
\center
\includegraphics[width=7cm]{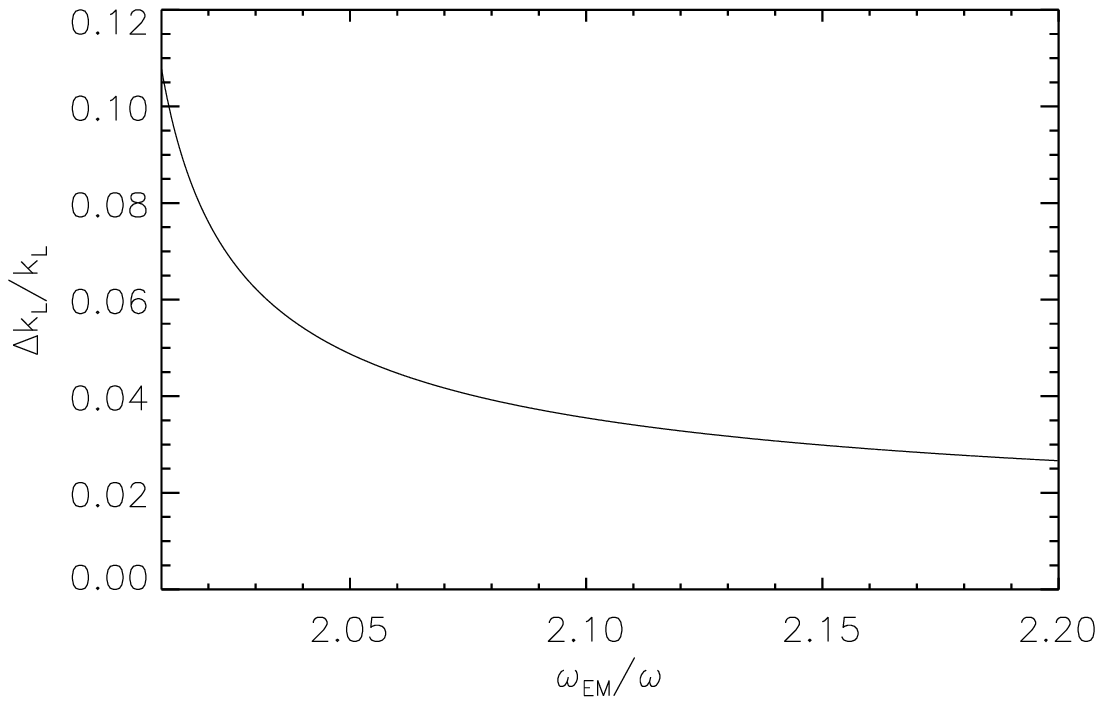}\includegraphics[width=7cm]{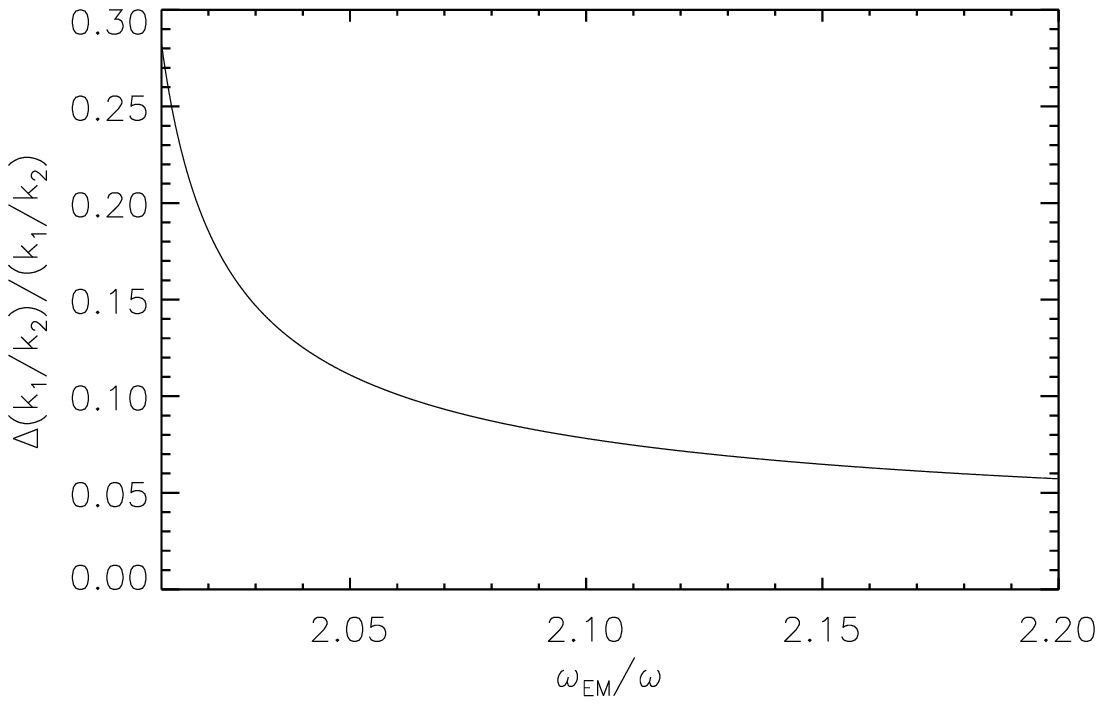}
\caption[Fractional variations in the participating wavenumbers for harmonic plasma emission.]{Fractional variation in $k_1$ and in the ratio $k_1/k_2$ due to variations in the angle of $k_1$ and $k_T$. We define $\Delta k/k =\left(k_{\pi/8}-k_{3\pi/8}\right)/k_{\pi/4}$, and plot this as a function of electromagnetic frequency $\omega_{EM}$  \label{fig:k_match}}
\end{figure}

For a particular electromagnetic wavenumber there will be a range of values for $k_1, k_2$ depending on the angle of emission $\theta_{LT}$, whereas our approximation assumes that $\theta_{LT}=\pi/4$. We now calculate the size of the variation in $k_1, k_2$ for emission within the angular range $[\pi/8, 3\pi/8]$. In Figure \ref{fig:k_match} we plot the difference in value at $\pi/8$ and $3\pi/8$, divided by the value at $\pi/4$, i.e. $\Delta k/k =\left|\left(k_{\pi/8}-k_{3\pi/8}\right)\right|/k_{\pi/4}$, as a function of the electromagnetic frequency. Clearly, for frequencies very close to $2\omega_{pe}$ the similarity in magnitude of $k_1$ and $k_T$ means that the angular effects become more important. However, above around $\omega_{EM}/\omega=2.02$ we have a reasonable approximation, with variation of less than $6\%$ in $k_1$. The ratio $k_1/k_2$ is even more dependent on the angle $\theta_{LT}$, but again, above $\omega_{EM}/\omega=2.02$ the variation is below $15\%$. 

The implications of this variation in $k_1,k_2$ may be summarised as follows. We consider emission to come from Langmuir waves at a wavenumber $k_L$, where in fact the emission is from a \emph{weighted average} over a band of $k_L \pm \Delta k_L$, the weighting in the average has a form similar to $\sin^2\theta \cos^2\theta$, and $\Delta k_L$ is of the order $5\%$ or less. On the other hand, for the purposes of our simulations, the average need only be taken over two, or perhaps a few grid points. For example, for a $k_L$ simulation grid of $250$ points covering $k_L$ from 0 to 0.4$k_{De}$ we require averaging over only $3$ grid points at $k=0.1k_{De}$ and for a slightly more coarse gridding no averaging is needed at all. 

The final consideration relates to the angular distribution of the Langmuir waves. In the previous paragraphs we assumed that a wave existed at the calculated wavevector $k_2$, regardless of the angle between $k_1$ and $k_2$. At a value of $\omega_{EM} = 2.02 \omega_{pe}$ and $\theta_{LT}=\pi/4$, the angle between $k_1$ and $k_2$ is $\pi/16$ and at $\omega_{EM}=2.1\omega_{pe}$ this has decreased to less than $\pi/32$. Thus, provided we have Langmuir waves with an angular spread of half-angle at least $\pi/16$ there are waves with the necessary wavevectors for coalescence to proceed. 

\begin{figure}
 \center
\includegraphics[width=7cm]{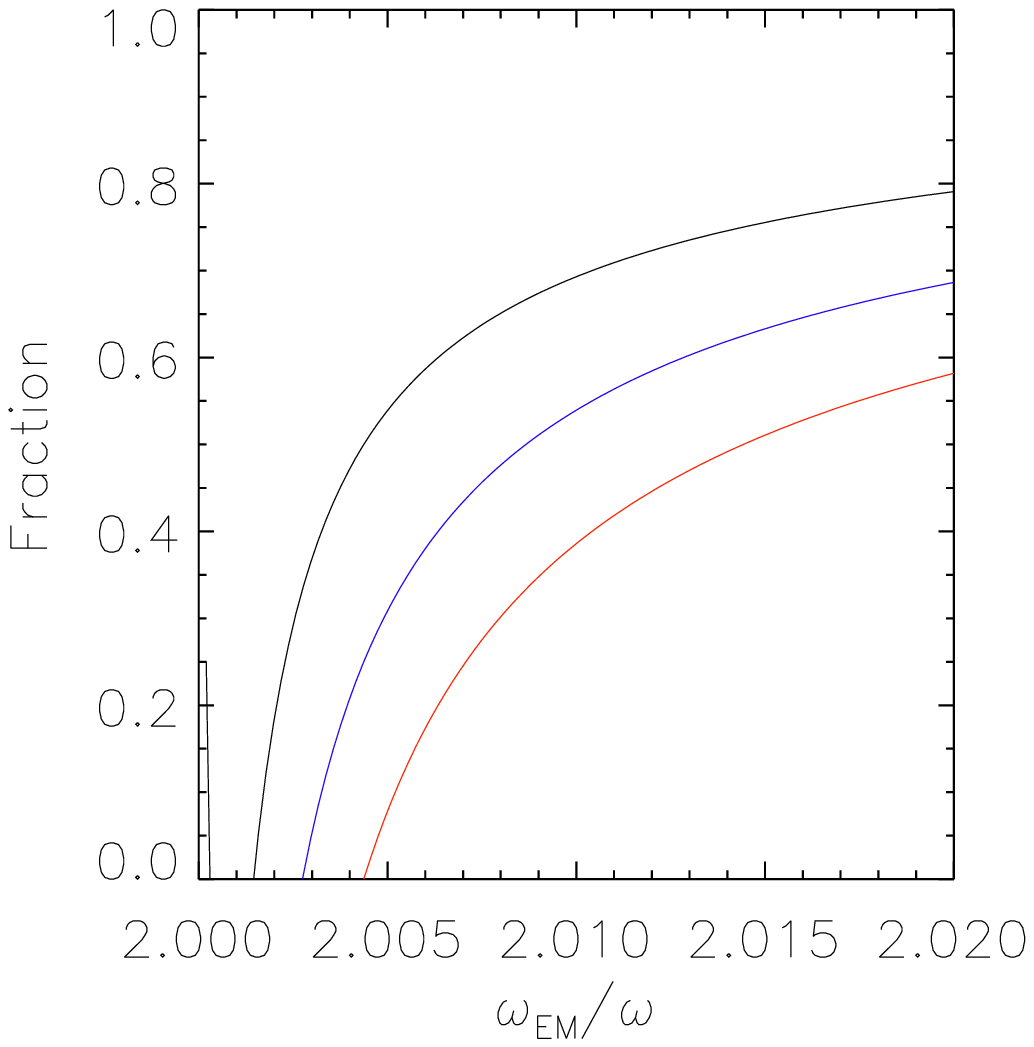} \includegraphics[width=7cm]{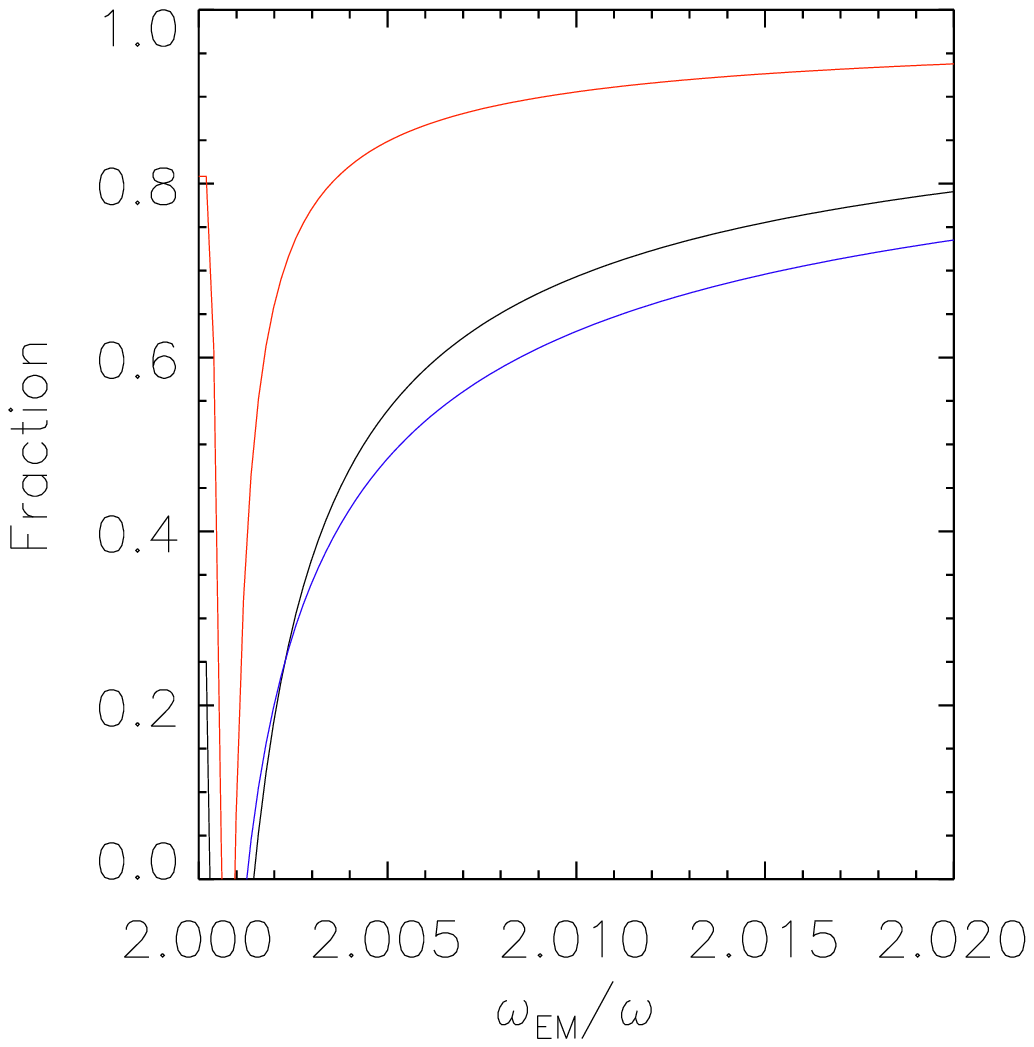}
\caption[The fraction of Langmuir waves contributing to harmonic plasma emission.]{The participating wave fraction as a function of electromagnetic frequency. Left: for $\theta_{LT}=\pi/4$ and Langmuir waves over a cone with half-angle of $\pi/6$ (black), $\pi/9$ (blue), and $\pi/12$ (red).  Right: for a Langmuir wave half-angle of $\pi/6$ and $\theta_{LT}$ of $\pi/4$ (black), $\pi/8$ (red) and $\pi/2$ (blue).}\label{Fig:angfrac}
\end{figure}

To further account for these angular effects, we can calculate the fraction of the waves within the angular spread that can take part in the coalescence, which will depend on the assumed angular spread, and the required angle between $k_1$ and $k_2$, and therefore the electromagnetic frequency $\omega_{EM}$. This is plotted in Figure \ref{Fig:angfrac} for various values of the angles involved. Above $2.02 \omega_{pe}$ this fraction is above approximately $0.5$ for the values considered, so a half angle  of at least $\pi/12$ in the Langmuir waves is sufficient to ensure our estimate is within a factor of around 2 of the true value. 

Taking all this into account, we can say that the approximation of taking $\theta_{LT} =\pi/4$ for energy and momentum conservation and using an angular averaged probability as found in the next section, we will obtain a good agreement in magnitude with the angular averaged true 3-D emission provided that the emission frequency is above approximately $\omega_{EM}/\omega_{pe} \simeq 2.01$, the Langmuir wave spectrum does not contain very narrow spikes, of width $< 5\%$ in Langmuir wavenumber, and the Langmuir waves are fairly isotropic over an angular spread with half angle at least $\pi/12$. The second two conditions are generally satisfied for our purposes as we consider the presence of ambient plasma density fluctuations, which will tend to smooth out any spikes and produce an angular spread, as described in Section \ref{Sec:angdiff}.  

\subsubsection{Harmonic Emission Equations}
Now we use the geometry just described to derive \AATxt equations for the coalescence process. For clarity, we label the Langmuir wave spectrum with wavenumbers parallel to the beam direction as the forwards spectrum, and those antiparallel as backwards. We assume these both cover a small angular spread $\Delta\Omega$, with a half angle of at least $\pi/12$. Then we define \begin{equation}\label{eqn:harmL} W_L^f(\vec{k}_1)=\frac{1}{\Delta\Omega k_1^2 } W_L^{\AAvL,f}(k_1)\;,\;W_L^b(\vec{k}_2)=\frac{1}{\Delta\Omega k_2^2 } W_L^{\AAvL,b}(k_2)\end{equation} within $\Delta\Omega$, and zero elsewhere, while the electromagnetic emission is assumed to cover a cone of half angle $\pi/4$ in each lobe of the emission probability, and therefore a solid angle of $\pi$, so we have \begin{equation} \label{eqn:harmT} W_T(\vec{k}_T)=\frac{1}{\pi k_T^2 } W_T^{\AAv}(k_T).\end{equation} 

Now we proceed exactly as for the case of fundamental emission, by defining the angle averaged probability as \begin{displaymath}
\langle P \rangle= \frac{1}{\Delta\Omega}\int\int_{\Delta\Omega}\frac{(k_2^2-k_1^2)^2 \sin^2\theta_{LT}}{k_2^2} \sin\theta  \rd\theta \rd\phi
\end{displaymath} and substituting the definitions in Equations \ref{eqn:harmL} and \ref{eqn:harmT} then considering in turn each term in the square brackets in Equation \ref{eqn:Harm3D}. 

The result is 
\begin{align}\frac{dW_T^{\AAv}(k_T)}{dt}=&\omega_{k_T}^T \int \rd k_1 \delta(\omega_{k_T}^{T}-\omega_{k_1}^L-\omega_{k_2}^L) \langle P\rangle \times \notag 
\\ & \left[\frac{k_T^2}{k_2^2}\frac{\pi}{\Delta\Omega}\frac{W_L^{\AAvL,f}(k_1)}{\omega_{k_1^L}}\frac{W_L^{\AAvL,b}(k_2)}{\omega_{k_2}^L}-\frac{W_T^{\AAv}(k_T)}{\omega_{k_T}^T} \left(\frac{W_L^{\AAvL,f}(k_1)}{\omega_{k_1}^L}-\frac{W_L^{\AAvL,b}(k_2)}{\omega_{k_2}^L}\right)\right]. 
\end{align} 
The final integral over $k_1$ is evaluated using the delta function $\delta(\omega_{k_T}^{T}-\omega_{k_1}^L-\omega_{k_2}^L)$, which leads to a factor of \begin{equation}\frac{\omega_{pe}}{3 v_{Te}^2}\frac{1}{(2k-k_T \cos \theta_{LT})}.\end{equation}

As a simple approximation to the probability, we can take the value assuming $\theta_{LT}=\pi/4$ to solve for the $k_1, k_2$ values, and then take the peak value of the probability multiplied by its fractional FWHM, which is $\simeq 1/2$. Taking $\sin\theta_{LT}= \sqrt{2}/2$ we have \begin{equation}\langle P\rangle = \frac{(k_2^2-k_1^2)^2}{4 k_2^2}.\end{equation}

Combining all of these factors we finally obtain
\begin{align}\label{eqn:HarmFinal}\frac{dW_T^{\AAv}(k_T)}{dt}=&\omega_{k_T}^T \frac{\pi \omega_{pe}^2}{48 m_e n_e v_{Te}^2} \frac{(k_2^2-k_1^2)^2}{4 k_2^2 (2k_1-k_T\sqrt{2}/2)}\times \notag 
\\ & \left[\frac{k_T^2}{k_2^2}\frac{\pi}{\Delta\Omega}\frac{W_L^{\AAvL,f}(k_1)}{\omega_{k_1^L}}\frac{W_L^{\AAvL,b}(k_2)}{\omega_{k_2}^L}-\frac{W_T^{\AAv}(k_T)}{\omega_{k_T}^T} \left(\frac{W_L^{\AAvL,f}(k_1)}{\omega_{k_1}^L}-\frac{W_L^{\AAvL,b}(k_2)}{\omega_{k_2}^L}\right)\right],
\end{align} which describes electromagnetic emission due to Langmuir wave coalescence in our approximation, with the \AATxt spectral energy densities given by Equations \ref{eqn:harmL} and \ref{eqn:harmT}. 

\section{Polarisation and the Weak Field Limit}\label{sec:Polaris}
In the preceding discussion we neglected the magnetic field, because in the weak field limit, $\Omega_{ce}\ll \omega_{pe}$, the emission probabilities are effectively the same as in the unmagnetised case. However, as we briefly mentioned earlier, even a weak magnetic field has significant effects on the polarisation of the electromagnetic emission. 

For fundamental emission, it was originally proposed that the polarisation should be 100$\%$ in the O-mode, because the X-mode has a cutoff at $\omega_X\simeq\omega_{pe}+\Omega_{ce}/2$. Considering the intrinsic bandwidth of the emission due to the dispersion, the necessary magnetic field strength to obtain 100$\%$ polarised emission can be estimated, and at 100~MHz this value is well below the inferred fields \citep[e.g][]{1980MelroseBothVols}.  

The observed polarisation values for lower frequency fundamental Type III emission of between 30 and 70$\%$ (Section \ref{sec:ClassicIIIs}) therefore require explanation. Scattering during propagation is certainly able to reduce the degree of polarisation, but cannot explain why fully polarised emission is never seen. \citet{1984SoPh...90..139W} therefore proposed that the emission was depolarised to some extent within the source region itself, and that this was inherent to the emission process. Further change during scattering can then explain the range of observed values. 

The case of harmonic emission is more complicated. \citet{1972AuJPh..25..387M, 1978A&A....66..315M} and the correction by \citet{1980PASAu...4...50M} treated the case of head-on Langmuir wave coalescence, and found that the angular spectrum of the Langmuir waves dictates whether the emission is O or X-mode. However, as was shown by \citet{1997SoPh..171..393W}, when the HOA is relaxed the percentage polarisation is very different, and when the difference in forwards and backwards Langmuir wavenumbers are accounted for, emission tends to favour the O-mode. However the Langmuir wavenumber and angular distributions are essential to decide both the mode and the degree of polarisation.
While some geometries of harmonic emission may allow weakly X-mode polarised emission to be produced, very strong X-mode polarisation is not possible. On the other hand, for emission occurring at an angle of $\pi/4$ to the ambient magnetic field, as we assume in the model of the previous section, the results of \citet{1997SoPh..171..393W} suggest polarisation of around 10 to 15 \% in the O-mode, consistent with the observations. 

\subsection{Polarisation Change During Transport}\label{sec:PolProp}
We also note that further changes in the degree of polarisation can occur during propagation from source to observer. For example, mode coupling due to magnetic fields \citep[e.g.][]{1964SvA.....7..485Z}, and scattering due to low-frequency waves \citep{1984SoPh...90..139W,1989SoPh..119..143M}, or kinetic Alfven waves \citep{2002A&A...390..725S} have all been considered, and can all lead to significant changes in polarisation degree. 

\section{Thermal Emission}\label{sec:ThEM}

In thermal plasma, there will be electromagnetic emission due to thermal bremsstrahlung, giving a thermal source term \begin{equation}P(\omega)=\frac{16 \pi e^6 n_i n_e n(\omega){\rm ln}\Lambda}{3 m_e^2 c^3 \rv_{Te}}\sqrt{\frac{2}{\pi}}\end{equation} \citep[e.g.][Equation 3.81]{1980gbs..bookQ....M}, for the power radiated per unit volume, per unit frequency over all angles. Now for electromagnetic waves we have $n(\omega)=\left(1- \omega^2/\omega_{pe}^2\right)^{1/2}= c k/\omega$ and in plasma we can approximate $n_i=n_e$ from quasineutrality. We have, assuming $P(\vec{k})$ is isotropic, \begin{equation}P(\vec{k})= P(\omega)/(4\pi k^2 dk/d\omega)\end{equation} and so with $ k^2 dk/d\omega = n(\omega) \omega^2/c^3$ we obtain \begin{equation}P(\vec{k})=\left[\frac{e^2 \omega_{pe}^4 {\rm ln}\Lambda}{3 \omega^2 \rv_{Te}}\sqrt{\frac{2}{\pi}}\right].\end{equation} Finally, assuming isotropy, we can change this to an \AATxt spectral energy density by multiplying by $4\pi k^2$, finding \begin{equation}\label{eqn:ThSrc}P^{\AAv}(k)=\frac{4 \pi k^2 e^2 \omega_{pe}^4 {\rm ln}\Lambda}{3 \omega^2 \rv_{Te}}\sqrt{\frac{2}{\pi}}.\end{equation} 

Now by Kirchoff's law, in optically thick plasma in thermal equilibrium the emission must be balanced by an absorption with coefficient $\gamma_d$ given by \begin{equation}P^{\AAv}(k)= \gamma_d(k) (4 \pi k^2 k_b T_e).\end{equation} Then we find \begin{equation}\label{eqn:brems}\gamma_d=\left[\frac{4\pi e^2 {\rm ln}\Lambda}{3 m_e \rv_{Te}^3}\sqrt{\frac{2}{\pi}}\right] \left(\frac{\omega_{pe}^4}{\omega^2}\right).\end{equation} As expected we then have that \begin{equation}\frac{d W_{th}(k)}{dt}= P(k) - \gamma_d W_{th}(k) =0 \end{equation} i.e. the emission and absorption are in equilibrium for the thermal spectrum. We add these two terms to our equations for electromagnetic evolution: a thermal source term $P(k)$ given by \ref{eqn:ThSrc} and a damping term $-\gamma_d W(k)$. 

\section{Observed Fluxes}

In order to relate the results of our simulations to observed emission, we must first convert the \AATxt spectral energy densities into fluxes, which requires that we specify the size of the emitting source. Then we must add the effects of absorption or scattering in the source and during propagation. For harmonic emission these effects are smaller than for the fundamental, and the time delay experienced by the radiation during propagation is also less, but these effects cannot be neglected.  

\subsection{Conversion of Energy Density to Flux} 
The total flux is given by \begin{equation}F= U S d \frac{1}{4\pi R_0^2} \frac{\rv_{gr}}{d}\end{equation} where $U$ is the energy density of the source, $S$ its area, $d$ its length (along the direction of the beam) and $R_0$ is its distance from the observer, and $\rv_{g}$ is the group velocity of the radiation. The flux as a function of frequency $\omega$ is then given by $dF/d\omega$. The energy density $U$ relates to the \AATxt spectral energy density by \begin{equation}U=\int \rd k W^{\AAv}(k)=\int \rd\omega W^{\AAv}(\omega)\end{equation} and we have \begin{equation}W^{\AAv}(\omega)=W^{\AAv}(k)\frac{\partial k}{\partial \omega}=\frac{W^{\AAv}(k)}{\rv_g}.\end{equation} Therefore \begin{equation}F(\omega)= W^{\AAv}(\omega) \frac{S}{4\pi R_0^2} \rv_{gr}=W^{\AAv}(k)\frac{S}{4 \pi R_0^2}.\end{equation} For emission at GHz frequencies the source is small (see Section \ref{sec:SrcSize}), and  located at approximately 1~AU from the observer, so we may apply the small angle approximation and find $S/(4\pi R_0^2)= \theta^2/4$ where $\theta$ is the half-angle of the source in radians. In this case we have \begin{equation}F(\omega)= W^{\AAv}(\omega) \frac{\theta^2}{4} \rv_{gr}=W^{\AAv}(k)\frac{\theta^2}{4}.\end{equation}

\subsection{Source Size} \label{sec:SrcSize}

Source size measurements  were discussed in Section \ref{sec:ClassicIIIs} where it was concluded that sizes observed at 1~GHz are around $1^\prime$. We can also make a reasonable estimate of the linear size of the emitting region, and calculate the angular size. This is sufficiently accurate for the purposes of estimating the observed flux from our simulations. Taking a  linear size of $10^9$~cm, located at a distance of 1~AU, and using the small angle approximation we obtain an angular extent of $0.2^\prime$, in reasonable agreement with the measurements, as some angular scattering may be expected. 

\subsection{Propagation Losses}\label{sec:propLoss}

Langmuir waves have a very small group velocity, and so the loss of waves from the emission region due to propagation is negligible. However, electromagnetic radiation has a group velocity given by \begin{equation}\rv_g=\frac{\partial}{\partial k}(\omega_{pe}^2 +c^2k^2)^{1/2}=\frac{c}{\omega}  (\omega^2-\omega_{pe}^2)^{1/2}\end{equation} where $\omega$ is the frequency of the emission and $\omega_{pe}$ the local plasma frequency, and this approaches $c$ for $\omega \gg \omega_{pe}$. 

For radiation in plasma of changing density this becomes \begin{equation}\rv_g(x)=\frac{c}{\omega_0}(\omega_0^2-\omega_{pe}^2(x))^{1/2}\end{equation} where $\omega_0$ is the frequency of the original emission, and $\omega_{pe}(x)$ is the local plasma frequency, which depends on position. 

We model the effects of radiation escape by adding a term \begin{equation} \left(\frac{\partial W^{\AAv}(k)}{\partial t}\right)_{esc}=- \frac{\rv_g}{d} W^{\AAv}(k)\end{equation} where $d$ is the size of the emitting region, to our equation for the evolution of the electromagnetic wave spectral energy densities.

\subsection{Collisional Damping}\label{sec:collisDamp}
In the corona, inverse bremsstrahlung is a significant cause of attenuation of electromagnetic radiation, with damping rate as given in Equation \ref{eqn:brems}, namely \begin{equation}\gamma_d=\left[\frac{4\pi e^2 {\rm ln}\Lambda}{3 m_e v_{Te}^3}\sqrt{\frac{2}{\pi}}\right] \left(\frac{\omega_{pe}^4}{\omega^2}\right)=:C \left(\frac{\omega_{pe}^4}{\omega^2}\right)\end{equation} which depends on position via the local plasma frequency. Assuming isothermal plasma between the radiation source and the observer\footnote{As will be seen, given an exponential density profile, the damping is only significant for a relatively short fraction of the total path length, and therefore this assumption is in general reasonable.} and writing the frequency and wavenumber of the emission at the source as $\omega_0$, $k_0$, we have the group velocity as in the previous section, and an optical depth given by definition as \begin{equation}\tau=\int_0^{x_0} \frac{\gamma_d(x)}{\rv_g(x)} dx.\end{equation} 

Substituting the expression for $\gamma_d$ the optical depth becomes \begin{equation}\tau(\omega_0)=\int_0^x dx  \frac{C}{c\omega_0}  \left(\frac{\omega_{pe}(x)^4}{(\omega_0^2-\omega_{pe}(x)^2)^{1/2}}\right).\end{equation} Clearly this diverges as $\omega_0 \rightarrow \omega_{pe}(0)$ because the group velocity goes to zero. 

Now the plasma frequency relates to the density via \begin{equation}\omega_{pe}^2(x)=\frac{4\pi e^2}{m_e} n_e(x)\end{equation} and so \begin{equation}\tau(\omega_0)=\frac{C}{c \omega_0} \left(\frac{4 \pi e^2}{m_e}\right)\int_0^x dx \left(\frac{n_e(x)^2}{(n_{0,k} -n_e(x))^{1/2}}\right).\end{equation}

Now we assume an exponential density profile with a fixed scale height $H$, so that $n_e(x)=n_0\exp(-x/H)$ where $n_0$ is the density at the source, and find \begin{equation}\label{eqn:optDepFin} \tau(\omega_0)=\frac{4 C H}{3c \omega_0}\left[ \omega_0^3 - \sqrt{(\omega_0^2-\omega_{pe}^2(0))}\left(\omega_0^2 +0.5\omega_{pe}^2(0) \right) \right]. \end{equation}

\begin{figure}
 \center
\includegraphics[width=0.5\textwidth]{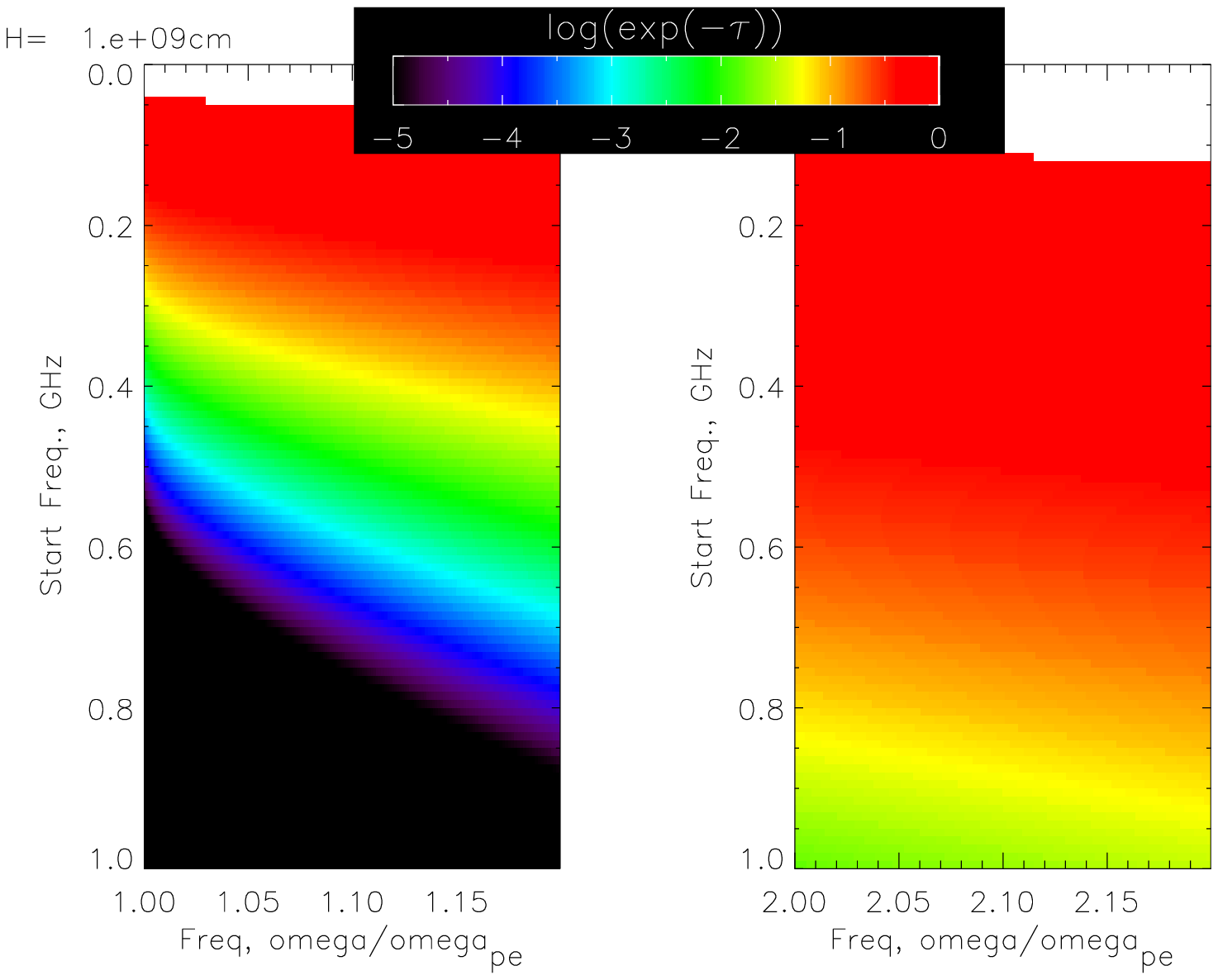}\includegraphics[width=0.5\textwidth]{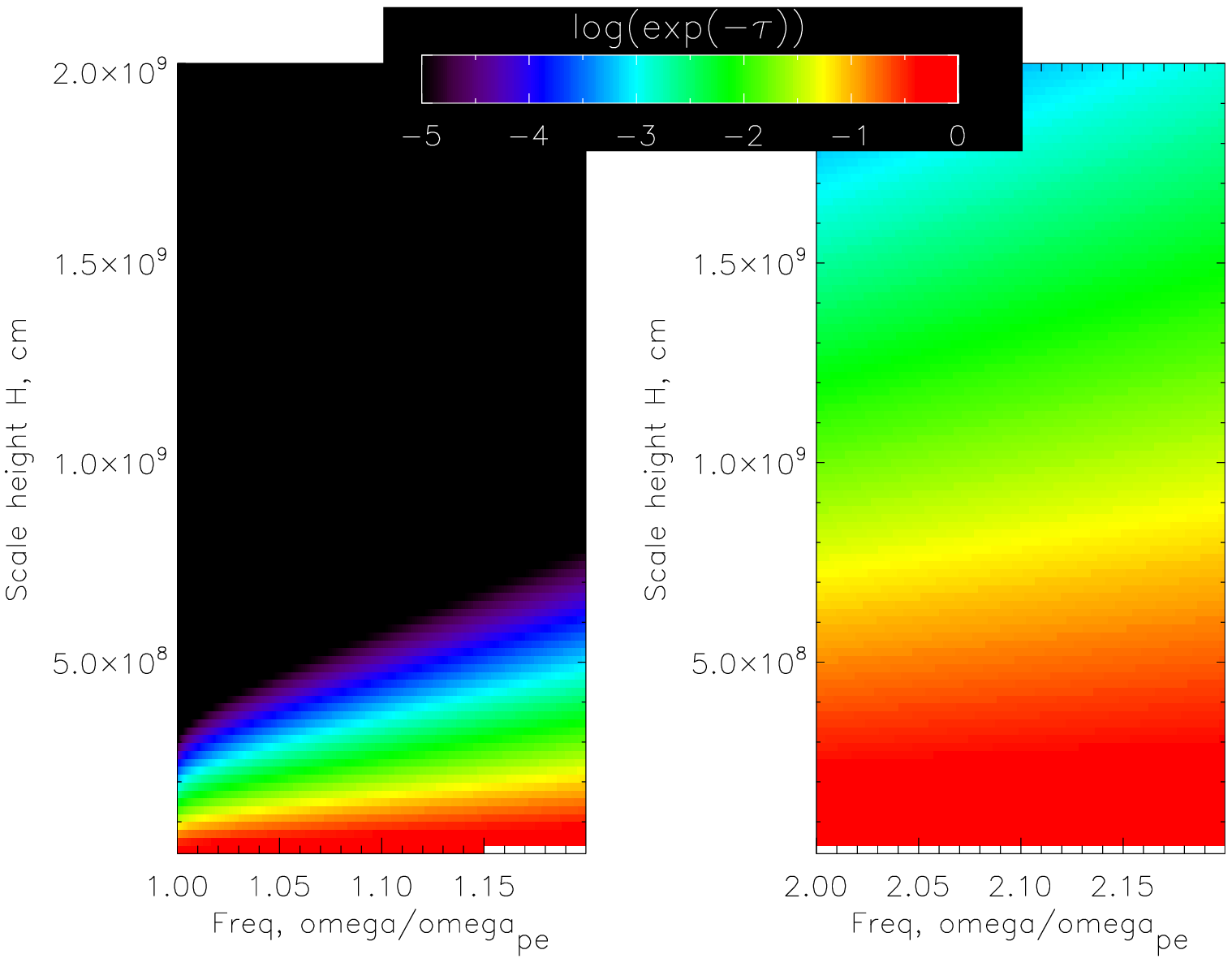}
\caption[The optical depth for radiation in the corona.]{Left: The optical depth for fundamental and harmonic plasma emission, assuming a exponential density profile with scale height $H=1\times10^9$~cm as function of the plasma frequency at height of emission. Right: The optical depth for fundamental and harmonic plasma emission from a local plasma frequency of $1$~GHz, as function of $H$ the scale height of the model exponential density profile.}\label{fig:optdepth}
\end{figure}

In Figure \ref{fig:optdepth} we plot the optical depth as a function of frequency for a given scale height, $H=1\times10^9$~cm, and also the optical depth as a function of scale height for a frequency of 1~GHz. Clearly, at frequencies much above 0.6~GHz there can be no escaping fundamental emission for this choice of scale height $H$, and at 1~GHz no escape of fundamental emission for any reasonable choice of $H$. For the harmonic the situation is better, but there will still be significant collisional damping, and at frequencies much larger than 1~GHz the escape fraction is rather small. 

A similar conclusion was reached by e.g. \citet{1992SoPh..141..335B} and \citet{1998SoPh..179..421K} for an exponential density profile, who suggest that harmonic emission can be the only escaping component, and moreover even this likely requires fibrous structures in the corona, with emission occurring in dense areas with small cross sectional area and escaping through the less dense surrounding plasma. 

For our purpose, which is an estimation of the observed flux due to a given spectral energy density of emission, this exponential density profile, with fixed scale length, is sufficient. For the corona, reasonable values of $H$ are around 1 to $5 \times 10^9$~cm. 

\subsection{Time Delay}
The time delay experienced by electromagnetic emission in propagating from where it is emitted to where it is observed also depends on frequency \begin{equation}\Delta t = \int _{0}^{x_{obs}} \frac{dx}{\rv_{gr}(x)},\end{equation} where $x_{obs}$ is the distance from the source to the observer. Substituting for the group velocity gives \begin{equation}\Delta t = \frac{\omega_0}{c}\left(\frac{m_e}{4\pi e^2}\right)^{1/2} \int _{0}^{x_{obs}} dx  (n_{0,k}-n_e(x))^{-1/2}.\end{equation} 

Again assuming an exponential density profile $n_e(x)=n_0 \exp(-x/H)$, we use the substitution, $s= (\omega_0^2-\omega_c^2 n_0 \exp(-x/H))^{1/2}$. Then $dx=ds (-2)H s (s^2-\omega_0^2)^{-1}$ with $\omega_c=4\pi e^2/m_e$. Now we have \begin{equation}\Delta t= \frac{-2 H \omega_0}{c} \int_{\sqrt{\omega_0^2-\omega_c^2n_0}}^{(\omega_0^2-\omega_c^2n_0\exp(-x_{obs}/H))^{1/2}}  \frac{ds}{s^2-\omega_0^2}.\end{equation} This may be evaluated using partial fractions. We assume \begin{equation}\frac{1}{s^2-\omega_0^2}=\frac{A s+B}{s-\omega_0}+\frac{Cs+D}{s+\omega_0},\end{equation} and find the constants $A,B,C,D$, which gives \begin{equation}\Delta t=\frac{-2 H\omega_0}{c} \int_{\sqrt{\omega_0^2-\omega_c^2n_0}}^{(\omega_0^2-\omega_c^2n_0\exp(-x_{obs}/H))^{1/2}} ds \left[ \frac{-s+\omega_0+\frac{1}{2\omega_0}}{s-\omega_0}+ \frac{s+\omega_0-\frac{1}{2\omega_0}}{s+\omega_0}\right].\end{equation} Performing this integral gives \begin{equation}\Delta t=\frac{-2H\omega_0}{c} \left[ \frac{1}{2\omega_0} \ln(|s-\omega_0|)-\ln(|s+\omega_0|)\right]_{\sqrt{\omega_0^2-\omega_c^2n_0}}^{(\omega_0^2-\omega_c^2n_0\exp(-x_{obs}/H))^{1/2}}. \end{equation}

Now the distance $x_{obs}$ for a coronal source is approximately 1AU and much larger than $H$, and so we can apply the binomial expansion to the upper limit, \begin{equation}(\omega_0^2-\omega_c^2n_0\exp(-x_{obs}/H))^{1/2}\simeq \omega_0 -\frac{1}{2}\omega_c^2n_0\exp(-x_{obs}/H)\end{equation} and use this in the argument of $\ln(|s-\omega_0|)$. 

The time delay is therefore finally \begin{equation}\Delta t=\frac{1}{c} \left[ x_{obs} +  H\left\{ \ln\left(\frac{4 \omega_0^2}{\omega_{pe,0}^2}\right) +\ln\left( \left|\frac{ \sqrt{\omega_0^2-\omega_{pe,0}^2}-\omega_0}{\sqrt{\omega_0^2-\omega_{pe,0}^2}+\omega_0}\right| \right)\right\}\right],\end{equation} with $\omega_0$ the frequency of the emission and $\omega_{pe,0}$ the local plasma frequency in the emitting region.

\subsection{Angular Scattering}

Radio observations at microwave and decimetre wavelengths rarely show fine structure on spatial scales below around $20^{\prime\prime}$ at 1.5~GHz and around $40^{\prime\prime}$ at 300~MHz, which can be explained by angular scattering of the radiation during propagation \citep{1994ApJ...426..774B}. Density fluctuations on a variety of scales have been measured in the corona at a few solar radii by various techniques \citep[e.g.][]{1989ApJ...337.1023C,1996AIPC..382..265S,1997JGR...102..263G}, and in the solar wind \citep[e.g.][]{1972ApJ...171L.101C,1983A&A...126..293C,1983PASAu...5..208R}. Various analytical \citep[e.g.][]{1994ApJ...426..774B,1998ApJ...506..456C,1999A&A...351.1165A} and numerical \citep[e.g.][]{1972PASAu...2...98R,1992A&A...253..521I} treatments of their effects have been attempted, with the general conclusion that at high MHz and GHz frequencies, scattering will produce sources of size at least $10$ to $20 ^{\prime\prime}$. The extent of this effect is far larger for emission at the fundamental, and tends to increase at lower frequencies, becoming very significant for IP bursts, where the source size can be approximately doubled \citep{1985A&A...150..205S}. 

However, for harmonic radiation the effects of angular scattering are likely to be small with respect to the size of the source itself, although any spatial structure within the source will be smeared out. Angular scattering is therefore unimportant for our purposes, although it will certainly occur due to the small-scale density fluctuations which we consider in our emission model. 

\section{Simulations of 2~GHz Plasma Emission}

Combining the model for plasma radio emission derived in the previous sections with the model of beam-wave interactions used in Chapters \ref{ref:Chapter2} and \ref{ref:Chapter3}, we now simulate the radio emission from a collisionally relaxing beam in inhomogeneous plasma. This work is currently begin prepared for publication as \citet{RK}. As discussed in Section \ref{sec:collRelax}, the formation of a reverse slope in the electron distribution will be mainly due to collisional effects for the case of dense plasma, considered over short spatial scales. Because of the high density we consider, the emission is purely at the harmonic, with the fundamental component produced inefficiently and unable to escape. 

The effects of plasma density fluctuations on the Langmuir waves in this model were considered in the previous chapter. In general, we found a decrease in the peak wave level, but significant increases at small wavenumbers. We also saw that collisional relaxation leads to a Langmuir wave spectrum which drifts to smaller wavenumbers over time. As mentioned in Section \ref{sec:TIIIs} and by e.g. \citet{1975A&A....39..107Z} the spectral evolution of the Langmuir waves is a significant factor in the rise and decay of the harmonic emission due to the strong constraints on participating wavenumber. We therefore expect the spectral evolution due to density fluctuations will play an important role in the emission.

\subsection{Complete Simulation Model}

\subsubsection{Equations for Electron and Wave Evolution}
We state here the entire set of equations which we simulate, dropping the superscripts $\AAv$ as all spectral energy densities are now \AATxt. Only plasma emission at the harmonic is considered, because the fundamental is very weak, and cannot escape from plasma of such high density (see Section \ref{sec:collisDamp}). 

For the electrons we have (Equation \ref{eqn:3ql1}) \begin{equation}\label{eqn:4ql1}
\frac{\partial f}{\partial t}= \frac{4\pi^2
e^2}{m_e^2}\frac{\partial}{\partial {\mr v}}\left( \frac{W_L}{{\mr
v}}\frac{\partial f}{\partial {\mr v}}\right) +
\Gamma\frac{\partial}{\partial \mr v}\left(\frac{f}{\mr
v^2}+\frac{\mr v_{Te}^2}{\mr v^3}\frac{\partial f}{\partial \mr
v}\right),
\end{equation} while for the Langmuir waves we have (Equation \ref{eqn:3ql2})
\begin{align}\label{eqn:4ql2}
\frac{\partial W_L}{\partial t} =&\frac{\omega_{pe}^3 m_e}{4\pi n_e}\rv \ln\left(\frac{\mr v}{\mr v_{Te}}\right)
+\frac{\pi\omega_{pe}^3}{n_ek^2}W_L\frac{\partial f}{\partial {\mr
v}}-\frac{\Gamma}{4 {\mr v}_{Te}^3}W_L+\notag \\&
\frac{\partial}{\partial
k}\left(D(k)\frac{\partial W_L}{\partial k}\right) + {\rm St}_{decay}(W_L,W_S(k_S))+{\rm St}_{ion}(W_L),
\end{align}
with the coefficient of diffusion, $D(k)$, given by Equation \ref{eqn:diffcoeffGauss}. The source term due to ion scattering is (Equation \ref{eqn:LIon}) \begin{align}\label{eqn:LIon2}{\rm St}_{ion}(W_L(k))=&\int dk_{L'} \frac{C_{ion}}{|k_L-k_{L'}|} \exp{\left(-\frac{(\omega_L-\omega_{L'})^2}{2|k_L-k_{L'}|^2 v_{Ti}^2}\right)} \times \notag\\&\left[\frac{\omega_{L'}}{\omega_L}W_L(k_L)-W_L(k_{L'})-\frac{(2\pi)^3}{T_i}\frac{\omega_{L'}-\omega_L}{\omega_L}W_L(k_{L'})W_L(k_L)\right]\end{align}  
and that due to ion-sound wave decays is (Equation \ref{eqn:3ql_sSrc})
\begin{align}\label{eqn:4ql_sSrc}
&{\rm St}_{decay}(W_L(k),W_S(k_S))=\alpha\omega_{k} \times \notag \\ &
\int\omega_{k_S}^S\left[ \left(
\frac{W_L(k_L)}{\omega^L_{k_L}}\frac{W_S(k_S)}{\omega^S_{k_S}}-
\frac{W_L(k)}{\omega^L_k}\left(\frac{W_L(k_L)}{\omega^L_{k_L}}+
\frac{W_S(k_S)}{\omega^S_{k_s}}\right)\right)\delta (\omega^L_{k}-\omega^L_{k_L}-\omega^S_{k_S})\right.
\notag \\&
-\left.
\left(
\frac{W_L(k_{L'})}{\omega^L_{k_{L'}}}\frac{W_S(k_S)}{\omega^S_{k_S}}-
\frac{W_L(k)}{\omega^L_k}\left(\frac{W_L(k_{L'})}{\omega^L_{k_{L'}}}-
\frac{W_S(k_S)}{\omega^S_{k_s}}\right)\right)
\delta(\omega^L_{k}-\omega^L_{k_{L'}}+\omega^S_{k_S}) \right.\biggr] \,dk_S 
\end{align} where we distinguish between $k_L= k-k_S$ and $k_{L'}=k+k_S$. For the ion-sound waves (where these can exist) we have (Equation \ref{eqn:3ql_s}) \begin{align}\label{eqn:4ql_s}
\frac{\partial W_S(k)}{\partial t}=&-2\gamma_S(k)
W_S(k)\notag \\&-\alpha ({\omega^S_k})^2\int
\left(
\frac{W_L(k_L)}{\omega^L_{k_L}}\frac{W_S(k)}{\omega^S_{k}}-
\frac{W_L(k_{L'})}{\omega^L_{k_{L'}}}\left(\frac{W_L(k_L)}{\omega^L_{k_L}}+
\frac{W_S(k)}{\omega^S_{k}}\right)\right)\times\notag \\&
\delta(\omega^L_{k_{L\prime}}-\omega^L_{k_L}-\omega^S_k)dk_{L\prime}
\end{align} where $k_L$ is the initial Langmuir wavenumber and  $k_L'$ is the scattered wave and the constants $\alpha, \beta$ and the sound wave damping rate are given by (Equations \ref{alphaL} and \ref{gam_sk} respectively):
\begin{equation}\alpha=\frac{\pi \omega^2_{pe}(1+3T_i/T_e)}{4n_ek_b T_e} \;\;
\beta=\frac{\sqrt{2\pi}\omega^2_{pe}}{4n_ek_b T_i(1+T_e/T_i)^2},
\end{equation}
\begin{equation}\gamma_S(k)=\sqrt{\frac{\pi}{8}}\omega^S_k\left[\frac{\rv_s}{\rv_{Te}}+\left(\frac{\omega
^S_k}{k\rv_{Ti}}\right)^3\exp\left[- \left(\frac{ \omega ^S_k}{
k\rv_{Ti} }\right)^2 \right]\right].
\end{equation}

For the harmonic emission we have \begin{equation}\frac{dW_T^{H}(k_T)}{dt} = \mr{St}_{harm}^{ll't}- \gamma_d W_T(k_T) + P(k_T) -\frac{\rv_g}{d}W_T(k_T)\end{equation}
where $\gamma_d$ is the collisional damping rate (Section \ref{sec:collisDamp}) \begin{equation}\gamma_d=\left[\frac{4\pi e^2 {\rm ln}\Lambda}{3 m_e \rv_{Te}^3}\sqrt{\frac{2}{\pi}}\right] \left(\frac{\omega_{pe}^4}{\omega^2}\right),\end{equation} the thermal emission is (Section \ref{sec:ThEM}) \begin{equation}P(k)=\frac{4 \pi k^2 e^2 \omega_{pe}^4 {\rm ln}\Lambda}{3 \omega^2 \rv_{Te}}\sqrt{\frac{2}{\pi}},\end{equation} and the term $\rv_g/d$, where $d$ is the size of the emitting region, accounts for propagation losses (Section \ref{sec:propLoss}).
The source term is given by \ref{eqn:HarmFinal}: 
\begin{align}\mr{St}_{harm}^{ll't} =&\omega_{k_T}^T \frac{\pi \omega_{pe}^2}{48 m_e n_e \rv_{Te}^2} \frac{(k_2^2-k_1^2)^2}{4 k_2^2 (2k_1-k_T\sqrt{2}/2)}\times \notag 
\\ & \left[\frac{k_T^2}{k_2^2}\frac{\pi}{\Delta\Omega}\frac{W_L(k_1)}{\omega_{k_1^L}}\frac{W_L(k_2)}{\omega_{k_2}^L}-\frac{W_T(k_T)}{\omega_{k_T}^T} \left(\frac{W_L(k_1)}{\omega_{k_1}^L}-\frac{W_L(k_2)}{\omega_{k_2}^L}\right)\right]. 
\end{align}

\subsubsection{Initial Conditions}
We take a plasma density of $n_e \simeq 10^{10}$~cm$^{-3}$, corresponding to a plasma frequency of $\nu_{pe}=\omega_{pe}/(2\pi)=1$~GHz. 

The initial beam and Langmuir wave distributions are as in Section \ref{sec:collRelax}, namely
\begin{equation}
f({\mr v}, t=0)= \frac{n_e}{\sqrt{2\pi} {\mr v}_{Te}} \exp\left(-\frac{{\mr v}^2}{2 {\mr v}_{Te}^2}\right) +\frac{2 n_{b}}{\sqrt{\pi}\, {\mr v}_b}\frac{\Gamma(\delta)}{ \Gamma(\delta-\frac{1}{2})}
\left[1+({\mr v}/{\mr v}_{b})^2\right]^{-\delta}
\end{equation} and \begin{equation}
W(k, t=0)= \frac{k_b T_e}{4 \pi^2}\frac{k^2\ln\left(\frac{1}{k\lambda_{De}}\right)}
{1+\frac{\ln \Lambda}{16\pi n_e}\sqrt{\frac{2}{\pi}}k^3 \exp\left(\frac{1}{2k^2\lambda_{De}^2}\right)},
\end{equation}
and the initial radio spectral energy density $W_T^{H}(k_T)$ is set to the thermal level, i.e.
\begin{equation}
W_T^{H}(k_T, t=0)= 4 \pi k_b T_e k_T^2.
\end{equation}

\begin{table}
\begin{tabular}{|l|c||l|c|}
\multicolumn{2}{c}{Plasma}&\multicolumn{2}{c}{Radio source}\\ \hline
Plasma frequency, $\nu_{pe}$& 1~GHz &Emission mode&Harmonic \\
Plasma density, $n_e$&$10^{10}$~cm$^{-3}$&Density scale height& $10^9$~cm\\
Electron temperature, $T_e$&1~MK&Source size&$10^9$~cm\\
Ion temperature, $T_i$&0.5-1~MK&Langmuir spread, $\Delta\Omega$&$\pi/16$\\ 
Debye wavenumber, $k_{De}$&16~cm$^{-1}$&&\\ 
Collisional time, $\tau_{coll}$&$10^{-4}$~s&&\\ 
\hline \multicolumn{2}{c}{Beam}&\multicolumn{2}{c}{Density fluctuations}\\ \hline
Spectral index, $\delta$&4&RMS fluctuation $\sqrt{\langle \tilde{n}^2\rangle}$&$10^{-5}$ to $10^{-3}$\\
Beam velocity, $\rv_b$&$5\times 10^{9}$~cm~s$^{-1}$&Velocity, $\rv_0$&$10^7$~cm~s$^{-1}$\\
Beam density, $n_b/n_e$&$10^{-3}$ to $10^{-2}$&Wavenumber, $q_0$&$10^{-4} k_{De}$\\
\hline
\end{tabular}
\caption{Simulation parameters.}\label{tab:1}
\end{table}

\subsubsection{Propagation Effects and Observed Fluxes}

From the \AATxt spectral energy density $W_T^{H}(k_T)$ for harmonic emission, we can estimate an observed flux, given by \begin{equation}\label{eqn:c4Flux}F(\nu)=2\pi F(\omega)=W_T(k_T) \frac{\pi\theta^2}{2} \exp{(-\tau(\omega))},\end{equation} where $\nu$ is the frequency, $\nu=\omega/2\pi$, $\theta$ is the observed angular extent of the source, which we take as $1^\prime$ (see Section \ref{sec:SrcSize}) and $\tau(\omega)$ is the optical depth from Equation \ref{eqn:optDepFin}. The density scale height is $H=10^9$~cm, as in Table \ref{tab:1}. 

\subsection{Scattering by Ions}

\begin{figure}
 \centering
\includegraphics[width=0.95\textwidth]{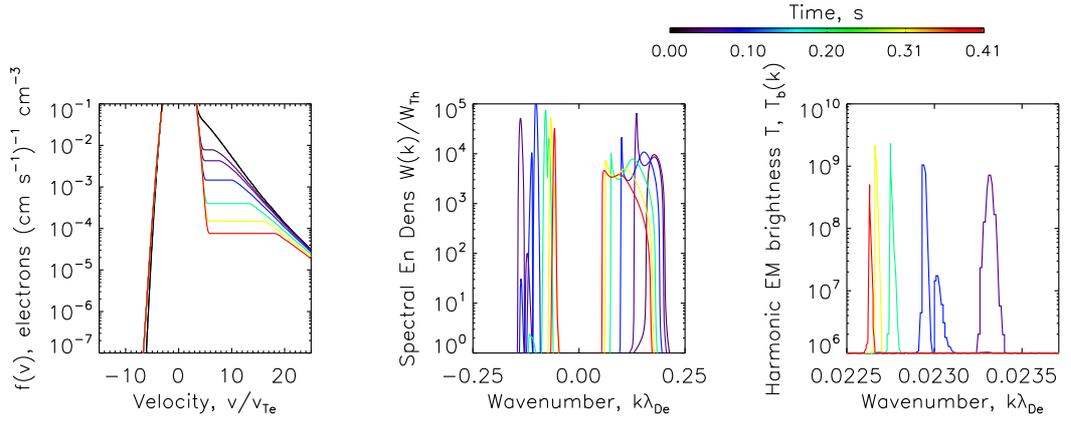}
\caption[Electron, Langmuir and harmonic electromagnetic wave distributions for a collisionally relaxing electron beam: homogeneous plasma.]{The electron distribution function $f({\mr v})$ (left); the spectral energy density of Langmuir waves $W(k)$ (middle); and the harmonic radio brightness temperature $T_b(k)$ (right) for a collisionally relaxing electron beam in homogeneous plasma. Each coloured line shows the distribution at a different time, as shown in the colour bar. Beam and plasma parameters are given in the text.}\label{fig:IShomO}
\end{figure}

We begin by considering the emission in plasma with equal ion and electron temperatures, $T_i=T_e=1$~MK. In this case, ion-sound wave interactions cannot occur, and the generation of backscattered Langmuir waves is due to scattering by individual ions, given by Equation \ref{eqn:LIon2}. We take a beam density of $n_b=10^8$~cm$^{-3} \simeq 10^{-2} n_e$, a velocity of $\rv_b=5\times 10^{9}$~cm~s$^{-1}$ and a velocity space power law index of $4$. Collisional relaxation proceeds as in the previous chapter on the collisional timescale $\tau_{coll}$, and we consider the evolution over one to two thousand $\tau_{coll}$, after which the levels of backscattered Langmuir waves have decreased and the radio emission peaked. 

\begin{figure}
 \centering
\includegraphics[width=0.95\textwidth]{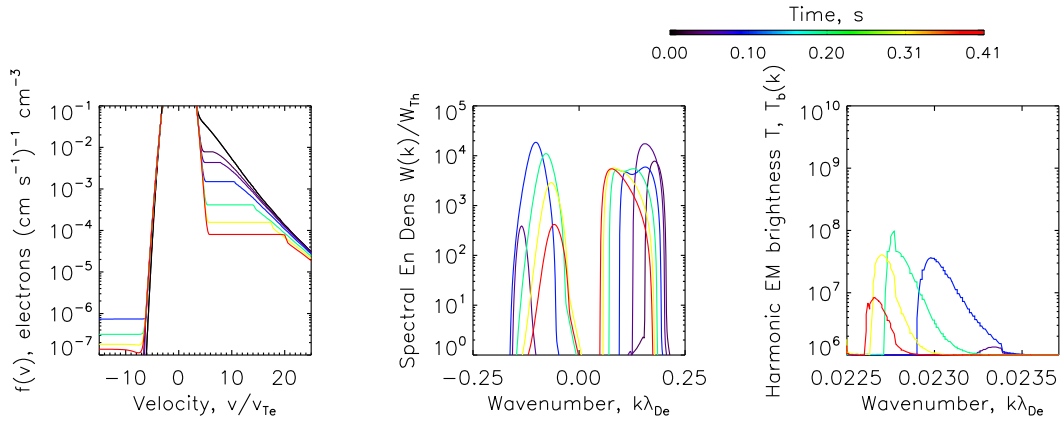}
\caption[Electron, Langmuir and harmonic electromagnetic wave distributions for a collisionally relaxing electron beam: weak inhomogeneity.]{As Figure \ref{fig:IShomO} for inhomogeneous plasma with $\sqrt{\langle \tilde{n}^2\rangle} =1\times 10^{-4}$.}\label{fig:ISinhO}
\end{figure}

In Figure \ref{fig:IShomO} we show the electron distribution, the Langmuir wave spectral energy density and the harmonic radiation brightness temperature for the case of homogeneous plasma. The exponential factor in the backscattering probability in Equation \ref{eqn:LIon2} means that only wavenumbers of magnitude close to the wavenumber under consideration can contribute to the backscattering. Spontaneous scattering, described by the first two terms in square brackets in Equation \ref{eqn:LIon}, proceeds fairly slowly, and gives a smooth backwards Langmuir wave spectrum. Once a small level of backscattered waves is present, stimulated emission can occur, as given by the third term in square brackets, involving the product $W_{L^\prime}W_L$. This proceeds much more rapidly but because of the factor of $\omega_{L^\prime}-\omega_L$ in this term, only regions with a locally positive gradient $d W(k)/dk$ can contribute. The resulting backscattered Langmuir wave spectrum is therefore irregular. 

\begin{figure}
 \centering
\includegraphics[width=0.49\textwidth]{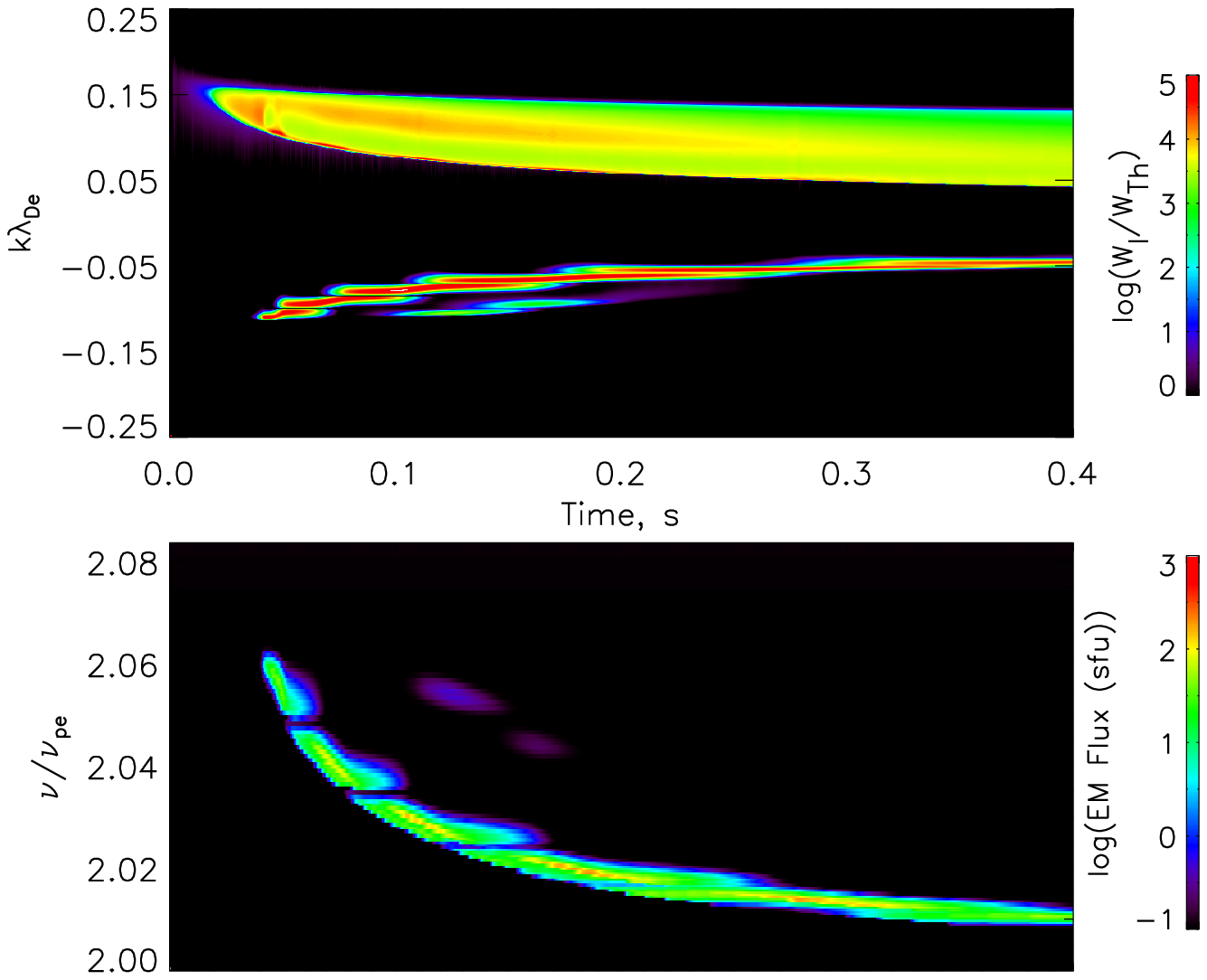}\includegraphics[width=0.49\textwidth]{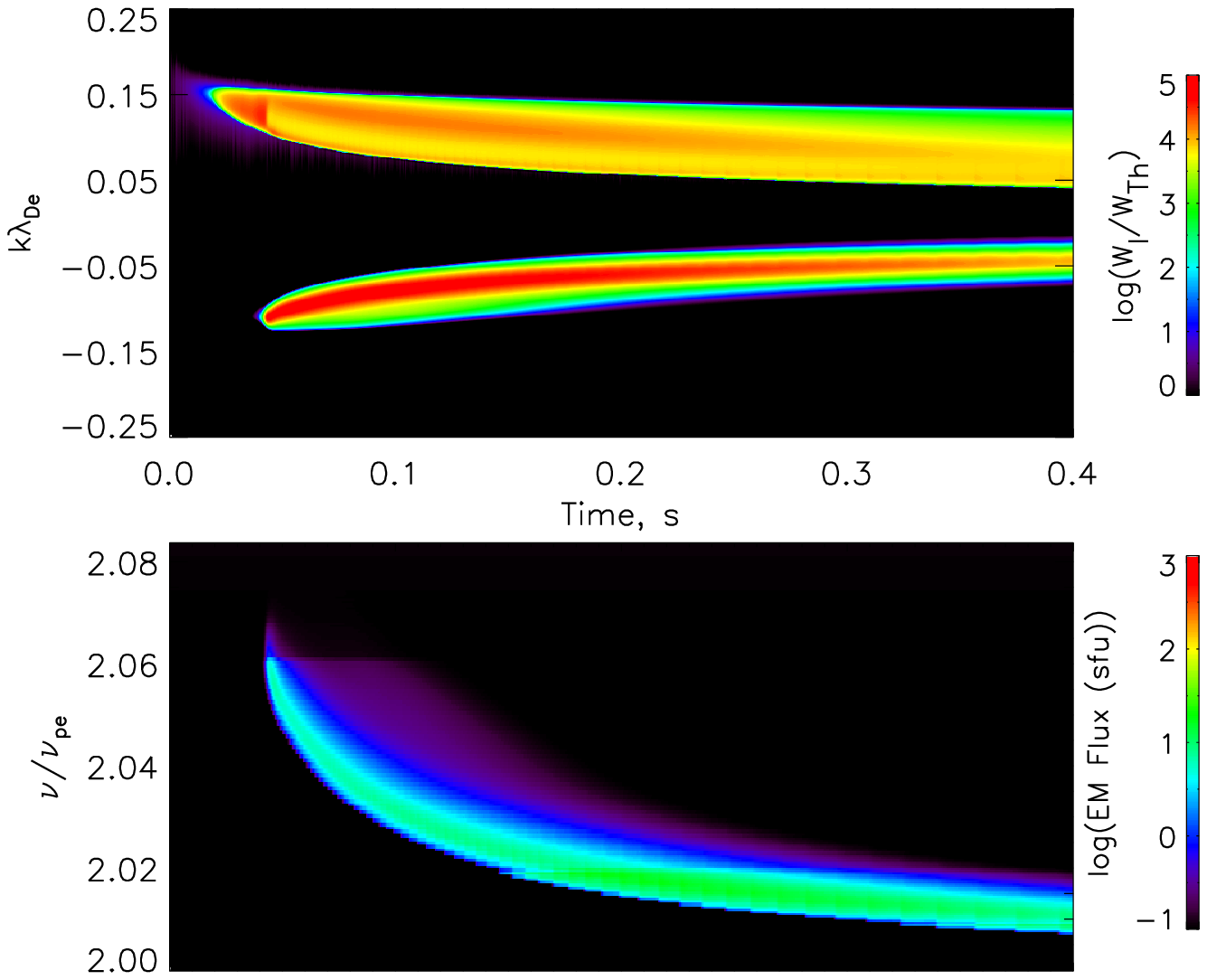}\\
\includegraphics[width=0.49\textwidth]{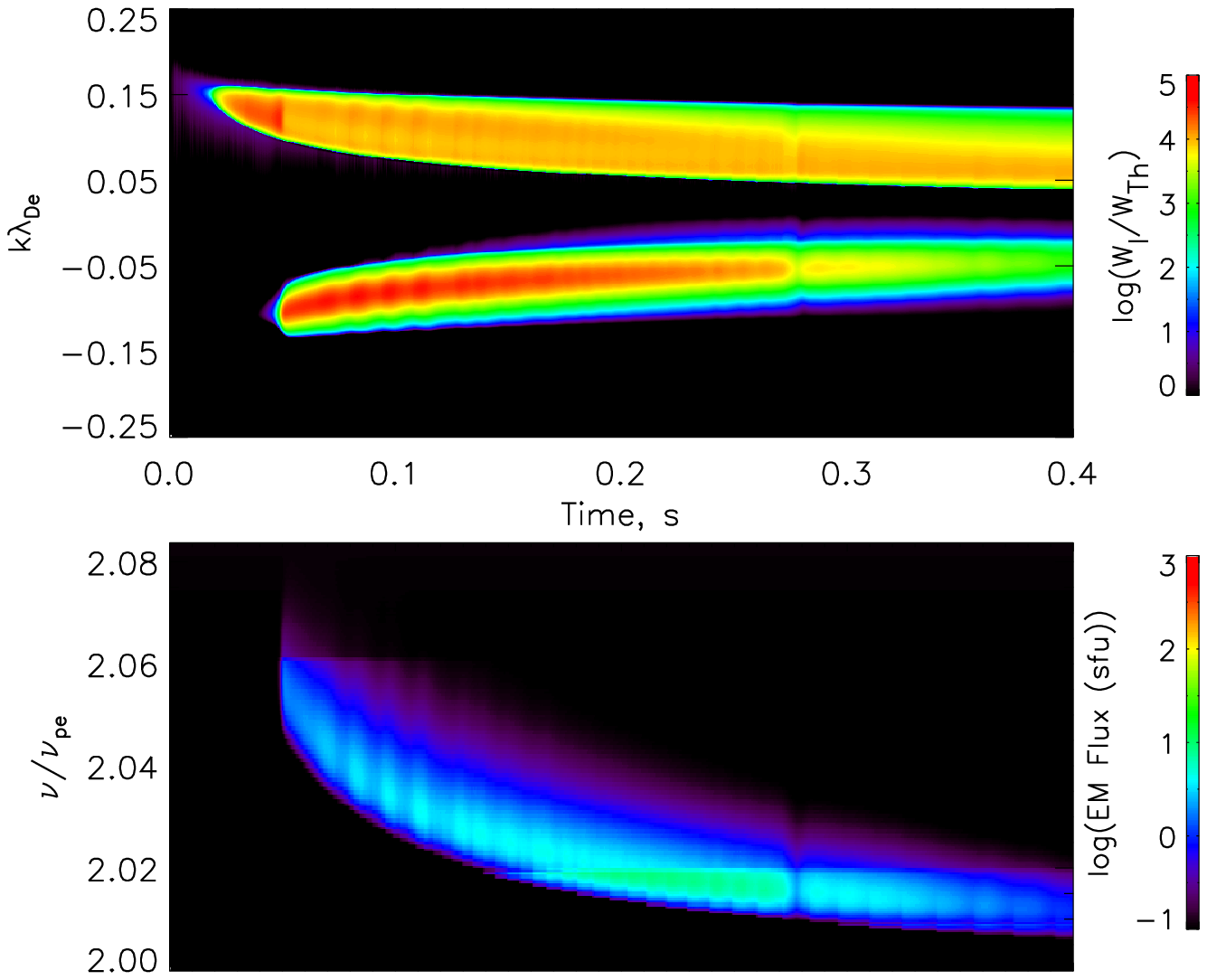}\includegraphics[width=0.49\textwidth]{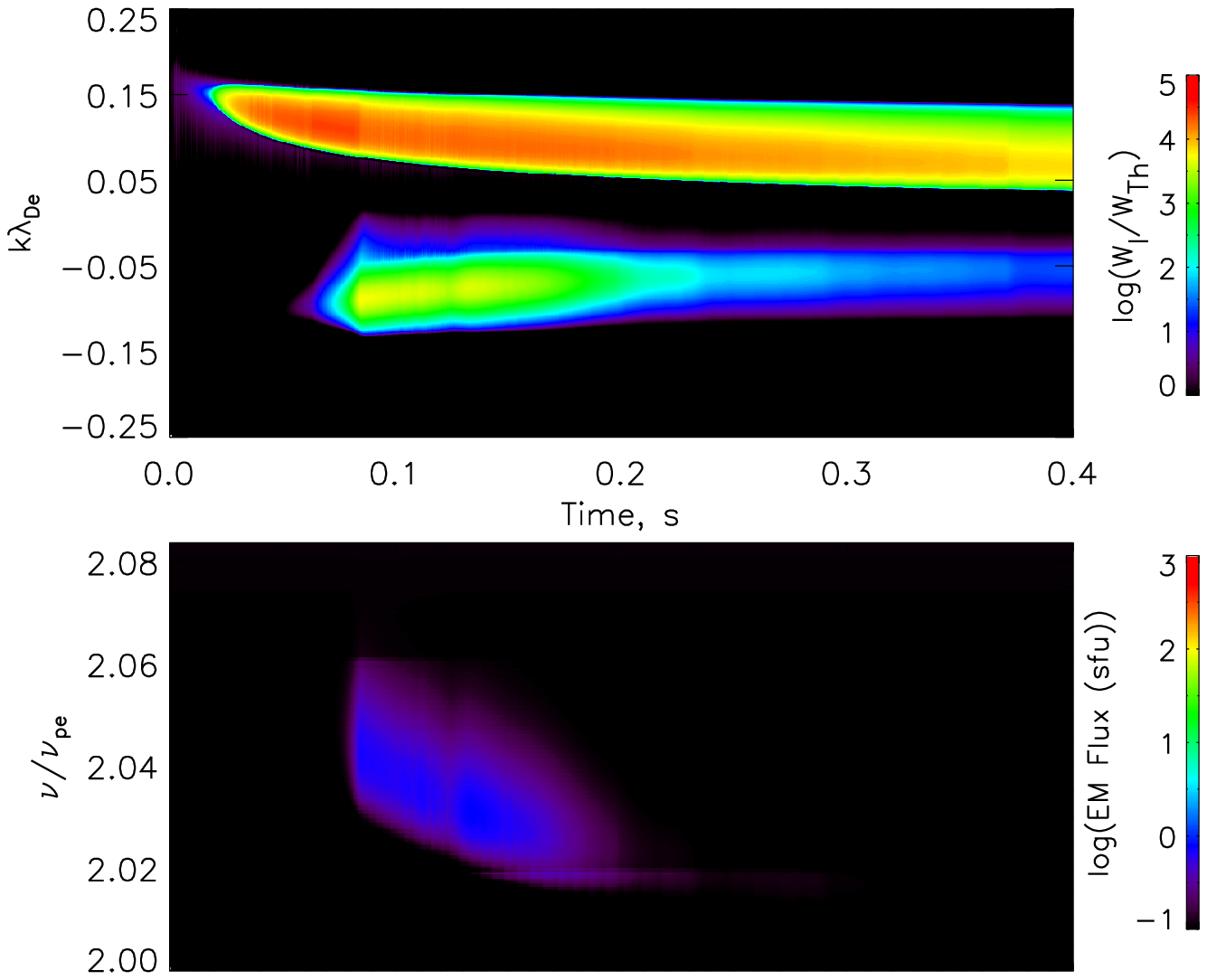}
\caption[The Langmuir wave and harmonic electromagnetic wave distributions for a collisionally relaxing electron beam: homogeneous and inhomogeneous plasma.]{The Langmuir wave spectral energy density and radio emission in sfu over the first 0.4~s of beam evolution. Top: Langmuir wave spectral energy density $W(k)$ normalised to the thermal level, against wavenumber $k$ on the vertical axis, and time on the horizontal axis. Bottom: the radio flux in sfu as a function of frequency, $\nu=\omega/(2\pi)$, including the effects of absorption during propagation. The source size and plasma density profile are as in Table \ref{tab:1}, and the observed background flux from a thermal source of this size is $\sim 10^{-2}$~ sfu. Left to right, top to bottom: homogeneous plasma, weak diffusion, moderate diffusion and strong diffusion (see definitions in text).}\label{fig:IShom}
\end{figure}

This effect is seen clearly in the plots of the Langmuir wave spectral energy density and radio flux shown in Figure \ref{fig:IShom}, giving the observed radio emission in sfu calculated using Equation \ref{eqn:c4Flux} with parameters as in Table \ref{tab:1}. The calculated fluxes of a few to a few hundred sfu in the strongest case are in good agreement with the observed fluxes of high frequency plasma emission (Section \ref{sec:ClassicIIIs}). On the other hand, we note that these fluxes are calculated assuming a simple exponential density profile, which for the parameters chosen gives a factor $\exp(-\tau)$ of approximately $1/50$. More efficient escape can therefore lead to fluxes up to $50$ times larger. 

In Figure \ref{fig:IShom} we show three cases of density fluctuations, chosen using the results of the previous chapter, as well as the homogeneous case. The characteristic velocity and wavenumber of the fluctuations are fixed at $\rv_0=10^7$~cm~s$^{-1}$ and $q_0=10^{-4} k_{De}$ respectively. For the RMS fluctuation magnitude, ``weak'' diffusion has $\sqrt{\langle \tilde{n}^2\rangle} =4\times 10^{-5}$, ``moderate'' diffusion has $\sqrt{\langle \tilde{n}^2\rangle} =1\times 10^{-4}$ and ``strong'' diffusion has $\sqrt{\langle \tilde{n}^2\rangle} =3\times 10^{-4}$. Although this range is small, it covers the interesting cases for these beam and plasma parameters. 

As the strength of the Langmuir wavenumber diffusion increases we see more and more smoothing of the backscattered emission, and a reduction in its peak magnitude. At the same time, the removal of waves from the forwards spectrum is also suppressed, and the levels of emission significantly reduced. 

\begin{figure}
 \centering
\includegraphics[width=0.49\textwidth]{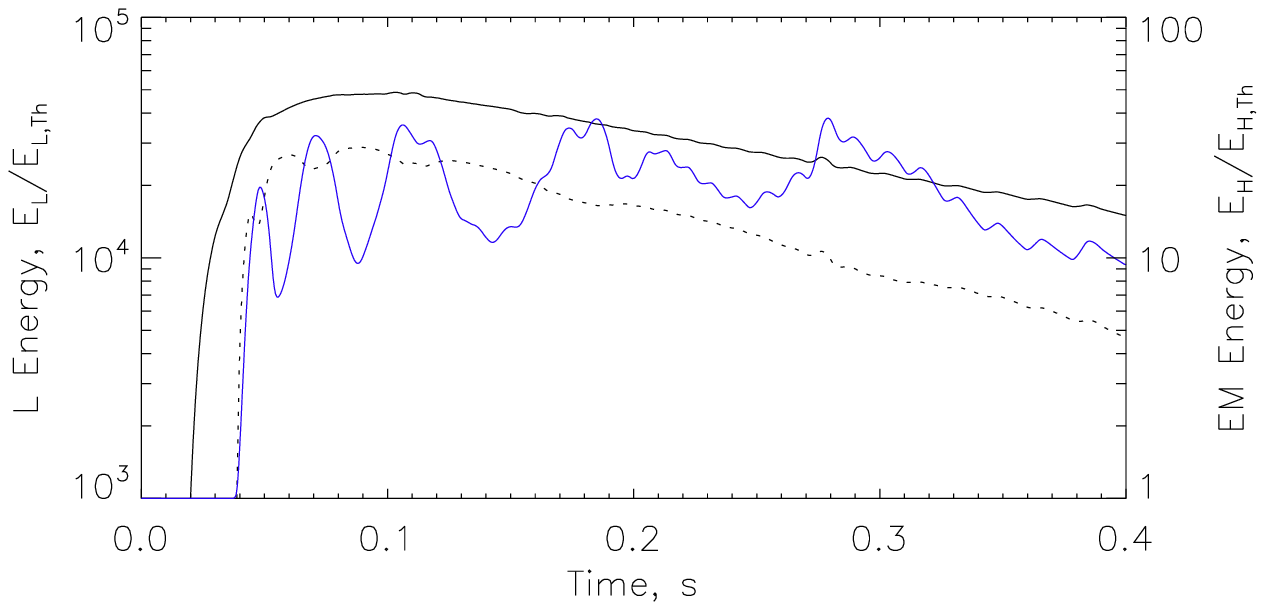}\includegraphics[width=0.49\textwidth]{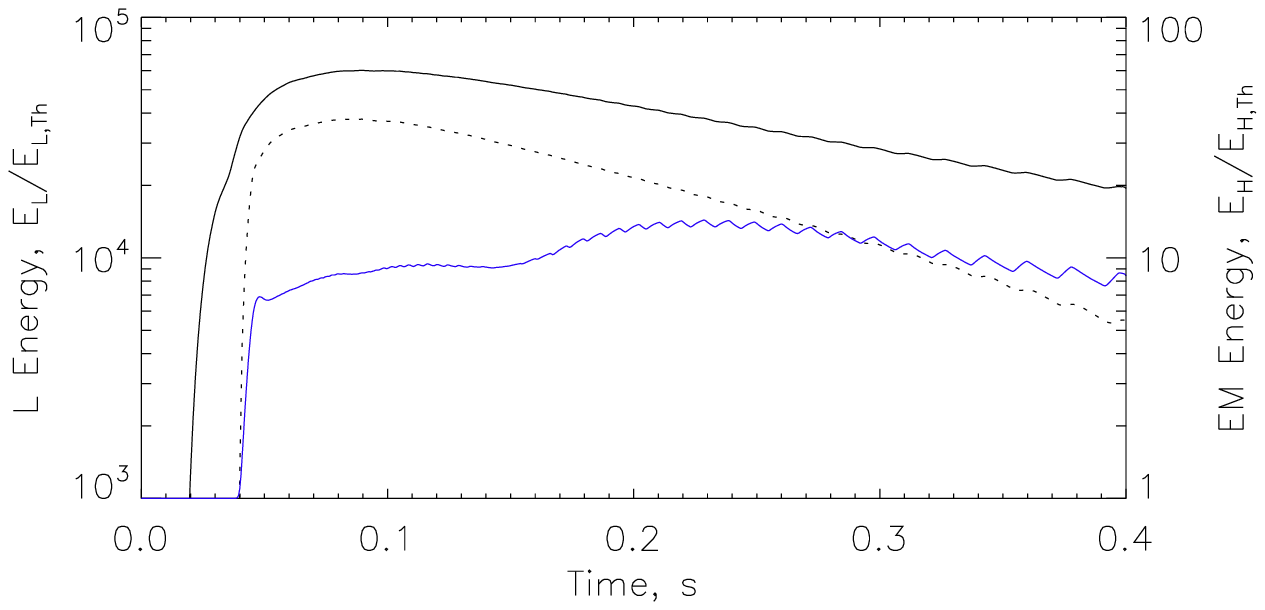} \\
\includegraphics[width=0.49\textwidth]{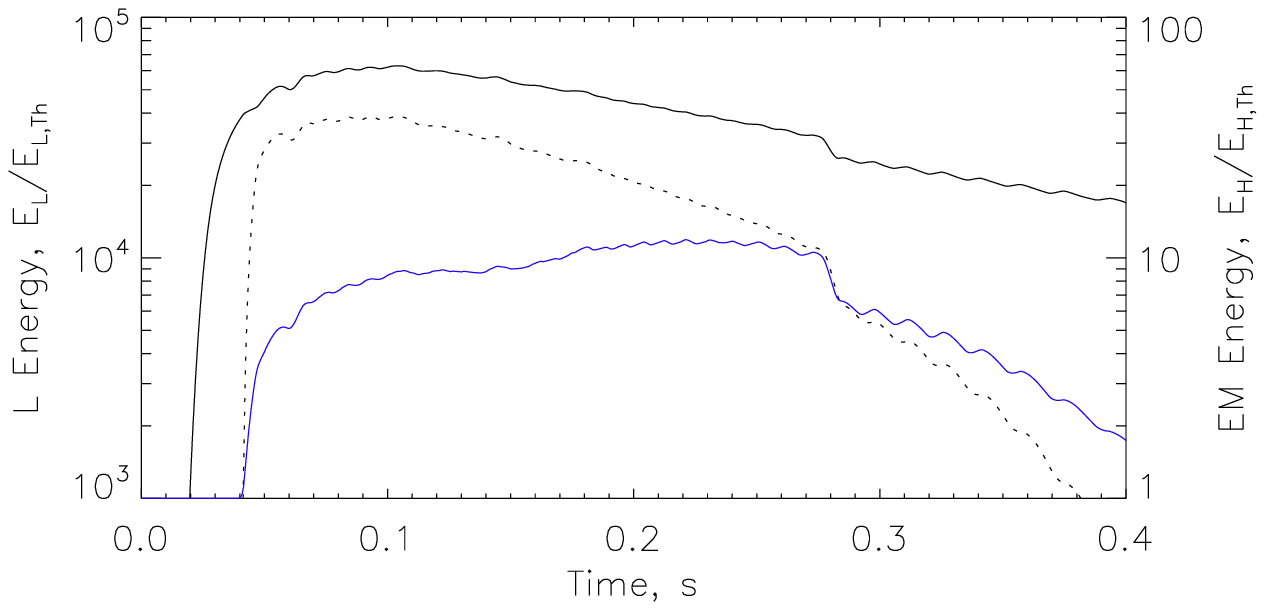}\includegraphics[width=0.49\textwidth]{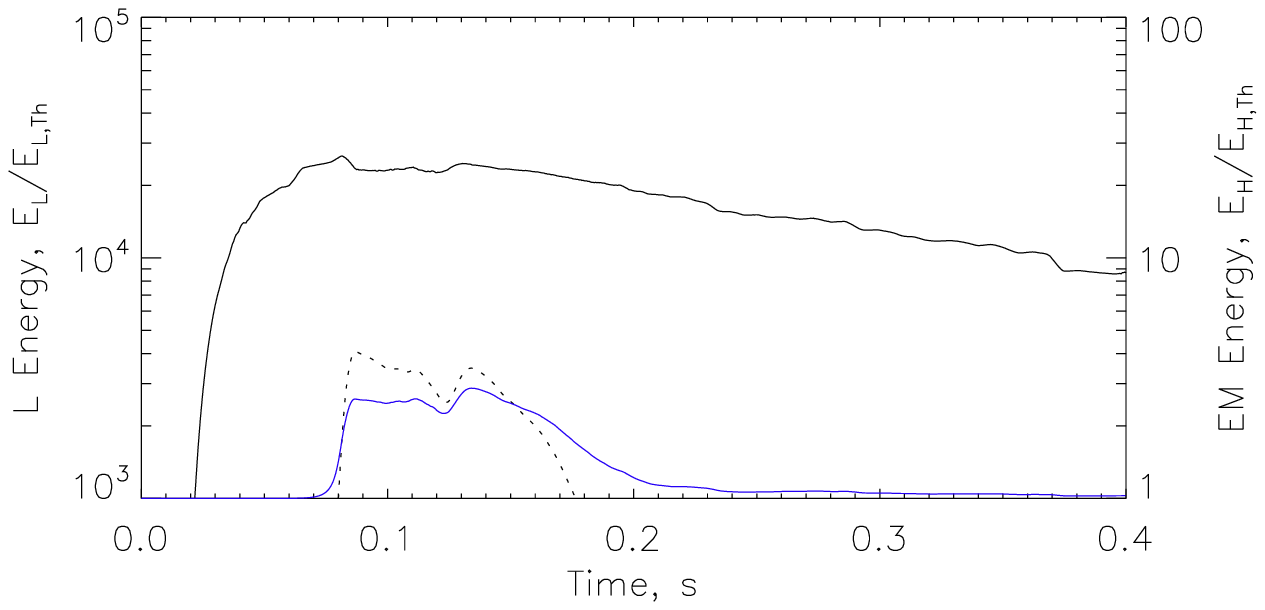}
EnsDiffMoreWeak
\caption[Time profiles of the Langmuir wave and radio energies.]{The energy above thermal (Equations \ref{eq:EnST}, \ref{eq:EnST2}) in Langmuir waves (solid line), backscattered Langmuir waves only (dotted line), and radio emission (blue line), normalised by the thermal levels, against time. Left to right, top to bottom: homogeneous plasma, and weak, moderate and strong inhomogeneity (see definitions in the text).}\label{fig:Ens1}
\end{figure}

To quantify the reductions, we calculate the non-thermal energy in the Langmuir waves, backwards Langmuir waves and electromagnetic emission, given by \begin{equation}\label{eq:EnST} E_L=\int_{-k_{De}}^{k_{De}} \left[W(k)-W_{Th}(k)\right] \,\mathrm{d}k\;,\;  E_L^b=\int_{-k_{De}}^0 \left[W(k)-W_{Th}(k)\right]\,\mathrm{d}k\end{equation} and 
\begin{equation}\label{eq:EnST2} E_H=\int_{\omega(k)=2.0}^{\omega(k)=2.1} \left[W(k)-W_{Th}(k)\right] \,\mathrm{d}k.\end{equation} We plot these in Figure \ref{fig:Ens1}. The saw-tooth oscillations seen around 0.3 to 0.4~s are a numerical effect due to the finite wavenumber grid used in the simulations. The relation between the harmonic emission and the backscattered Langmuir waves is clearly seen, as the rise of $E_H$ closely tracks $E_L^b$. However, at later stages, the drift of the maximum Langmuir wave level towards smaller wavenumbers, where the conversion to harmonic emission occurs more rapidly, can become important and the emission level can increase despite continuing decrease of the total and backwards Langmuir wave energy. This is clearly seen around 0.15 to 0.3~s for fluctuations of weak or moderate strength. 

\begin{figure}
 \centering
\includegraphics[width=0.49\textwidth]{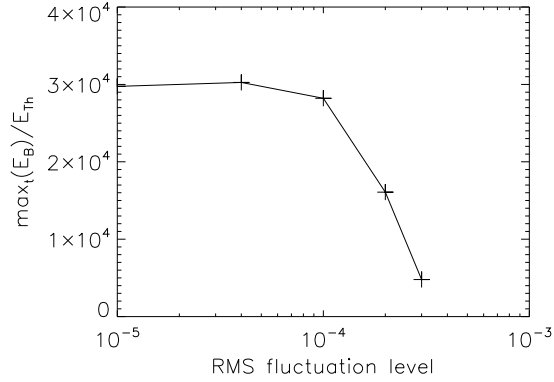}
\caption[The peak energy in backscattered Langmuir waves, against the magnitude of plasma inhomogeneity.]{The peak energy in backscattered Langmuir waves, $\mathrm{max}(E_L^b)$ (see Equation \ref{eq:EnST}), against the RMS magnitude of plasma inhomogeneity, $\sqrt{\langle \tilde{n}^2\rangle}$.}\label{fig:peakBack}
\end{figure}

On the other hand, for the strongest fluctuations shown, we see that the onset of rapid stimulated scattering is strongly suppressed and the level of backwards waves is decreased by more than an order of magnitude. Figure \ref{fig:peakBack} shows the peak energy in backscattered waves as a function of the RMS fluctuation magnitude, and displays a very sharp reduction for a level of $\sqrt{\langle \tilde{n}^2\rangle} =3\times 10^{-4}$. The energy in radio emission is similarly decreased, and remains above the thermal level for only a very short time, in contrast to the 0.4~s or more duration in the less inhomogeneous cases. Thus, despite having the same energy in initial electrons, and a very similar energy in the forwards Langmuir waves, the Langmuir wavenumber diffusion can suppress the backscattering and therefore the emission.

Finally we note that the intrinsic band width of the emission due to ion-scattering is rather large, 50~MHz at 1~GHz plasma frequency, and the emission shows an intrinsic frequency drift due to the drift of the Langmuir wave peak to lower frequencies during the collisional relaxation process.

\subsection{Wave-wave Interactions}
\begin{figure}
 \centering
\includegraphics[width=10cm]{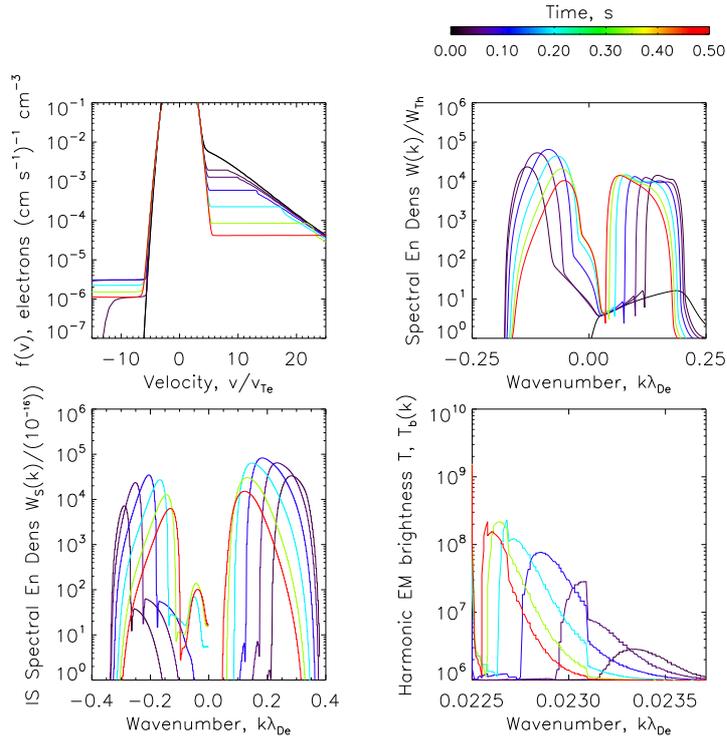}
\caption[Electron, Langmuir and harmonic electromagnetic wave distributions for a collisionally relaxing electron beam: wave-wave interactions.]{The electron distribution function $f({\mr v})$ (top left); the spectral energy density of Langmuir waves $W(k)$ (top right); the spectral energy density of ion-sound waves $W_s(k)$ (bottom left); and the harmonic radio brightness temperature $T_b(k)$ (bottom right) for a collisionally relaxing electron beam in inhomogeneous plasma. Each coloured line shows the distribution at a different time, as shown in the colour bar. We set $T_i=0.5 T_e$ and $n_b=10^7$~cm$^{-3}$. The effects of ion-sound wave interactions are included, as well as density fluctuations with RMS magnitude of $\sqrt{\langle \tilde{n}^2\rangle} = 10^{-3}$.}\label{fig:SndStrO}
\end{figure}

For plasma with a larger electron temperature than ion temperature, the backscattering of Langmuir waves can occur due to interactions with ion sound waves. We consider here a temperature ratio of $T_i=0.5 T_e$, for which ion-sound waves are still heavily damped, but the 3-wave decay process are allowed. The decay $L+s \rightleftarrows L'$ gives more efficient Langmuir wave backscattering than the ion-scattering process and so for simplicity we switch off the latter and consider only the former. 

\begin{figure}
 \centering
\includegraphics[width=10cm]{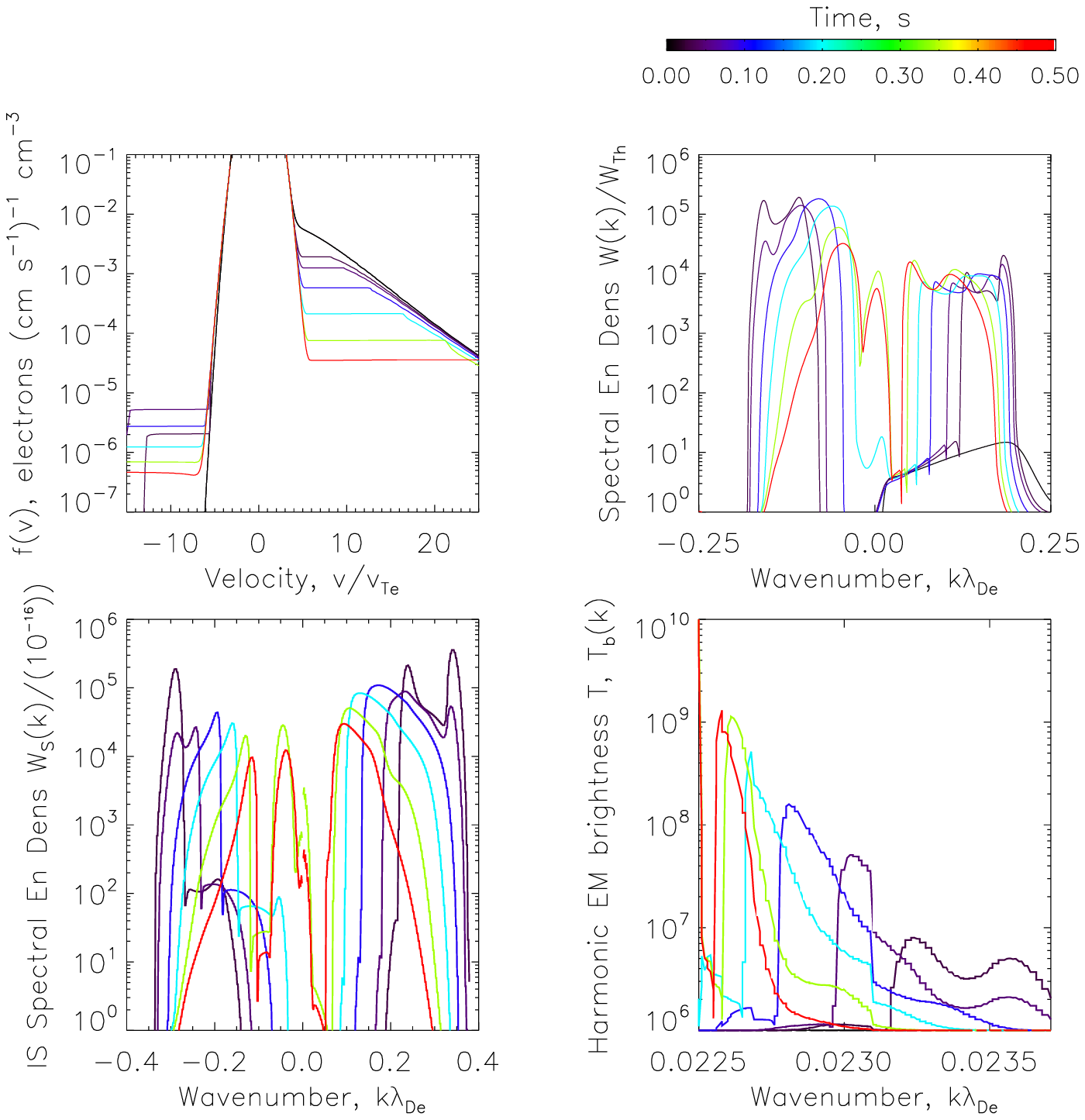}
\caption[Electron, Langmuir and harmonic electromagnetic distributions for collisionally relaxing electron beam: wave-wave interactions.]{As Figure \ref{fig:SndStrO} for density fluctuations with RMS magnitude of $\sqrt{\langle \tilde{n}^2\rangle} = 10^{-4}$.}\label{fig:SndWkO}
\end{figure}

In Figures \ref{fig:SndStrO} and \ref{fig:SndWkO} we show the resulting electron distributions, and the Langmuir, ion-sound and electromagnetic wave distributions over time for a beam density of $n_b=10^7$~cm$^{-3}$, including the effects of decay $L+s \rightleftarrows L'$, for plasma inhomogeneities of RMS magnitude $\sqrt{\langle \tilde{n}^2\rangle} = 10^{-3}$ and $\sqrt{\langle \tilde{n}^2\rangle} =10^{-4}$ respectively. As before, we see a slow drift of the Langmuir wave peak from large to small wavenumbers due to the collisional relaxation, but now there is additional scattering of the Langmuir waves to small wavenumbers. The Langmuir waves backscatter to smaller wavenumbers, $k-\Delta k$ (see Section \ref{sec:wave_wave}) and so the radio emission occurs for slightly smaller wavenumbers, giving an intrinsic bandwidth of only about 25~MHz, half that seen in the ion-scattering case. 

On the other hand, for ion-sound wave decays, even much stronger inhomogeneity, with $\sqrt{\langle \tilde{n}^2\rangle} = 10^{-3}$, does not drastically decrease the backscattering, and we obtain emission reaching a brightness temperature of $2\times 10^8$~K, corresponding to a flux of around 20~sfu. 

\section{Conclusions}
A collisionally relaxing electron beam of initial density $10^7$ or $10^8$~cm~s$^{-3}$ will produce harmonic radio emission due to the plasma mechanism. Our simulations show emission at a single value of the local plasma frequency with brightness and duration in good agreement with observations. The above quoted fluxes are derived by assuming a simple model for the escape of emission, and so could be increased by up to perhaps 10 times if the escape were more efficient. Using the results of \citet{1997SoPh..171..393W} we infer the emission will be weakly O-mode polarised, up to around 15 \%, also in good agreement with observations. 

For equal temperature plasma, $T_i=T_e$, the only source of backscattered Langmuir waves is the process of scattering by plasma ions. Spontaneous scattering is a slow process, but stimulated scattering can occur rapidly. However, as we have seen, plasma density fluctuations lead to the diffusion of the Langmuir waves in wavenumber space, and can prevent the onset of rapid stimulated scattering. In this case, very few backscattered Langmuir waves are produced and consequently very little radio emission. This is seen to occur for a very low level of fluctuations, $\sqrt{\langle \tilde{n}^2\rangle} = 3\times 10^{-4}$ for the parameters considered here. 

The magnitude of relative density fluctuations $\sqrt{\langle \tilde{n}^2\rangle}$ commonly observed in the corona and solar wind range from $10^{-4}$ and $10^{-2}$ \citep[e.g.][and Section \ref{sec:flucs}]{1972ApJ...171L.101C, 1979ApJ...233..998S,1989ApJ...337.1023C}. Assuming equal ion and electron temperatures, we infer that only in the least inhomogeneous case will radio emission be seen. Clearly this is only proven in cases where our model of a collisionally relaxing beam adequately captures the Langmuir waves dynamics: however the effects of density inhomogeneity on the onset of stimulated ion scattering may extrapolate to other situations. 

On the other hand, if $T_i=0.5 T_e$ we can consider scattering due to ion-sound waves, which occurs much more rapidly, and is less affected by inhomogeneities. However, the ion temperature in the corona at the frequencies we consider is generally thought to be equal to the electron temperature, and so the ion sound waves cannot exist and scattering must be due to ions. 

Thus, in the case where $T_i=T_e$, our model suggests that it is possible to obtain a large number of Langmuir waves at positive wavenumbers, but, due to their diffusion in wavenumber caused by plasma inhomogeneity, not obtain very high levels of waves at negative wavenumber and thus not produce visible levels of harmonic radio emission.

\chapter{Conclusions}
\label{ref:Chapter5}
During solar flares, vast amounts of energy are released from the Sun's magnetic field, a fraction of which goes into accelerating electrons. This thesis has considered the propagation of such a fast electron beam in the plasma of the solar corona, and its visible signatures in the form of hard X-ray and radio emission. When a beam propagates along a magnetic field line downwards into the dense chromosphere, it can produce hard X-rays via bremsstrahlung, and the spectrum of these will be affected by the evolution of the electron distribution during transport. Radio emission can be produced during propagation via the generation of Langmuir waves, and the spectrum of this will therefore depend strongly on the electron and Langmuir wave distributions. 

It has long been known that a propagating beam can become unstable to the generation of high levels of Langmuir waves, either due to transport effects, where faster electrons outpace slower ones, or to collisional effects due to the velocity dependence of the collisional operator. It is also well known that Langmuir waves are strongly affected by density inhomogeneities, and that this can potentially lead both to electron-self acceleration in certain experimental setups, and to suppression of the beam-plasma instability, and thus the relaxation of the electron beam to a plateau. 

\subsubsection{Langmuir Wavenumber Diffusion}
We have considered here primarily the effects of a fluctuating plasma density, with long wavelength and small relative density change, such as is observed in the solar corona and solar wind. Chapter \ref{ref:Chapter2} outlined the development of a mathematical model for the interaction of such fluctuations with Langmuir waves, previously published in \citet{RBK}, both in the full 3-D treatment of inelastic scattering, an extension of the previously considered elastic scattering treatment in the literature, and also for the 1-D situation of beam aligned density fluctuations. 

We found that the Langmuir waves will diffuse in wavenumber space. In 3-D this wavenumber diffusion occurs both in angle, as for the elastic scattering case, and also in magnitude, resulting in a significant modification of the parallel wavenumber, that is the wavenumber projected onto the beam direction. When the back reaction of Langmuir waves on the beam electrons is considered, this will result in electron acceleration. 

\subsubsection{Electron Acceleration}
We also expect to see an electron acceleration effect due to beam-aligned density fluctuations. Chapter \ref{ref:Chapter3} presented quasi-linear simulations of the beam-Langmuir wave interactions, including the Langmuir wavenumber diffusion, and confirmed that this indeed occurs. We considered first the test case of a Maxwellian initial beam, as in \citet{RBK}, where the beam distribution is given by $f(\rv)\propto \exp(-(\rv-\rv_b)^2/\Delta \rv_b^2)$ with $\rv_b$ the beam velocity and $\Delta \rv_b$ its width. This is unstable to Langmuir wave generation and thus quickly relaxes to form a plateau. We simulated the effects of Langmuir wavenumber diffusion over a very broad range of parameters for the beam, the background plasma, and the density fluctuations, and considered both random fluctuations, and those obeying a power-law distribution with wavenumber, as observed in the corona.  

Overall, we found that the acceleration depends primarily on the ratio of the timescales for the wavenumber diffusion and for the beam-plasma interaction, with most effect when these are approximately equal. When the wavenumber diffusion is the faster process, we see suppression of the beam-plasma instability, as the waves are removed very rapidly from resonance with the beam electrons, and so no acceleration can occur. The exact spectral shape of the density fluctuations and their characteristic velocity, $\rv_0$, were found to have small additional effects on the extent of the electron acceleration. For example, for a given level of fluctuation intensity, the strongest acceleration occurs when $\rv_0 \sim \rv_{Te}/10$. These effects are not accounted for simply by the ratio of the diffusion and beam-wave interaction timescales. 

\subsubsection{Beam Generated Hard X-Ray Emission}
Next, we considered the case of a power-law electron beam, relaxing due to collisional interactions, as in \citet{KRB}. In very dense plasma, this relaxation can produce instability to Langmuir wave generation significantly more rapidly than the electron time-of-flight effects that dominate in less dense plasma. Our model therefore applies to, for example, a beam in a very dense coronal loop, with background plasma density $n_b \sim 10^{10}$~cm$^{-3}$, over timescales of around 0.1~s or so. Langmuir waves are generated by the beam, but later reabsorbed, and so the electron distribution averaged over the electron's collisional lifetime is unaffected. 

However, when the evolution of the Langmuir waves due either to a density gradient, a fluctuating plasma density, or to wave-wave interactions is considered, the time-averaged electron flux is strongly affected. Energy is shifted from electrons at energies below around 20~keV to the region between 20 and 200~keV. Above 200~keV the effects are small. The time-averaged electron flux is directly related to the hard X-ray emission generated by the beam, which is generally observed only above around 20~keV, as thermal emission dominates below this energy. Thus, the Langmuir wave evolution is seen to shift significant numbers of electrons in energy space from the unobservable part of the distribution to the HXR emitting part. 

The time-averaged electron flux can be increased by an order of magnitude or more at around 100~keV. Thus if we ignore the effects of Langmuir wave generation and evolution and assume an accelerated electron beam evolving purely collisionally, the number of initially accelerated electrons derived from hard X-ray observations will be overestimated by up to this amount. Moreover, it was found that the acceleration effect was strongly dependent on the initial beam density, and so this overestimate is more significant the more electrons there are in the initial beam.

\subsubsection{Radio Emission}
The high levels of Langmuir waves generated by fast electron beams in plasma can also lead to very bright radio emission. The mechanism for this was discussed in the Introduction, and in Chapter \ref{ref:Chapter4} we developed a simulation model for this emission at the second harmonic of the local plasma frequency. Emission also occurs at the plasma frequency, but in dense plasma, as considered here, this is generated inefficiently, and cannot escape as the optical depth is very large. 

Emission at the second harmonic relies on the generation of an initial Langmuir wave population by the electron beam, and the subsequent backscattering of these to negative wavenumbers. A wave from the forwards and backwards population may then coalesce to produce radio emission. However, the emission probability is strongly angularly dependent, and so we must average over assumed distributions of the angles of the Langmuir wavenumbers to the direction of beam propagation. The processes may then be included in our simulations from Chapter \ref{ref:Chapter3}. 

We used this model to simulate the radio emission at 2~GHz for a collisionally relaxing beam in very dense plasma, considering both the backscattering of Langmuir waves due to individual plasma ions, which is important in plasma of equal ion and electron temperatures, and also the decay of a Langmuir wave to an ion-sound wave and a backscattered Langmuir wave, as in Chapter \ref{ref:Chapter3}. This work is currently being prepared for publication \citep{RK}.

We found that the brightness and duration of the simulated radio emission for both homogeneous and weakly inhomogeneous plasma are in good agreement with observations, and moreover the theoretical predictions for the polarisation of the emission in our model also agree well with observed values for high frequency plasma emission. 

However, the process of scattering by plasma ions was seen to be strongly affected by the wavenumber diffusion of Langmuir waves due to plasma density inhomogeneities. Significant suppression of the backscattering, and consequently the radio emission, was seen to occur for fluctuations of very low magnitude. The observed strengths of inhomogeneity in the corona often exceed this limit, and in this case our simulations suggest it is possible to have high levels of beam generated Langmuir waves without generating significant radio emission, as the backscattered wave population remains small. 

On the other hand, when the assumption of equal temperatures is relaxed, the additional backscattering process due to the decay to ion-sound waves can operate, and this leads to high levels of backwards Langmuir waves even in strongly inhomogeneous plasma, and corresponding levels of radio emission. 

\subsubsection{Closing Remarks}
To conclude, we have considered the effects of plasma density inhomogeneities on Langmuir waves, and consequently on propagating fast electron beams, and their hard X-ray and radio emission. Strong effects of electron acceleration and enhancement of the hard X-ray emission were seen. We have also simulated beam-generated radio emission, finding results that are in good agreement with observations. The inclusion of density inhomogeneities in these simulations show that in inhomogeneous plasma, this radio emission can be suppressed, despite the presence of high levels of Langmuir waves. Density inhomogeneities are generally agreed to commonly exist in the corona, and so the generation of Langmuir waves by fast electrons and their subsequent evolution have been shown to be vitally important considerations in the evolution of coronal fast electron beams and their emission.

\newpage

\bibliographystyle{aa1}
\bibliography{refs}

\appendix
\chapter{An Angle-averaged  Model for Fundamental Radio Emission}\label{ref:App1}

The mechanism behind plasma emission at the fundamental was discussed in Section \ref{sec:plasEmm}. Firstly, we must consider the evolution of the Langmuir waves due to their decay $L\rightleftarrows L' + s$ and scattering by ions $L + i\rightleftarrows L' + i'$, as in Chapter 4. In addition we have an analog of this ion-scattering, in which a Langmuir wave is scattered into an electromagnetic wave, $L + i \rightleftarrows t + i'$, and two processes involving ion-sound wave interactions, namely $L + s \rightleftarrows t$, $L \rightleftarrows t + s$. 

Ion-sound waves have been directly observed in the solar wind in correlation with Langmuir waves and Type III bursts \citep[e.g.][]{1977JGR....82..632G,1978JGR....83...58G}. \citet{1986ApJ...308..954L} suggested that the decay to electromagnetic waves $L \rightleftarrows t + s$ was the most likely generating process for this turbulence, but it is now generally thought that Langmuir wave decay $L \rightleftarrows L' + s$ produces the ion-sound waves and these drive the production of fundamental electromagnetic emission \citep{1982SoPh...79..173M,1994ApJ...422..870R}. \citet{1993ApJ...416..831T} show that ion-scattering cannot explain Type III emission at kHz wavelengths, but \citet{2003SoPh..215..335M} show that the brightness of coronal bursts can be explained by the ion-scattering process alone. 

Thus, in a complete theory of fundamental plasma emission in both the corona and solar wind, or more specifically in regions with either $T_i\simeq T_e$ or $T_i \ll T_e$, we must consider both processes. In general, the dominant one is determined by the existence of ion-sound waves, but due to the different functional forms of the two probabilities, their relative efficiencies are also dependent on wavenumber. In this case the two can both be significant, or there can be a transition from one to the other during the emission process.

\section{Ion-Sound Wave Processes}
First we consider the process $L + s \rightleftarrows t$. For fundamental emission, the frequency of the daughter electromagnetic wave must be close to the plasma frequency, and therefore from the dispersion relation $\omega=\left(\omega_{pe}^2 +c^2 k_T^2\right)^{1/2}$ we know the electromagnetic wavenumber must be small. Then considering the energy and momentum conservation conditions, we see that the wavenumbers must satisfy $k_T \ll k_L \simeq -k_S$, and so the Langmuir and ion-sound wave will coalesce very nearly head-on. 

Supposing in addition that the beam-generated Langmuir waves are confined to a small range in angle around the beam direction, the geometry becomes quite simple. The emission probability appearing in Equations \ref{eqn:3Wv3D1} to \ref{eqn:3Wv3D3} for the process $L + s \rightleftarrows t$ is \citep[e.g.][]{1995lnlp.book.....T} \begin{equation}\label{eqn:fundSprob} w^{LST}(\vec{k}, \vec{k}_L,\vec{k}_T)=:C_{fund} \omega_k^S \frac{|\vec{k}_T\times\vec{k}|^2}{k_T^{2}|\vec{k}_L|^2}\delta(\omega_{k_T}^T-\omega^L_{k_L}-\omega^S_k)\end{equation}
\begin{equation}
C_{fund}=\frac{\pi \omega_{pe}^3\left(1+\frac{3T_i}{T_e}\right)}{\omega_{k_T}^T 4n_eT_e} 
\end{equation} where the wavevectors are denoted $\vec{k}_T$ for the electromagnetic wave, $\vec{k}$ for the ion-sound wave and $\vec{k}_L$ for the Langmuir wave. This probability is proportional (via the cross product) to $\sin^2 \theta_{ST}$, where $\theta_{ST}$ is the angle between the parent ion-sound or Langmuir waves and the resulting electromagnetic waves, and is therefore zero for an electromagnetic wave propagating in the same direction as the initial Langmuir wave, and maximised for perpendicular emission. 

On the other hand, given the relatively broad emission probability, the smallness of the electromagnetic wavenumber, and the expected angular spread of the Langmuir waves, we can make the approximation that the electromagnetic emission is almost isotropic. Therefore we combine wedge shaped Langmuir and ion-sound wave spectra with an isotropic electromagnetic wave spectrum. We define \begin{equation} \label{eqn:L1d}W_L(\vec{k})=\frac{1}{\Delta\Omega k^2 } W_L^{\AAvL}(k)\end{equation} within $\Delta \Omega$ the small solid angle occupied by the parent waves, and zero elsewhere, with $W_L^{\AAvL}(k)$ defined by \begin{equation}W_L^{\AAvL}(k)=\int\int k^2 \sin\theta W_L(\vec{k}) \rd\theta \rd\phi,\end{equation} and \begin{equation} W_T(\vec{k})=\frac{1}{4\pi k^2 } W_T^{\AAv}(k)\end{equation} where again we define $W_T^{\AAv}(k)$ by \begin{equation} \label{eqn:T1d} W_T^{\AAv}(k)=\int\int k^2 \sin\theta W_T(\vec{k}) \rd\theta \rd\phi.\end{equation}

Now we write the participating frequencies as $\omega_k^S, \omega^{T}_{k_T},\omega^{L}_{k_L}$ respectively, and rewrite the integral in the general equation \ref{eqn:3Wv3D1} using spherical coordinates, $\vec{k}=(k,\theta,\phi)$ finding
\begin{align}\label{eqn:fundS1}
\frac{dW_T(\vec{k}_T)}{dt}=&\omega_{k_T}^T \int\int\int w^{LST}(\vec{k}, \vec{k}_L,\vec{k}_T)k^2 \sin\theta\times \notag \\&\left[\frac{W_S(\vec{k})}{\omega_{k}^S}\frac{W_L(\vec{k}_L)}{\omega_{k_L}^L}-\frac{W_S(\vec{k})}{\omega_{k}^S}\frac{W_T(\vec{k}_T)}{\omega_{k_T}^T}-\frac{W_T(\vec{k}_T)}{\omega_{k_T}^T} \frac{W_L(\vec{k}_L)}{\omega_{k_L}^L}\right] \rd k\rd\theta \rd\phi
\end{align}
with probability given above by Equation \ref{eqn:fundSprob}. The momentum conservation condition $\vec{k}_L+\vec{k}=\vec{k}_T$ has been used to implicitly perform the integral over $\vec{k}_L$, and we note also the energy conservation condition $\omega_{k_T}^T-\omega^L_{k_L}-\omega^S_k=0$ in the probability.

Now we substitute our definitions of the \AATxt spectral energy densities, Equations \ref{eqn:L1d} to \ref{eqn:T1d} and perform the angular integrals. For the first term in the square brackets we require
\begin{equation}\int \int k^2 \sin \theta  \sin^2\theta_{ST} \frac{W_S(\vec{k})}{\omega_k^S} \frac{W_L(\vec{k}_L)}{\omega_{k_L}^L}  \rd\theta \rd \phi.\end{equation} Substituting for $W_L,W_S$ gives \[\frac{W_S^{\AAvL}}{\Delta\Omega \omega_{k}^S}\frac{W_L^{\AAvL}}{k_L^2 \Delta\Omega \omega_{k_L}^L}\int \int_{\Delta\Omega} \sin\theta  \sin^2\theta_{ST} \rd\theta \rd\phi, \] and we assume the average value of $\sin^2\theta_{ST}$ is well defined and given by \begin{equation}\int \int_{\Delta\Omega} \sin\theta  \sin^2\theta_{ST} \rd\theta \rd\phi=\Delta\Omega \langle \sin^2 \theta_{ST}\rangle\end{equation} to obtain \[\frac{W_S^{\AAvL}(k)}{\omega_{k}^S} \frac{W_L^{\AAvL}(k_L)}{k_L^2\Delta\Omega \omega_{k_L}^L} \langle \sin^2 \theta_{ST}\rangle \] assuming that the Langmuir and ion-sound wave vectors both lie within $\Delta\Omega$.

Similarly, for the second term in square brackets we find
\[\int \int k^2 \sin \theta  \sin^2\theta_{ST} \frac{W_S(\vec{k})}{\omega_{k}^S} \frac{W_T(\vec{k}_T)}{\omega_{k_T}^T} \rd\theta \rd \phi= \frac{W_S^{\AAvL}}{\omega_k^S} \frac{W_T^{\AAv}(k_T)}{4\pi k_T^2 \omega_{k_T}^T} \langle \sin^2\theta_{ST}\rangle \] and similar for the third.

We have \begin{equation}\frac{d}{dt}W^{\AAv}(k_T)=4\pi k_T^2 \frac{d}{dt}W(\vec{k}_T)  \end{equation} and so substituting the results just found for the terms in square brackets in Equation \ref{eqn:fundS1} we get
\begin{align}\frac{dW^{\AAv}_T(k_T)}{dt}=& \omega_{k_T}^T C_{fund}\langle\sin^2\theta_{ST}\rangle \int \rd k \omega_k^S \delta(\omega_{k_T}^T-\omega^L_{k_L}-\omega^S_k)\times \notag\\& \left[\frac{W_S^{\AAvL}(k)}{\omega_{k}^S} \frac{k_T^2}{k_L^2\Delta\Omega} \frac{W_L^{\AAvL}({k}_L)}{\omega_{k_L}^L}-\frac{W_T^{\AAv}({k}_T)}{\omega_{k_T}^T}\left( \frac{ W_S^{\AAvL}(k)}{\omega_k^S}  +\frac{W_L^{\AAvL}(k_L)}{\omega_{k_L}^L}\right)\right]. \end{align} 

Now we perform the integral over $k$ using $\delta(\omega_{k_T}^T-\omega^L_{k_L}-\omega^S_k)$\footnote{For a delta function of a function, the integral is \[\int dx g(x) \delta(f(x)) = \frac{g(a)}{|f'(a)|} \;, \; f(a)=0\]}, which gives a factor of $\omega_{pe}/(3\rv_{Te}^2 k)$ assuming that $k_T \ll k$ and  $k\simeq k_L$. 

Finally, we evaluate the average $\langle\sin^2\theta_{ST}\rangle$ over a sphere, which gives a value of $1/2$ and so we obtain
\begin{align}\label{eqn:FundFinal}\frac{dW^{\AAv}_T(k_T)}{dt}=&\frac{\pi \omega_{pe}^4\rv_s \left(1+\frac{3T_i}{T_e}\right)}{24 \rv_{Te}^2n_eT_e}\times\notag \\ &\left[\frac{W_S^{\AAvL}(k)}{\omega_{k}^S} \frac{k_T^2}{k_L^2 \Delta\Omega} \frac{W_L^{\AAvL}({k}_L)}{\omega_{k_L}^L}-\frac{W_T^{\AAv}({k}_T)}{\omega_{k_T}^T}\left( \frac{ W_S^{\AAvL}(k)}{\omega_k^S}  +\frac{W_L^{\AAvL}(k_L)}{\omega_{k_L}^L}\right)\right] \end{align} with the \AATxt spectral energy densities given by Equations \ref{eqn:L1d} to \ref{eqn:T1d}.

A few restrictions should be noted on the equation just derived. Firstly, some angular spread in \emph{either} the Langmuir or ion-sound wave spectrum is required to supply the small transverse momentum of the electromagnetic wave. Secondly, the assumption of isotropy of the electromagnetic emission is improved as this angular spread increases.

\section{The Crossed Process}
In addition to the process $L + s \rightleftarrows t$ just considered, there is a related process $L \rightleftarrows s + t$, called the ``crossed'' process because the $S$ wave crosses from one side of the equation to the other. Alternately, we can write both together as $L\pm s \rightleftarrows t$. In this case, the probabilities are the same, but there are some sign changes. Firstly, in the wavenumber matching we have now $\vec{k}_L=\vec{k}_T+\vec{k}$  and so the Langmuir and ion-sound wave are now approximately parallel and in the same direction, secondly the frequency matching condition becomes $\delta(\omega_{k_T}^T-\omega^L_{k_L}+\omega^S_k)$, and finally the general equation now reads \begin{equation} \left[\frac{W_S(\vec{k})}{\omega_k^S}\frac{W_L(\vec{k}_L)}{\omega_{k_L}^L}-\frac{ W_S(\vec{k})}{\omega_k^S} \frac{ W_T(\vec{k}_T)}{\omega_{k_T}^T} +\frac{W_T(\vec{k}_T)}{\omega_{k_T}^T} \frac{W_L(\vec{k}_L) }{\omega_{k_L}^L}\right]\end{equation}

Following the exact same procedures as in the previous section gives us the equation 
\begin{align}\label{eqn:FundFinalCross}\frac{dW^{\AAv}_T(k_T)}{dt}=&\frac{\pi \omega_{pe}^4\rv_s\left(1+\frac{3T_i}{T_e}\right)}{ 24 \rv_{Te}^2n_eT_e}\times\notag \\ &\left[\frac{W_S^{\AAvL}(k)}{\omega_{k}^S} \frac{k_T^2}{k_L^2\Delta\Omega} \frac{W_L^{\AAvL}({k}_L)}{\omega_{k_L}^L}-\frac{W_T^{\AAv}({k}_T)}{\omega_{k_T}^T}\left( \frac{ W_S^{\AAvL}(k)}{\omega_k^S}  -\frac{W_L^{\AAvL}(k_L)}{\omega_{k_L}^L}\right)\right] \end{align} where the final term has changed sign with respect to Equation \ref{eqn:FundFinal}.

The most significant difference between this process and the previous one is the condition for saturation, $d W(k_T)/dt =0 $. Because of the sign change, saturation can occur only if \begin{equation}W_T=\frac{k_T^2}{k_L^2 \Delta\Omega}\frac{W_L W_S}{W_S-W_L}\end{equation} which requires that $W_S > W_L$. Because in general the energy in Langmuir wave decay produced ion-sound waves is a small fraction of that in the parent L waves, this condition is unlikely to be satisfied, and thus the electromagnetic waves can continue to grow. 

Also of note is the fact that, due to the differences in the wave-number conservation equations, a population of ion-sound waves can satisfy the matching for either or both of the processes $L\pm s \rightleftarrows t$. In the Type III bursts in the solar wind, decay of a Langmuir wave to an electromagnetic wave plus an ion-sound wave is likely to be dominant \citep{1994ApJ...422..870R}.

\section{Equations for L and S wave evolution}
In general, the energy in the electromagnetic waves is a very small fraction of that in the parent Langmuir waves, and so the effects of plasma emission at either the fundamental or the harmonic on their evolution can be neglected. For ion-sound waves, the situation is less clear, as their generation due to $L\rightleftarrows t + s$ and due to Langmuir wave scattering $L\rightleftarrows L' + s$ can be comparable. Therefore we must generally include the effects of the $L \pm s\rightleftarrows t$ processes on the ion-sound waves. In 3-D we have:

\begin{align}
\frac{dW_S(\vec{k})}{dt}= &- \omega_{k}^S \int\int\int w^{LST}(\vec{k}, \vec{k}_L,\vec{k}_T)k_L^2 \sin\theta_L\times\notag \\ &\left[\frac{W_S(\vec{k})}{\omega_{k}^S}\frac{W_L(\vec{k}_L)}{\omega_{k_L}^L}-\frac{W_S(\vec{k})}{\omega_{k}^S}\frac{W_T(\vec{k}_T)}{\omega_{k_T}^T}-\frac{W_T(\vec{k}_T)}{\omega_{k_T}^T} \frac{W_L(\vec{k}_L)}{\omega_{k_L}^L}\right] \rd k_L\rd\theta_L \rd\phi_L
\end{align} which we treat exactly as for the electromagnetic equation and obtain \begin{align}\frac{dW^{\AAvL}_S(k_S)}{dt}=&\frac{\pi \omega_{pe}^4\rv_s\left(1+\frac{3T_i}{T_e}\right)}{ 24 \rv_{Te}^2n_eT_e} \times\notag \\ &\left[\frac{W_S^{\AAvL}(k)}{\omega_{k}^S} \frac{W_L^{\AAvL}({k}_L)}{\omega_{k_L}^L}-\frac{W_T^{\AAv}({k}_T)}{\omega_{k_T}^T}\left( \frac{ W_S^{\AAvL}(k)}{\omega_k^S}  \pm\frac{W_L^{\AAvL}(k_L)}{\omega_{k_L}^L}\right)\right] \end{align} 
where the sign is ``$+$'' for the $L + s \rightleftarrows t$ process and ``$-$'' for $L \rightleftarrows s + t$. 

\section{Ion Interactions}

In the originally proposed plasma emission mechanism \citep{1958SvA.....2..653G} fundamental emission was produced via the interaction with plasma ions, $L+i\rightleftarrows t+i'$. The equation for evolution of the electromagnetic waves in 3-D is:
\begin{align}\label{eqn:IS3d}\frac{d W_T(\vec{k}_T)}{d t}=&\int w_i^{LT}(\vec{k}_T, \vec{k}_L)\times\notag\\&\left[\frac{\omega_T}{\omega_L}W_L(\vec{k}_L)-W_T(\vec{k}_T)-\frac{(2\pi)^3}{T_i}\frac{\omega_T-\omega_L}{\omega_L}W_T(\vec{k}_T)W_L(\vec{k}_L)\right] \frac{\rd \vec{k}_L}{(2\pi)^3}\end{align} with probability \begin{equation}w_i^{LT}(\vec{k}_T, \vec{k}_L)= \frac{C_{ion} \sin^2\theta_{LT}}{|\vec{k}_T-\vec{k}_L|} \exp{\left(-\frac{(\omega_T-\omega_L)^2}{2|\vec{k}_T-\vec{k_L}|^2 \rv_{Ti}^2}\right)} \end{equation} and \begin{equation}C_{ion}=\frac{\sqrt{\pi} \omega_{pe}^2}{2 n_e \rv_{Ti} (1+T_e/T_i)^2}\end{equation}
where $\theta_{LT}$ is the angle between the initial Langmuir wave and the scattered electromagnetic wave. 

In general this process is slower than the 3-wave processes involving ion-sound waves, but if we consider a regime where $T_i\simeq T_e$ and ion-sound waves cannot exist, then the ion-scattering process is the only possibility. The ion-scattering probability is greatest when $T_i \ge T_e$, and falls rapidly for $T_i\ll T_e$. 

The ion-scattering probability is dipolar and in this case we make the approximation that scattering tends to produce electromagnetic waves perpendicular to the initial Langmuir wave. Like the previous section, we take the L waves as constant over a small solid angle, and assume the electromagnetic emission is also constant over the same area, a reasonable assumption given the slow variation of the probability around $\theta_{LT}\simeq \pi/2$. We write \begin{equation}\label{eqn:IonL}W_L(\vec{k})=\frac{1}{\Delta\Omega k^2 } W_L^{\AAvL}(k) \;\mathrm{with}\; W_L^{\AAvL}(k)=\int\int k^2 \sin\theta W_L(\vec{k}) \rd\theta \rd\phi\end{equation} for $\vec{k}$ within $\Delta\Omega$ and zero elsewhere, and \begin{equation}\label{eqn:IonT} W_T(\vec{k}_T)=\frac{1}{\Delta\Omega k_T^2 } W_T^{\AAv}(k_T).\end{equation}  

Due to the exponential factor in the probability, scattering is negligible unless $\omega_L\simeq \omega_T$ and so we have $k_T\ll k_L$ and $|\vec{k}_T-\vec{k}_L|\simeq k_L$. Then the only angle dependent factor in the probability $w^i_{LT}$ is from $\sin^2\theta_{LT}$. Writing $\vec{k}_L=(k_L, \theta, \phi)$ and $\vec{k}_T=(k_T,\theta_T,\phi_T)$ we require \begin{equation}\label{eqn:require}\int\int \rd\phi \rd\cos\theta \sin^2\theta_{LT}\left[\frac{\omega_T}{\omega_L}W_L(\vec{k}_L)-W_T(\vec{k}_T)-\frac{(2\pi)^3}{T_i}\frac{\omega_T-\omega_L}{\omega_L}W_T(\vec{k}_T)W_L(\vec{k})\right]. \end{equation} Substituting the definitions in Equations \ref{eqn:IonL} and \ref{eqn:IonT} and defining the angle average of $\sin^2\theta_{LT}$ as in the previous section by \begin{equation}\int \int_{\Delta\Omega} \rd\cos\theta \rd\phi \sin^2\theta_{LT}=\Delta\Omega \langle \sin^2 \theta_{LT}\rangle, \end{equation} Equation \ref{eqn:require} becomes \begin{equation} \left[\frac{\omega_T}{\omega_L}\frac{W_L^{\AAvL}(k_L)}{k_L^2}-\frac{W_T^{\AAv}(k_T)}{k_T^2}-\frac{(2\pi)^3}{T_i}\frac{\omega_T-\omega_L}{\omega_L}\frac{W_T^{\AAv}(k_T)W_L^{\AAvL}(k)}{\Delta\Omega k_L^2k_T^2}\right]\langle \sin^2 \theta_{LT}\rangle\end{equation}

Now we substitute the definition of $W_T^{\AAv}(k_T)$ into the LHS of Equation \ref{eqn:IS3d} and take $\langle \sin^2 \theta_{LT}\rangle =1/2$ to find
\begin{align}\label{eqn:FundFinalIon}\frac{d W_T^{\AAv}(k_T)}{d t}=&\int \rd k_L \frac{C_{ion}}{2 k_L} \exp{\left(-\frac{(\omega_T-\omega_L)^2}{2k_L^2 \rv_{Ti}^2}\right)}\times \notag\\&\left[\frac{\omega_T}{\omega_L}\frac{W_L^{\AAvL}(k_L)}{k_L^2}-\frac{W_T^{\AAv}(k_T)}{k_T^2}-\frac{(2\pi)^3}{T_i}\frac{\omega_T-\omega_L}{\omega_L}\frac{W_T^{\AAv}(k_T)W_L^{\AAvL}(k_L)}{\Delta\Omega k_L^2k_T^2}\right]\end{align}  with the \AATxt spectral energy densities given by Equations \ref{eqn:IonL} and \ref{eqn:IonT}.

As for the Langmuir wave case, we have an integral to evaluate numerically, but again we can restrict to only a small range in $k_L$, in this case defined by $\omega_L \simeq \omega_T$.

\end{document}